\documentclass[a4paper,11pt]{article}
\pdfoutput=1 
\RequirePackage{snapshot}  

\usepackage{jheppub} 

\usepackage{bm,amsmath,amssymb,slashed,graphicx,%
            enumerate,alltt,xspace,multirow,xcolor,mathrsfs}
\usepackage{fancyvrb}
\usepackage{booktabs}
\usepackage{graphicx}
\usepackage{subcaption}
\usepackage{xspace}
\usepackage[utf8]{inputenc}
\usepackage[export]{adjustbox}
\usepackage{ulem}

\makeatletter
\g@addto@macro\bfseries{\boldmath}
\makeatother

\definecolor{labelkey}{rgb}{0,0.5,0.0}

\usepackage{listings}
\definecolor{semiblue}{rgb}{0.3,0.3,0.8}
\newcommand{\logbook}[2]{}

\definecolor{darkgreen}{rgb}{0,0.7,0}
\definecolor{ddarkgreen}{rgb}{0,0.5,0}
\definecolor{grey}{rgb}{0.5,0.5,0.5}
\definecolor{orange}{rgb}{1.0,0.4,0.4}
\definecolor{cyan}{rgb}{0.0,1.0,1.0}
\definecolor{magenta}{rgb}{1.0,0.0,1.0}

\newcommand{\blue}{\color{blue}}
\newcommand{\black}{\color{black}}

\newcommand{\cut}{\text{cut}}
\newcommand{\kT}{k_t}

\newcommand{\itilde}{{\tilde \imath}}
\newcommand{\jtilde}{{\tilde \jmath}}

\newcommand{\obs}{\text{obs}}
\newcommand{\PS}{\text{PS}}

\newcommand{\mus}{\mu\text{s}}

\newcommand{\order}[1]{{\cal O}\left(#1\right)}
\newcommand{\as}{\alpha_s}

\newcommand{\ptilde}{{\widetilde p}}

\newcommand{\nc}{N_\text{\textsc{c}}}

\newcommand{\qbar}{{\bar q}}

\newcommand{\DL}{DL\xspace}
\newcommand{\DLLC}{DL-LC\xspace}
\newcommand{\DLNLC}{DL-NLC\xspace}

\newcommand{\DLFC}{DL-FC\xspace}

\newcommand{\NDL}{NDL\xspace}
\newcommand{\NDLLC}{NDL-LC\xspace}

\newcommand{\NDLFC}{NDL-FC\xspace}

\newcommand{\NNDL}{NNDL\xspace}
\newcommand{\NNDLLC}{NNDL-LC\xspace}
\newcommand{\NNDLNLC}{NNDL-NLC\xspace}

\newcommand{\LL}{LL\xspace}
\newcommand{\LLLC}{LL-LC\xspace}
\newcommand{\LLNLC}{LL-NLC\xspace}

\newcommand{\LLFC}{LL-FC\xspace}

\newcommand{\NLL}{NLL\xspace}
\newcommand{\NLLLC}{NLL-LC\xspace}

\newcommand{\NLLFC}{NLL-FC\xspace}

\newcommand{\NNLL}{NNLL\xspace}

\newcommand{\meas}{NODS\xspace}

\title{Colour and logarithmic accuracy in final-state parton
  showers}

\newcommand{\OXaff}{Rudolf Peierls Centre for Theoretical Physics, Clarendon Laboratory, Parks Road,
  University of Oxford, Oxford OX1 3PU, UK}
\newcommand{\ASCaff}{All Souls College, Oxford OX1 4AL, UK}

\newcommand{\IPhTAff}{IPhT, Universit\'{e} Paris-Saclay, CNRS UMR 3681,
  CEA Saclay, F-91191 Gif-sur-Yvette, France}

\author[a]{Keith Hamilton,}%
\author[b]{Rok Medves,}%
\author[b,c]{Gavin P.~Salam,}%
\author[b]{Ludovic Scyboz,}%
\author[d]{Gregory Soyez}%

\affiliation[a]{Department of Physics and Astronomy, University College London, London, WC1E 6BT, UK}
\affiliation[b]{\OXaff}
\affiliation[c]{\ASCaff}
\affiliation[d]{\IPhTAff}

\date{Received: date / Accepted: \today}

\abstract{
  Standard dipole parton showers are known to yield incorrect
  subleading-colour contributions to the leading (double) logarithmic
  terms for a variety of observables.
  In this work, concentrating on final-state showers, we present two
  simple, computationally efficient prescriptions to correct this
  problem, exploiting a Lund-diagram type classification of emission
  regions.
  We study the resulting effective multiple-emission matrix elements
  generated by the shower, and discuss their impact on subleading
  colour contributions to leading and next-to-leading logarithms (NLL)
  for a range of observables.
  In particular we show that the new schemes give the correct full
  colour NLL terms for global observables and multiplicities.
  Subleading colour issues remain at NLL (single logarithms) for
  non-global observables, though one of our two schemes reproduces the
  correct full-colour matrix-element for any number of energy-ordered
  commensurate-angle pairs of emissions. 
  While we carry out our tests within the PanScales shower framework,
  the schemes are sufficiently simple that it should be
  straightforward to implement them also in other shower frameworks.
}

\begin{document}
\normalem

\maketitle


\section{Introduction}
\label{sec:intro}

Parton showers are ubiquitous tools in high-energy collider physics.
In recent years, however, it has become clear that differences between
parton showers are among the limiting systematics in many collider
physics applications.
This has motivated multiple efforts to better understand the
consequences of the approximations contained within parton showers,
and to exploit that understanding to guide their further development.

The majority of today's most commonly used showers, in particular
those of the dipole family~\cite{Gustafson:1987rq}, make use of the
idea of approximating QCD as if it had a large number of colours
($\nc$).
Within this approximation one can view each event as a collection of
independent colour dipoles:
each gluon in an event functions as the colour-triplet end of one
dipole and the colour anti-triplet end of another, while each
(anti-)quark is the colour (anti-)triplet end of a single
dipole.
Those dipoles then radiate independently (and incoherently) from each
other.
This makes it relatively straightforward, at each stage of the
showering, to generate a radiation pattern that is correct across
small and large angles in the large-$\nc$ limit.

There are several ongoing efforts to include subleading-$\nc$
corrections, for example
Refs.~\cite{Platzer:2012np,Nagy:2012bt,Nagy:2015hwa,Platzer:2018pmd,Nagy:2019pjp,Forshaw:2019ver,DeAngelis:2020rvq,Hoeche:2020nsx,Holguin:2020oui}.
Including subleading colour corrections in full generality turns out
to be computationally  challenging, because as the parton multiplicity
increases one should keep track of a rapidly-growing number of
possible colour
configurations, with contributions from higher-dimensional colour
representations.
Here we explore a complementary approach, one which connects the
questions of subleading colour and subleading logarithmic
contributions.

To help make things concrete, let us recall the PanScales
criteria~\cite{Dasgupta:2018nvj,Dasgupta:2020fwr} for assessing the
logarithmic accuracy of a shower
\begin{enumerate}
\item We should identify the kinematic configurations for
  which a shower correctly reproduces tree-level squared matrix elements.
  Typically it is useful to discuss this as a function of the
  separation between emissions in a Lund
  diagram~\cite{Andersson:1988gp}.
  We return to this in more detail in
  section~\ref{ref:angular-ordering+lund}.
  
\item We should evaluate the logarithmic accuracy of the shower's
  predictions for a range of common observables.
  Suppose we calculate some property $P(\as,L)$ of an event, where
  $\as$ is the strong coupling at a scale close to the hard scale,
  $Q$, of the event, and $L$, which we take negative throughout this
  paper, is the logarithm of a ratio of scales.
  For example this might be the cross section for events whose thrust
  is larger than $1-e^{-|L|}$, or it might be the number of subjets 
  found when clustering the event with resolution scale
  $Q^2 e^{-2|L|}$.
  There are two ways of classifying logarithmic
  accuracy.
  \begin{enumerate}
  \item For observables that exponentiate (typically event shapes and
    some jet rates), one can organise logarithmically enhanced terms
    as follows~\cite{Catani:1992ua}:
    \begin{equation}
      \label{eq:accuracy-exponentiating}
      \mbox{}\hspace{-1em}
      P(\as,L) = P(\as,0) \exp\!
      \Bigg(\!
        \underbrace{\as^{-1}g_1(\as L)}_\text{\LL}
        \,+\,
        \underbrace{g_2(\as L)}_\text{\NLL}
        \,+\,
        \underbrace{\as g_3(\as L)}_\text{\NNLL} + \cdots
      \!\!\Bigg)
      + \order{e^{-|L|}}.
    \end{equation}
    \LL stands for leading-logarithmic accuracy, \NLL for
    next-to-leading logarithmic, and so forth.
    The N$^k$\LL functions, $\as^{k-1}g_{k+1}(\as L)$, resum terms
    $\as^n L^{n+1-k}$ and may in some cases involve operators rather
    than numbers.
    The \LL function, $\as^{-1}g_1(\as L)$, starts off with a double
    logarithmic term $\as L^2$.
    Certain observables, such as fragmentation functions and energy
    flow into a limited angular region, start only from the $g_2$
    function, (in much of the literature, the $g_2$ function is then
    called \LL; for consistency across the full set of observables, here we
    still call it \NLL).
  \item For other observables, for example subjet multiplicities and
    certain other jet rates, there is no simple
    exponentiation of
    double logarithmic (DL) terms, and one may instead write
    \begin{equation}
      \label{eq:accuracy-non-exponentiating}
      \mbox{}\hspace{-1em}
      P(\as,L) = P(\as,0)
      \Bigg(\!\!
        \underbrace{h_1(\as L^2)}_\text{\DL}
        \,+\,
        \underbrace{\as^{1/2} h_2(\as L^2)}_\text{\NDL}
        \,+\,
        \underbrace{\as h_3(\as L^2)}_\text{\NNDL} + \cdots
      \!\!\Bigg)
      + \order{e^{-|L|}}\!,
    \end{equation}
    where the N$^k$\DL
    function, i.e.\ $\as^{k/2} h_{k+1}(\as L^2)$, resums terms
    $\as^n L^{2n-k}$.
    In other work, this classification is often called N$^k$LL (or
    occasionally N$^k$LL$_\Sigma$). We adopt the N$^k$DL nomenclature
    here to avoid confusion with the N$^k$LL of
    Eq.~(\ref{eq:accuracy-exponentiating}).
  \end{enumerate}
\end{enumerate}
For both the matrix element and observable-resummation logarithmic
accuracy criteria, one may keep track of powers of the number of
colours.
Making the number of colours explicit, the leading-colour (LC) part of
the \LL function, \LLLC, involves terms $\as^n \nc^n L^{n+1}$, while
the next-to-leading colour (NLC) part, \LLNLC, involves terms
$\as^n \nc^{n-2} L^{n+1}$ and so forth.
When needed, we will use the abbreviation FC to explicitly denote
contributions that include the full colour structure.

Standard dipole showers correctly capture the full set of \LLLC terms
(or \DLLC terms, as appropriate for the event property being
measured).
For exponentiating event properties, it is natural to consider values
of the logarithm down to $L\sim -1/\as$ where NLL terms are of order 1
(cf.\ Eq.~(\ref{eq:accuracy-exponentiating})).
Keeping in mind that, numerically, $\as \sim 1/\nc^2 \sim 0.1$, one then
concludes that \LLNLC and \NLLLC terms are of comparable
importance.\footnote{Strictly, the expansion parameters that should be
  compared in the large-$\nc$ limit are $\as \nc/\pi$, which would be
  held constant under the operation of taking $\nc \to \infty$, and
  $1/\nc^2$.
  However, for $\nc=3$, the numerical similarity between the two
  expansion parameters remains.  }
For observables that do not exponentiate, one instead considers values
of the logarithm down to $L\sim -1/\sqrt{\as}$, and with the same
$\as \sim 1/\nc^2$ equivalence, one may take DL-NLC terms to be
comparable to \NNDLLC terms.

Recently there has been significant progress in designing classes of
showers that are \NLL and \NDL (LC) accurate aside from spin
correlations, and in numerically demonstrating that accuracy in
practice~\cite{Dasgupta:2020fwr}
(Refs.~\cite{Nagy:2020rmk,Nagy:2020dvz} instead examine an analytical
approach).
An approach that bears similarities to one of those shower classes was
discussed in Ref.~\cite{Forshaw:2020wrq}.
At this point, to consistently control all first subleading aspects
beyond the \LLLC approximation, it becomes essential to identify
approaches to construct showers that are correct not just at \NLLLC,
but also \LLNLC (such approaches tend also to bring \DLNLC
accuracy).

In this paper, we present two related, simple approaches whose colour
handling goes beyond \LLLC/\DLLC accuracy.
Both approaches are based on the observation that colour coherence (or
equivalently, angular ordering) provides an understanding of the
colour structure for emissions in phase-space regions that involve
disparate angles.
Specifically, when angles are disparate, one can use colour coherence
to identify the colour factor for radiation, either $C_F$ if all
emissions at smaller angles form a net colour (anti-)triplet, or $C_A$
if they form a net colour octet.%
\footnote{Angular ordering is a key feature of the Herwig
  family of showers~\cite{Marchesini:1987cf,Corcella:2000bw,Bellm:2019zci}, which should generate the correct \LLFC terms by
  construction, as well as NLL-FC terms for global observables (though internal cuts in phase-space can complicate the
  picture~\cite{Bewick:2019rbu}).
  However, there are certain classes of \NLLLC terms, those associated
  with non-global logarithms~\cite{Dasgupta:2001sh}, that cannot be
  accounted for in angular-ordered showers~\cite{Banfi:2006gy}, and so
  angular ordered showers do not at this stage appear to provide the
  foundations needed for systematic improvements beyond \LL across all
  classes of logarithmically enhanced terms.
}

The relevant information can be organised with the help of Lund
diagrams.
The potential to use Lund diagrams and colour coherence to understand
the structure of colour assignment in dipole branching was pointed out
long ago by Gustafson~\cite{Gustafson:1992uh}, with a concrete scheme
proposed in Ref.~\cite{Friberg:1996xc} (section 4.2).
However, subsequent dipole showers adopted different schemes, which
have since been found to generate spurious \LLNLC
terms in some cases~\cite{Dasgupta:2018nvj} (cf.\ also
Refs.~\cite{BryanUnpublished,NagySoperUnpublished}). 

The two schemes that we develop, which are computationally efficient,
will achieve \NDLFC and \NLLFC accuracy for multiplicities
and global event shapes respectively,
thus going beyond the accuracy of the scheme proposed in
Ref.~\cite{Friberg:1996xc}.
For non-global logarithms (which start at \NLL in our counting), they
are \NLLFC accurate only up to some fixed order ($\order{\as}$ or
$\order{\as^2}$, depending on the scheme), however they are in good
numerical agreement with the full-colour all-order computation of
Hatta and Ueda~\cite{Hatta:2013iba}, to within the latter's
few-percent accuracy.
For many practical purposes, therefore, it seems that our colour
schemes have sufficient accuracy not just for NDL/NLL showers, but
even as a basis for use in potential future NNDL/NNLL showers (where
the colour terms left out by our schemes will be commensurate with
N$^3$LL leading-colour terms).

This article is structured as follows.
In section~\ref{ref:angular-ordering+lund} we will recall how Lund
diagrams can be used to understand the assignment of $C_F$ and $C_A$
colour factors and then, in sections
\ref{sec:transition-points} and \ref{sec:ME-solution}, introduce two
concrete schemes that can be straightforwardly applied to a range of
21st century dipole showers.
For reference, in section~\ref{sec:cffe-algs} we will briefly review the
standard colour scheme in modern dipole showers. 
Then in section~\ref{sec:ME-tests} we shall carry out a set of
numerical tests, comparing the effective tree-level matrix elements
being generated by our schemes to known exact results in
energy-ordered limits.
In section~\ref{sec:numerical-tests} we will examine a range of
observables, using both standard colour assignment schemes and our new
schemes, comparing the results to known \DLFC and \NDLFC, as well as
\LLFC and \NLLFC expectations.
%

\section{Angular ordering and Lund diagrams}
\label{ref:angular-ordering+lund}

Let us start by elaborating on the first of our two criteria for
logarithmic accuracy, i.e.\ the reproduction of matrix elements in
suitably ordered limits.
It is convenient to use Lund diagrams as a way of visualising the
phase-space (see Ref.~\cite{Dreyer:2018nbf} for a concrete prescription
to construct the Lund diagram from an event's kinematics).
At \LL accuracy, the tree-level matrix elements should be correct for
any number of emissions that are well separated in a Lund diagram in
both the logarithm of transverse momentum ($k_t$) and in rapidity
($\eta = -\ln \tan \theta /2$), which one might call double strong
ordering;
at \NLL accuracy, the tree-level matrix elements should be correct for
any number of emissions that are well separated in at least one
direction in the Lund diagram.
Well-separated means that the distance $D$ between points in the Lund diagram
corresponding to any given pair of emissions should satisfy
$e^{-D} \ll 1$.
The correctness of the matrix element should hold no matter what that
direction is, e.g.\ some may be well separated in rapidity but have
similar $\ln k_t$, while others may be well separated in $\ln k_t$ but
have similar $\eta$ values.
In this article, the only respect in which we will relax this
requirement on (leading-colour) logarithmic accuracy concerns the
treatment of azimuthal correlations in 
collinear splittings, as induced by spin correlations, a topic that we
defer to future work.

Throughout this section and the next ones, we will discuss how we
attribute the correct colour factor for real emissions.
The reader should keep in mind that virtual contributions are also
being implicitly corrected at the same time, a consequence of the unitary
nature of the showers that we consider in this paper.\footnote{The
  discussion of the relation between real and virtual corrections is
  simple until one has four or more partons at commensurate angles;
  from that point onwards one should worry about amplitude-level
  evolution~\cite{Botts:1989kf}, which is beyond the accuracy and
  scope of this article.
  Additionally, when considering both initial and final-state
  emitters, non-trivial $i\pi$ terms enter at amplitude level,
  associated with Coulomb gluons, and these are a source of
  super-leading logarithms and coherence
  violation~\cite{Forshaw:2006fk,Catani:2011st}.
  They have been addressed in the case of initial-final showers in
  Ref.~\cite{Nagy:2019rwb}, but are not relevant for the final-state
  showers considered here.}

For subleading colour effects at \LL accuracy, one only needs to obtain
the correct tree-level matrix element in regions where emissions are
all well separated in rapidity.
In this limit, for radiation at an angle $\theta$, the question of
colour reduces to that of examining the set of partons contained
within a cone of aperture $\theta$ around the dipole end that is
  closer in angle.
If that cone contains a single net quark (or anti-quark), i.e.\
$|n_q - n_\qbar|=1$, then the radiation is associated with a
$C_F = (\nc^2-1)/(2\nc)$ colour factor, while if the cone contains zero
net quarks, the radiation is associated with a $C_A = \nc$ colour
factor.\footnote{%
  Recall that in dipole showers, the $C_A$ colour factor will be
  shared equally between two dipoles.
}
In the limit where all emissions are well separated in rapidity, these
are the only two possible situations, and there will also never be any
partons close to the edge of a given cone (because then the radiation
would end up close in rapidity to an existing parton, keeping in mind
that $g \to q\qbar$ splittings are implicitly collinear up to and
including NLL accuracy).

\begin{figure}
  \centering
  \begin{subfigure}{0.48\textwidth}
    \includegraphics[width=\textwidth]{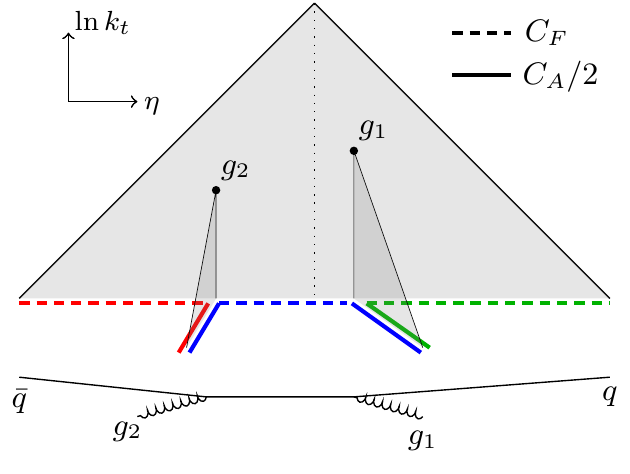}
    \caption{}
    \label{fig:lund-diagrams-simple}
  \end{subfigure}%
  \begin{subfigure}{0.48\textwidth}
    \includegraphics[width=\textwidth]{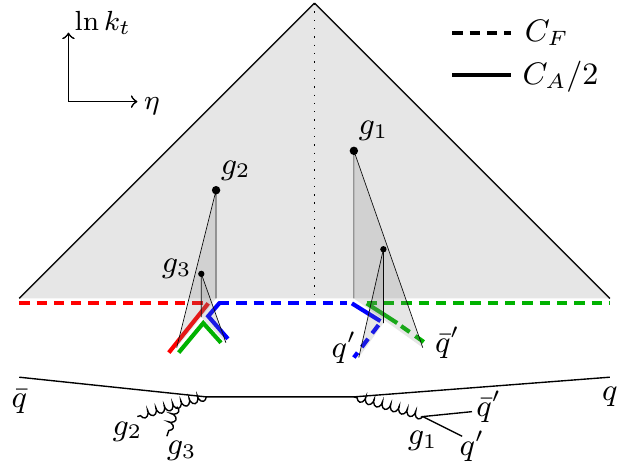}
    \caption{}
    \label{fig:lund-diagrams-complex}
  \end{subfigure}
  \caption{Two example events and associated Lund diagrams.
    Points in the Lund diagram represent branchings, the triangles represent
    the two-dimensional logarithmic phase-space, while the coloured
    lines at the base of the diagram represent the phase-space for
    individual colour-dipoles (at the corresponding  $k_t$ value), with dashed
    and solid segments indicating the parts of the dipole that should be
    associated respectively with a $C_F$ versus a $C_A/2$ colour factor.
  }\label{fig:lund-diagrams}
\end{figure}

Although it is informative to refer to angular cones, it is more
practical, going forwards, to use Lund diagrams.
The angular-ordered colour-assignment prescription is illustrated with
Lund diagrams in Fig.~\ref{fig:lund-diagrams} for two
$e^+e^- \to \bar q q$ events, each dressed with additional radiation.
Recall that the main large triangle corresponds to the phase-space
associated with the original $e^+e^- \to \bar q q$ event (primary Lund
plane), while each ``leaf'' that comes away from the main plane
represents the additional phase-space that becomes available following
emissions from that $q\bar q$ system (for example radiation collinear
to the gluon in a $\qbar q g$ system).
For concreteness we imagine a transverse momentum-ordered shower and
consider the state of the event at a value of the ordering variable
$k_t$ corresponding to the lower edge of the diagram (though our
arguments apply to a range of shower ordering choices).
The phase-space for emission at that $k_t$ is given by the base
of the Lund diagram.

Let us first consider Fig.~\ref{fig:lund-diagrams-simple}.
In a normal leading-$\nc$ picture, this event consists of three colour
dipoles: $\qbar g_2$, $g_2 g_1$ and $g_1 q$, represented as red, blue
and green solid/dashed lines at the base of the Lund diagram.
The dashed and solid styles indicate the colour factor based on
angular ordering, which we work through in the rest of this paragraph.
Along the dashed part of the (red) $\qbar g_2$ dipole, i.e.\ the part on the primary
Lund plane, any subsequent gluon emission is closer in angle to the $\qbar$ than to
$g_2$ and a cone drawn around the $\qbar$ contains just the $\qbar$,
so the colour factor is $C_F$.
Along the solid part of the dipole, i.e.\ the part on the leaf
associated with $g_2$, the cone should be drawn around the gluon
$g_2$, since that is the dipole end that is closer in angle.
The only particle contained within the cone around $g_2$ is the gluon
$g_2$ itself, and one should use a colour factor of $C_A/2$. 

Next, we consider the (blue) $g_2 g_1$ dipole.
In the solid blue region, along $g_2$'s leaf, the cone is to be drawn
around $g_2$ and the only particle that is contained is a gluon, so we
have a $C_A/2$ colour factor; the situation is analogous for the solid
blue region along $g_1$'s leaf.
For the part of the dipole that is dashed, along the primary Lund
plane, we need to separately examine the parts to the left and the right
of the vertical dotted line.
To the left, $g_2$ is the end of the dipole that is closer in angle.
Writing the angle of the radiation, $r$, with respect to $g_2$ as
$\theta_{r g_2}$, and keeping in mind that we are in a region of the
Lund plane were $\theta_{r g_2} \gg \theta_{\qbar g_2}$, the cone of
angle $\theta_{rg_2}$ that we draw around $g_2$ automatically contains
$\qbar$ as well as $g_2$, so the colour factor is $C_F$.
The situation is similar to the right of the vertical dotted line, but
with a cone containing $g_1$ and $q$.
Finally the (green) $g_1 q$ dipole can be understood in analogy with
the (red) $\qbar g_2$ dipole.

Based on the above reasoning it is relatively straightforward to see
that if one has an arbitrarily large number of gluon emissions
(including secondary gluon emissions, but no $g\to q\bar q$
splittings), then to determine the colour factor, one should identify
whether an emission is on the primary Lund plane and if it is assign a
$C_F$ colour factor, otherwise assign a $C_A/2$ colour factor.
That was the key observation made long ago by
Gustafson~\cite{Gustafson:1992uh}.
A specific scheme (based on boosts) for achieving this was
incorporated~\cite{Friberg:1996xc} as a modification of the Ariadne
parton shower~\cite{Andersson:1988gp,Lonnblad:1992tz}.

The next question one may ask is what happens if one allows for $g\to
\qbar q$ splittings.
Whether one should consider this to be part of the \LL terms is a matter
that one may debate: on one hand $g\to \qbar q$ splittings can exist
in double-strongly ordered configurations.
On the other hand if one integrates over their phase-space (as is
relevant for thinking about logarithmically enhanced contributions to
common observables), each $g\to \qbar q$ splitting contributes only a
single logarithm, i.e.\ an effect that is at most \NLL in our counting.
Here, we take the view that we should favour the (slightly)
more ambitious goal and so aim to obtain the correct colour factors
for double-strongly ordered configurations including $g\to \qbar q$
splittings.
Such a configuration is shown in Fig.~\ref{fig:lund-diagrams-complex} and
observing the (blue) $g_3 q'$ dipole, one sees
that it consists of a sequence of solid ($C_A/2$) and dashed ($C_F$) segments,
with the final dashed segment along the $q'$ leaf, which is not part of the
primary Lund plane.
In general the number of segments on a dipole can be anywhere between
$1$ (e.g.\ the solid green $g_2g_3$ dipole) and infinity, though in
practice the average number of segments per dipole turns out to be of
order $1$ even in high-multiplicity events, a consequence of the
smaller number of quarks than gluons produced in the parton shower.
This picture differs from the standard approach within dipole showers
of breaking every dipole into two parts, each associated with one of
the two ends.
Note that in this paper we will apply our new segmentation just for
the purposes of colour assignment, retaining the default two-part
segmentation for the kinematic maps of each shower that we examine.%
\footnote{
  One could also imagine constructing a shower with global recoil
  whose kinematic map follows the pattern of the colour map.
  Indeed, we wonder whether this might be the most physical approach
  to constructing a global recoil map, avoiding certain ugly features
  of existing global
  maps~\cite{Nagy:2008eq,Dasgupta:2020fwr,Forshaw:2020wrq}.
  However we leave the investigation of this question to future work.
}

\section{A solution with segments and transition points}
\label{sec:transition-points}

Based on the reasoning in section~\ref{ref:angular-ordering+lund},
here we propose the first of our concrete schemes for achieving \LL
full-colour accuracy.
We start by introducing the key ideas with the help of a worked
example, section~\ref{sec:transition-worked}.
We then discuss choices we can make that affect aspects beyond \LLFC
accuracy, section \ref{sec:transition-eta-defs}, or that arise in
occasional special cases, section \ref{sec:special-cases} (some
readers may prefer to skip these parts).
In section~\ref{sec:transition-algorithm} we give our full algorithm.
Finally in section \ref{sec:use-in-pythia} we show how it can be
adapted to the Pythia~8 shower~\cite{Sjostrand:2004ef,Sjostrand:2014zea}.

\subsection{A worked example}
\label{sec:transition-worked}

The key insight from section~\ref{ref:angular-ordering+lund} is that to
obtain \LL full colour accuracy it is enough to break every dipole into
a suitable sequence of $C_F$ and $C_A$ colour segments.
We label those segments by their extremities in rapidity and the
colour factor along the segment.
For an initial $e^+e^- \to \qbar q$ event we start with a single
dipole consisting of one segment stretching from a rapidity of
$\eta = -\infty$ ($\qbar$ end) to $\eta = +\infty$ ($q$ end),
associated with a $C_F$ colour factor.
We denote this as
\begin{equation}
  \label{eq:2}
  [-\infty,C_F,\infty]_{\bar q q}\,,
\end{equation}
where our notation consists of a sequence of segment boundaries (or,
equivalently, transition points) and segment colour factors.
When a dipole emits a gluon $g_1$, we define the gluon's rapidity
within the dipole, $\eta_{g_1} = \pm|\ln \tan \theta/2|$, in terms of
its angle $\theta$ with respect to the dipole end that is closer in
the event frame.\footnote{For \LL accuracy, shifts of $\eta$ by an
  amount of order $1$ do not have an impact; below in
  section~\ref{sec:transition-eta-defs} we will discuss the
  constraints that arise for certain aspects of \NLL accuracy, notably
  for parton and jet multiplicities.}
We assign a positive (negative) rapidity if it is closer to the
triplet (anti-triplet) end.
Since there is only a $C_F$ segment in the $\bar q q$ dipole, the
gluon is necessarily emitted with a $C_F$ colour factor.

\begin{figure}
  \centering
  \begin{subfigure}{0.33\textwidth}
    \includegraphics[width=\textwidth]{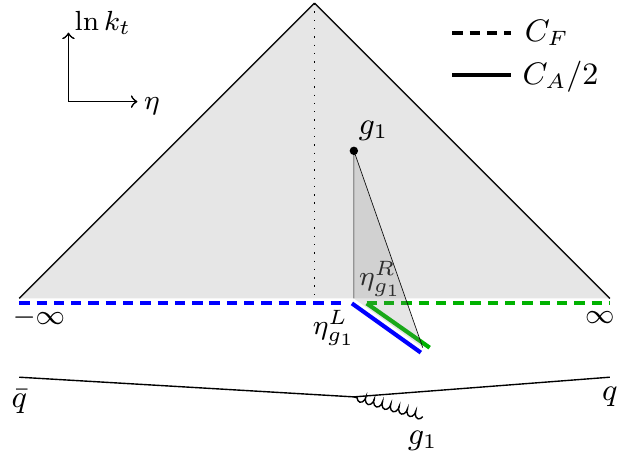}
    \caption{}
    \label{fig:lund-diagram-g1}
  \end{subfigure}%
  \begin{subfigure}{0.33\textwidth}
    \includegraphics[width=\textwidth]{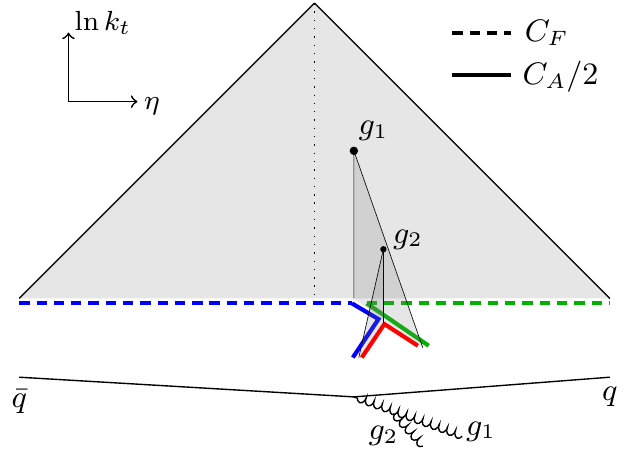}
    \caption{}
    \label{fig:lund-diagram-g2}
  \end{subfigure}%
  \begin{subfigure}{0.33\textwidth}
    \includegraphics[width=\textwidth]{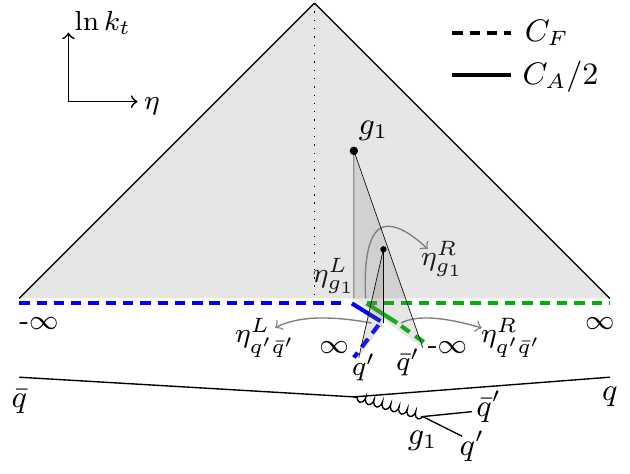}
    \caption{}
    \label{fig:lund-diagram-qqbar}
  \end{subfigure}
  \caption{Examples of colour segments and transition points for
    representative configurations: (a) a $q\to qg$ splitting (i.e.\
    the emission of a gluon from a quark segment), (b) a $g\to gg$
    splitting (i.e.\ a gluon emission from a gluon segment), and (c) a
    $g\to q\bar q$ splitting.}
  \label{fig:lund-diagrams-extra}
\end{figure}

When we radiate a gluon $g_1$ from the $\qbar q$ dipole, we end up with
two dipoles, $\qbar g_1$ and $g_1 q$, each of which now has one $C_F$
and one $C_A$ region, cf.\ Fig.~\ref{fig:lund-diagram-g1}.
The segmentation for the two dipoles is as follows
\begin{equation}
  \label{eq:3}
  [-\infty,C_F,
  \blue \eta_{g_1}^L, C_A, \infty]_{\bar q g_1}
    + [-\infty, C_A, \eta_{g_1}^R, \black
  C_F, \infty]_{g_1 q}\,,
\end{equation}
where we have highlighted in blue the extra segments that are
a consequence of the gluon emission and, for now, we take
$\eta_{g_1}^L \equiv \eta_{g_1}^R \equiv \eta_{g_1}$.
We see here that both new dipoles have a transition point at the
rapidity of the emission itself (see Fig.~\ref{fig:lund-diagram-g1}).
Note that since gluons are shared across two dipoles, the second
segment should really have a colour factor $C_A/2$.
For compactness, we suppress the explicit factor $1/2$ in our
notation.

Next, we consider a second emission, $g_2$.
It is sufficient to examine the situation where it is emitted from the
$\qbar g_1$ dipole.
We use the same procedure to evaluate the rapidity $\eta_{g_2}$ of
$g_2$ as given above, but now with respect to the $\qbar g_1$ dipole,
and we examine where $\eta_{g_2}$ lies in the $\qbar g_1$ sequence of
segments.
Since there are two segments, there are two possible cases:
(a) $-\infty < \eta_{g_2} < \eta_{g_1}^L$, where radiation occurs with a
$C_F$ colour factor and (b) $\eta_{g_1}^L < \eta_{g_2} < \infty$ where it
occurs with a $C_A/2$ colour factor.
As well as differing in the colour factor for emission, the two cases
also differ in terms of the resulting set of new segments. 
Case (a), emission from the $C_F$ region, gives the following
segmentation of the resulting three dipoles
\begin{equation}
  \label{eq:4}
  [-\infty,C_F, \color{blue} \eta_{g_2}^L, C_A, \infty]_{\bar q g_2}
  + [-\infty, C_A, \eta_{g_2}^R, \color{black} C_F, \eta_{g_1}^L, C_A, \infty]_{g_2 g_1}
  + [-\infty, C_A, \eta_{g_1}^R, C_F, \infty]_{g_1 q},
\end{equation}
i.e.\ one splits the $\qbar g_1$ sequence of segments in
Eq.~(\ref{eq:3}) into two separate sequences: one is for the
$\qbar g_2$ dipole, which keeps everything to the left of the $C_F$
segment where the radiation occurred, plus a closing
$\color{blue} \eta_{g_2}^L, C_A, \infty]$ right segment; and one for
the $g_2 g_1$ dipole, which keeps everything to the right of the $C_F$
segment, plus a closing $\blue [-\infty, C_A, \eta_{g_2}^R$ left
segment.
This pattern corresponds exactly to what we see in
Fig.~\ref{fig:lund-diagrams-simple}. 
Case (b), insertion into the $C_A$ region, is shown in
Fig.~\ref{fig:lund-diagram-g2} and gives
\begin{equation}
  \label{eq:5}
  [-\infty,C_F, \eta_{g_1}^L, C_A, \blue \infty]_{\bar q g_2}
  + [-\infty,  \black C_A, \infty]_{g_2 g_1} 
  + [-\infty, C_A, \eta_{g_1}^R, C_F, \infty]_{g_1 q}\,.
\end{equation}
Again we have highlighted the new part in blue.
Since we are inserting a gluon into a $C_A$ region there are no
additional $C_F{-}C_A$ transition points.

The final case that we need to consider is the splitting of a gluon to
$ q'\qbar'$, cf.\ Fig.~\ref{fig:lund-diagram-qqbar}.
In a strong ordering (i.e.\ collinear) limit, as relevant for
accuracies up to and including NLL, the $|\eta|$ associated with the
$g \to q'\bar q'$ splitting will always be larger than the last
non-infinite $\eta$ transition point in each dipole's sequence, i.e.\
the splitting will always be associated with a $C_A$ segment that is
at the extreme left or right ends of a sequence.
We consider a $\qbar g_1 q$ event and the branching of $g_1\to q'\bar q'$ in the
$\qbar g_1$ dipole, i.e.\ we have Eq.~(\ref{eq:3}) as our starting point.
Defining
\begin{equation}
  \label{eq:etaLRqqbar}
  \eta^L_{q'\bar q'} = -\eta^R_{q'\bar q'} = |\eta_{q'\bar q'}|\,,
  \qquad
  \eta_{q'\bar q'} \equiv -\ln\tan\theta_{q'\bar q'}/2\,,
\end{equation}
where $\theta_{q'\bar q'}$ is the opening angle of the $q'\bar q'$ pair in the event frame,
we then obtain the following segments
\begin{equation}
  \label{eq:6}
  [-\infty,C_F,
  \eta_{g_1}^L, C_A, \blue \eta^L_{q'\bar q'}, C_F,  \infty]_{\bar q q'}
    + [-\infty, C_F, \eta^R_{q'\bar q'},\black C_A, \eta_{g_1}^R, C_F,
    \infty]_{\bar q' q}.
\end{equation}
i.e.\ the $\qbar g_1$ and $g_1
q$ dipoles have respectively become $\qbar q'$ and $\qbar' q$ dipoles
and each of those dipoles has additional transition points to a $C_F$
colour factor at $\eta^L_{q'\bar q'} = |\eta_{q'\bar q'}|$ and
$\eta^R_{q'\bar q'} = -|\eta_{q'\bar q'}|$ respectively.
The opposite signs for $\eta^L_{q'\bar q'}$ and $\eta^R_{q'\bar q'}$
arise because the transition needs to be at angles close to the $q'$
for the $\qbar q'$ dipole, i.e.\ positive rapidity, and at angles
close to the $\qbar'$ end of the $\qbar' q$ dipole, i.e.\ negative
rapidity.

Before formulating a full, concrete algorithm, some details need to be
specified, related to $\order{1}$ choices for the rapidity boundaries
(section~\ref{sec:transition-eta-defs}), and the handling of cases
where two emissions are close in rapidity, with resulting ambiguities
in how to construct the segments (section~\ref{sec:special-cases}).
%

\subsection{Specific rapidity definitions and \NDL accuracy}
\label{sec:transition-eta-defs}

There are two aspects to examine concerning the exact choices of
rapidity transition points, both relevant for configurations where two
branchings occur at similar angles.
One involves identifying a choice that can provide full-colour \NDL
 (\NDLFC) accuracy
for basic quantities like jet and particle
multiplicities, i.e.\ control of terms $\as^n
L^{2n-1}$.
The control of this class of terms at \NDLFC
 accuracy has
long been one of the strong arguments in favour of angular ordered
showers~\cite{Marchesini:1983bm}.
The other question is more practical: in the segment
algorithm we implicitly assume that all transition points are ordered,
with a consistent alternating set of $C_F$ and $C_A$ segments.
Those properties are trivially maintained when all branchings are at
disparate angles, but that is no longer necessarily the case when two
or more branchings are at commensurate angles.

\begin{figure}
  \centering
  \includegraphics[width=0.5\textwidth]{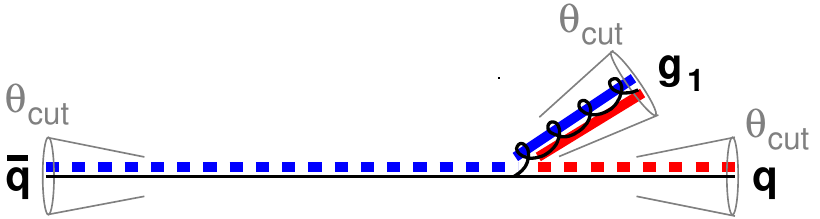}
  \caption{Illustration of the configuration that we consider to
    examine the equivalence between exact angular ordering and our
    partitioning of dipoles, as relevant for \NDL accuracy
    in particle and jet multiplicities.
    We will consider the emission of a further soft gluon, $g_{2}$,
    with respect to this configuration, at an angle of at least
    $\theta_{\text{cut}}$ away from the $\qbar$, $q$ and $g_{1}$.
    The dashed/solid coloured lines indicate schematically where we
    use a $C_F$ v.\ $C_A/2$ colour factor for each of the two dipoles
    (blue: $\qbar g_1$, red $g_1 q$).}
  \label{fig:subleading-config}
\end{figure}

We first consider what is required for \NDLFC
accuracy.
We start with a system $\qbar g_1 q$, with $\theta_{g_1q} \ll 1$, and
examine the emission of a second much softer gluon, $g_2$, in a limited
range of energy, $\Delta\!\ln E$, and with the following angular 
constraints: 
\begin{equation}
  \label{eq:angular-constraints}
  \{\theta_{g_2q}, \theta_{g_2\qbar}, \theta_{g_2 g_1}\} > \theta_\text{cut}\,.
\end{equation}
The angular limits for the emission of $g_{2}$ are represented
schematically in Fig.~\ref{fig:subleading-config} and we work
in a limit where $\theta_\text{cut} \ll \theta_{g_1 q}$.
The integrated full-colour rate of $g_2$ emission in this region is
well known to be~\cite{Dokshitzer:1991wu,Ellis:1991qj}
\begin{subequations}
  \label{eq:ang-ordered}
  \begin{align}
    I_\text{FC} &= \frac{2\as\, \Delta\!\ln E}{\pi} \left[
                       2 C_F \ln \frac{2}{\theta_\text{cut}}
                       + C_A \ln \frac{\theta_{g_1q}}{\theta_\text{cut}} \right],
    \\ 
                     &= \frac{2\as\, \Delta\!\ln E}{\pi} \left[
                       2C_F \eta_\text{cut} + 
                       C_A (\eta_\text{cut} - \eta_{g_1})
                       \right],
  \end{align}
\end{subequations}
where $\eta_{g_1} = -\ln\tan\theta_{g_1 q}/2$,
$\eta_\text{cut} = -\ln \tan \theta_\text{cut}/2$.
The parton shower's ability to reproduce this result is a key
requirement for \NDL full-colour accuracy.
In particular, if the evaluation of this integrated emission rate for
the shower has an extra $\order{1}$ constant in the square bracket, \NDL accuracy
will not be achieved.

To provide a concrete demonstration of how to satisfy
Eq.~(\ref{eq:ang-ordered}), we consider the PanScales showers~\cite{Dasgupta:2020fwr}.
The kinematics for branching in these showers is
parameterised in terms of the shower ordering
variable $\ln v$, a longitudinal variable $\bar \eta$ (linearly
related to the logarithm of a light-cone momentum component) and an
azimuthal angle $\phi$.
The emission density (working at fixed coupling for illustrative
purposes) is
\begin{equation}
  \label{eq:etabar-density}
  C \frac{2\as}{\pi} d\bar \eta \, d\ln v \frac{d\phi}{2\pi}\,,
\end{equation}
where $C$ is the colour factor ($C_F$ or $C_A$ with $C_F=C_A/2$ in the
large-$\nc$ limit).%
\footnote{Recall, e.g., that the PanLocal family of
  mappings~\cite{Dasgupta:2020fwr} for emission of $p_k$ from a dipole
  $\{\ptilde_i,\ptilde_j\}$ is
  \begin{equation} \label{eq:IIc-split-map} 
      p_{k} =a_{k}\tilde{p}_{i}+b_{k}\tilde{p}_{j}+k_{\perp}\,,\qquad
      p_{i} =a_{i}\tilde{p}_{i}+b_{i}\tilde{p}_{j}-fk_{\perp}\,,\qquad
      p_{j} =a_{j}\tilde{p}_{i}+b_{j}\tilde{p}_{j}-(1-f)k_{\perp}\,,
  \end{equation}
  where $k_{\perp}=\kT\left[{n}_{\perp,1}\cos\phi+{n}_{\perp,2}\sin\phi\right]$,
  with ${n}_{\perp,m}^2 = -1$,
  ${n}_{\perp,m}\cdot \ptilde_{i/j}
  = 0$ ($m=1,2$),
  ${n}_{\perp,1}\cdot{n}_{\perp,2} = 0$
  and
  \begin{equation}
    \label{eq:evolution-variable}
    \kT = \rho v e^{\beta|\bar \eta|}\,,
    \qquad
    \rho = \left(\frac{s_{\itilde} s_{\jtilde}}{Q^2 s_{\itilde\jtilde}}\right)^{\frac{\beta}{2}}\,.
  \end{equation}
  Here the parameter $\beta$ sets the choice of ordering variable, $s_{\itilde \jtilde} =2\ptilde_{i}\cdot \ptilde_{j}$,
  $s_{\itilde}=2\ptilde_{i} \cdot Q$, and $Q$ is the total event
  momentum.
  The light-cone components of $p_{k}$ are given by $a_{k}=\sqrt{\frac{s_{\jtilde}}{s_{\itilde\jtilde}s_{\itilde}}}\,
    \kT{e}^{+\bar \eta}$ and $b_{k}=\sqrt{\frac{s_{\itilde}}{s_{\itilde\jtilde}s_{\jtilde}}}\,
    \kT{e}^{-\bar \eta}$.
  The quantity $f$ in Eq.~(\ref{eq:IIc-split-map}) determines how
  transverse recoil is shared between $p_i$ and $p_j$, cf. below.
  The $a_i,b_i,a_j,b_j$ are fully specified by the requirements
  $p^2_{i/j}= 0$, $(p_i+p_j+p_k)=(\ptilde_i+\ptilde_j)$ and
  $p_i = \tilde p_i$ for $\kT \to 0$.
  These and the shower-dependent choices for $f$ are given in the
  supplemental material to Ref.~\cite{Dasgupta:2020fwr}.
  The PanGlobal shower drops the $-f k_\perp$ and $-(1-f)k_\perp$
  terms in Eq.~(\ref{eq:IIc-split-map}), sets $b_i=a_j=0$, and boosts
  and rescales the event after each emission so as to conserve momentum.
}
To integrate the shower over the region of
Eq.~(\ref{eq:angular-constraints}), one needs to relate $\bar \eta$ to
the actual rapidity of a soft and collinear emission from a dipole
$ij$ with respect to the closer of $i$ and $j$.
The following approximate formula (specific to the PanScales showers)
\begin{equation}
  \label{eq:eta-approx}
  \eta_\text{approx}^{(ij)} = \left\{
    \begin{array}{lll}
      \bar \eta - \frac12 \ln \left(\frac{1- \cos\theta_{ij}}{2}\right)\,,
      && \bar\eta > 0\,,
      \\
      \bar \eta + \frac12 \ln \left(\frac{1- \cos\theta_{ij}}{2}\right)\,,
      && \bar\eta < 0\,,
    \end{array}
  \right.
\end{equation}
provides the (signed) rapidity with respect to the closer of $i$ and
$j$ (when the dipole for which we evaluate $\eta_\text{approx}$ is
obvious, we omit the $(ij)$ superscript).
It is approximate in the following sense: for an emission $k$ from an
$ij$ dipole, the true rapidity with respect to the closer of $i$ and
$j$ has additional $\order{1}$ corrections when the smaller of the
$\theta_{ik}$ and $\theta_{jk}$ is commensurate with $\theta_{ij}$.
For a soft emission, those corrections vanish when the ratio of
$\min(\theta_{ik},\theta_{jk})$ to $\theta_{ij}$ is small.
This is the case notably close to the $\theta_\text{cut}$ limits in
Eq.~(\ref{eq:angular-constraints}).

For the configuration in Fig.~\ref{fig:subleading-config},
with $\theta_{g_1 q} \ll 1$ ($\theta_{\qbar g_1} \simeq \pi$), we have
\begin{subequations}
  \label{eq:eta-approx-qbar-g1-q-config}
  \begin{align}
    \text{$\bar q g_1$ dipole:}\quad & \eta_\text{approx}^{(\qbar g_1)} = \bar \eta\,,
    \\
    \text{$g_1 q$ dipole:}\quad &
                                  \eta_\text{approx}^{(g_1 q)} =
                                  \left\{
                                  \begin{array}{lll}
                                    \bar \eta + \eta_{g_1} && \bar\eta > 0\,,
                                    \\
                                    \bar \eta - \eta_{g_1} && \bar\eta < 0\,.
                                  \end{array}
                                                              \right.
  \end{align}
\end{subequations}
Note that $\eta_\text{approx}$ depends linearly on $\bar \eta$ except
at $\bar \eta = 0$, where it is discontinuous for the $g_1 q$ (i.e.\
small-angle) dipole.
With these results one can translate the constraints of
Eq.~(\ref{eq:angular-constraints}) into constraints on $\bar \eta$ for
each of the two dipoles,
\begin{subequations}
  \label{eq:eta-approx-qbar-g1-q-constraints}
  \begin{align}
    \text{$\bar q g_1$ dipole:}\quad
          & -\eta_{\cut} < \bar \eta < \eta_{\cut}\,,
    \\                                       
    \text{$g_1 q$ dipole:}     \quad
          & -\eta_{\cut} + \eta_{g_1} <
                                       \bar \eta < \eta_{\cut} - \eta_{g_1}\,.
  \end{align}
\end{subequations}
For each of the dipoles, the procedure of
section~\ref{sec:transition-worked} splits the dipole's rapidity range
into a $C_F$ piece and a $C_A/2$ piece.
The choice that we make for the transition points is $\bar\eta=\eta_{g_1}$
for the $\qbar g_1$ dipole (using $C_F$ for $\bar\eta < \eta_{g_1}$),
and $\bar\eta = 0$ for the $g_1 q$ dipole (using $C_F$ for
$\bar\eta > 0$).
Together with the limits in
Eq.~(\ref{eq:eta-approx-qbar-g1-q-constraints}), one then finds that
the $C_A/2$ pieces from the two dipoles add up to give a total
$\bar \eta$ range of $2(\eta_{\cut} - \eta_{g_1})$, while the $C_F$
pieces add up to give $2\eta_{\cut}$, giving a total
result for the parton shower of
\begin{equation}
  \label{eq:integral-shower}
  I_\text{PS} = \frac{2\as\, \Delta\!\ln E}{\pi} \left[
    2C_F \eta_{\cut} + 
    C_A (\eta_{\cut} - \eta_{g_1})
  \right],
\end{equation}
in agreement with the full-colour result of
Eq.~(\ref{eq:ang-ordered}).

While the discussion in terms of $\bar \eta$ is useful for locating an
emission with respect to a dipole with a single transition point, to
handle more complicated events, it becomes simpler to reason directly
in terms of $\eta_\text{approx}$.
Thus the transition point for the $\qbar g_1$ dipole is at
$\eta_\text{approx} = \eta_{g_1}$.
For the $g_1 q$ dipole, since $\eta_\text{approx}$ is discontinuous at
$\bar \eta = 0$, any transition between $-\eta_{g_1}$ and $+\eta_{g_1}$
is equivalent.
One simple convention is to set the transition for the $g_1 q$ dipole
to be at $\eta_\text{approx}=0$, i.e.\ along the dipole's angular bisector
in the event frame.

Together, these observations bring us to the following recipe for soft
emission from a configuration with an existing collinear branching.
Working in the event centre-of-mass frame, when inserting a gluon emission $i$ into a $C_F$ segment, determine
its angle $\theta_i$ to the triplet (anti-triplet) end of the dipole
if $\eta_{\text{approx},i} > 0$ ($<0$) and define
$\eta_i=\mp\ln \tan \theta_i/2$, using a negative (positive) sign when the
$\theta_i$ is determined with respect to the triplet (anti-triplet)
end.
In constructing new transition points, e.g.\ as in Eqs.~(\ref{eq:3})
or (\ref{eq:4}), use 
\begin{equation}
  \label{eq:7}
  \eta^L_i = \max\left(0, \eta_i \right)\,,\quad
  \eta^R_i = \min\left(0, \eta_i \right)\,.
\end{equation}
When considering the generation of a new emission $k$, determine its
$\eta_{\text{approx},k}$ from Eq.~(\ref{eq:eta-approx}) and compare that
$\eta_{\text{approx},k}$ to the dipole's existing transition points to
determine the segment to which the emission belongs and the associated
$C_F$ v.\ $C_A/2$ colour factor.

Similar considerations about the total integral for soft gluon
emission are relevant for a $g \to \qbar q$ splitting.
One can demonstrate that the
prescription given in Eqs.~(\ref{eq:etaLRqqbar}) and (\ref{eq:6})
already ensures the correct integral, an analogue of Eq.~(\ref{eq:ang-ordered}).

\subsection{Special cases}
\label{sec:special-cases}

When inserting a gluon $g$ into a $C_F$ segment, and that gluon's
angle with respect to the dipole end is close to an existing
transition point, it can sometimes happen that the insertion
$\eta_g^L$ or $\eta_g^R$ lies outside the segment range (recall that
$\eta_\text{approx}$ and $\eta_\text{exact}$ can differ by an order
$1$ amount when the emission angle is commensurate with the dipole
opening angle).
From a \DL logarithmic point of view, e.g.\ for particle or jet
multiplicities, such a situation is beyond \NDL.
Specifically, to trigger it, one must have two energy-ordered gluons
at similar angles.
In such a situation, any subleading-$\nc$ issue induced by
mis-ordering of transition points only affects emission  of a
subsequent, third, soft gluon that is once again at a similar angle
(it also affects related virtual corrections).
The requirement of three energy-ordered gluons at commensurate angles
corresponds to \NNDL terms for quantities such as multiplicities,
because one loses a logarithm relative to \DL for each of the second
and third gluons when requiring its angle to be similar to the first.

Still, we need a definite prescription to handle such cases.
We choose to give priority to the requirement that the transition
points in the colour-segment sequence should always remain ordered.
Consider the configuration
$[\ldots, C_A, \eta_{i-1}, C_F, \eta_i, C_A\ldots]_{ab}$ and a
situation where we have located a new emission, according to its
$\eta_\text{approx}$, as belonging to the central $C_F$ segment.
We would normally expect to split the sequence as
\begin{subequations}
  \label{eq:new-transition-CF-gone-bad}
  \begin{multline}
    [\ldots, C_A, \eta_{i-1}, C_F, \eta_i, C_A\ldots]_{ab} \to
    [\ldots, C_A, \eta_{i-1}, C_F, \blue \eta_g^L, C_A, \infty]_{ag}
    +\\\blue + [-\infty, C_A, \eta_g^R,\black C_F, \eta_i, C_A,
    \ldots]_{gb}\,,
  \end{multline}
  but if $\eta_g^L \le \eta_{i-1}$,
  we discard the last two segments of the new $ag$ sequence
  ($\eta_{i-1}, C_F, \eta_g^L$ and $\eta_g^L, C_A, \infty$) and extend
  the remaining rightmost $C_A$ segment to $+\infty$,
  \begin{equation}
    \label{eq:new-transition-CF-gone-bad-soln}
    [\ldots, C_A, \eta_{i-1}, C_F, \eta_i, C_A\ldots]_{ab}
    \to 
    [\ldots, C_A, \blue \infty]_{ag} +
    [-\infty, C_A, \eta_g^R,\black C_F, \eta_i, C_A, \ldots]_{gb}\,.
  \end{equation}
\end{subequations}
We use an analogous procedure when $\eta_g^R \ge \eta_i$, dropping the
two first segments of the $gb$ sequence and extending its leftmost
$C_A$ segment to $-\infty$. 
Since $\eta_g^L \ge \eta_g^R$, and since the transition points are
always ordered prior to insertion, this adaptation is needed at most
on one side, never both.

Similar considerations apply to $g \to q\bar q$ splittings, though the
impact on accuracy is now only expected to be N$^3$\DL, because
$g \to q\bar q$ splittings start one logarithm down.
Where normally we would have
\begin{subequations}
  \label{eq:new-transition-qgg-gone-bad}
  \begin{multline}
    [\ldots, C_F, \eta_i, C_A, \infty]_{ag} + [-\infty, C_A, \eta_j,
    C_F, \ldots]_{gb}
    \to\\
    [\ldots, C_F, \eta_i, C_A \blue, \eta^L_{q'\bar q'}, C_F,
    \infty]_{aq'} + [-\infty, C_F, \eta^R_{q'\bar q'}, \black C_A,
    \eta_j, C_F,\ldots]_{\qbar' b}\,,
  \end{multline}
  if $\eta_{q'\qbar'}^L \le \eta_i$, then we remove the last two
  segments of the new $aq'$ sequence and extend its remaining rightmost
  ($C_F$) segment to infinity,
  \begin{multline}
    \label{eq:new-transition-qgg-gone-bad-soln}
    [\ldots, C_F, \eta_i, C_A, \infty]_{ag} + [-\infty, C_A, \eta_j,
    C_F, \ldots]_{gb}
    \to\\
    [\ldots, C_F, \blue \infty]_{aq'}
    + [-\infty, C_F, \eta^R_{q'\bar q'}, \black C_A, \eta_j, C_F,\ldots]_{\qbar'b}\,.
  \end{multline}
\end{subequations}
There is an analogous adaptation when $\eta^R_{q'\bar q'} \ge
\eta_j$. 

In practice, for physical $\as$ values, these adaptations are used in
at most a few percent of gluon insertions in $C_F$ segments and for up
to $10{-}20\%$ of $g \to q\qbar$ splittings.

\subsection{The algorithm}
\label{sec:transition-algorithm}

With the help of the above reasoning we are now ready to formulate a
full segment-based algorithm for the PanScales showers.
The event should start with Eq.~(\ref{eq:2}) for an initial $e^+e^-
\to \qbar q$ event and analogues with $C_A$ for the two dipoles of an
initial $H \to gg$ event.
Splitting functions and their integrals for Sudakov factors should by
default all be evaluated with the replacement $C_F \to C_A/2 = \nc/2$.
Dipoles are always labelled by their anti-triplet end followed by
their triplet end.

For emission of a gluon $g$ from an $ab$ dipole, use the parent dipole
momenta and PanScales kinematic generation variable $\bar \eta$ to
determine $\eta_{\text{approx}}^{(ab)}$ for the gluon, according to
Eq.~(\ref{eq:eta-approx}). 
Identify the location of $\eta_{\text{approx}}^{(ab)}$ within the $ab$
dipole's sequence of segments.
If it is in a $C_F$ segment, reject the emission with probability
$(1 - 2C_F/C_A)$ (or, alternatively, multiply the event weight by
$(1 - 2C_F/C_A)$).
If it is rejected, continue by generating the next value for the
shower ordering variable, starting from the value that was been
rejected. 
If the emission is accepted, and $\eta_{\text{approx}}^{(ab)} > 0$ ($<0$)
evaluate $\eta_g = -\ln \tan \theta_{gb}/2$ ($\ln \tan \theta_{ag}/2$).
Determine $\eta_g^L$ and $\eta_g^R$ from $\eta_g$ according to
Eq.~(\ref{eq:7}), and replace the $ab$ dipole's sequence of segments
with new sequences for the $ag$ and $gb$ dipoles as follows,
\begin{equation}
  \label{eq:new-transition-CF}
  [\ldots, C_F, \ldots]_{ab}
  \to 
  [\ldots, C_F \blue, \eta_g^L, C_A, \infty]_{ag} +
  [-\infty, C_A, \eta_g^R,\black C_F, \ldots]_{gb}\,.
\end{equation}
The left and right ``$\ldots$'' represent the full sequences to the
left and right ends of the original $C_F$ factor.
They are assigned respectively to the left and right ends of the $ag$
and $gb$ dipoles' sequences.
If $\eta_g^L$ is such that its insertion would create a disordered
sequence of transition rapidities, remove the last two segments in the
$ag$ dipole, i.e.\ use
Eq.~(\ref{eq:new-transition-CF-gone-bad-soln}). 
Proceed analogously for $\eta_g^R$.

If the emitted gluon's $\eta_{\text{approx},g}$ is in a $C_A$ segment,
do not apply any additional rejection factor (i.e. generate the gluon
with its original $C_A/2$ colour factor) and replace the $ab$ dipole
sequence of segments as follows.
\begin{equation}
  \label{eq:new-transition-CA}
  [\ldots, C_A, \ldots]_{ab}
  \to 
  [\ldots, C_A \blue, \infty]_{ag} +
  [-\infty, \black C_A, \ldots]_{gb}\,,
\end{equation}
where the $C_A$ on the left corresponds to the segment associated with
$\eta_{\text{approx},g}$.

Finally we consider $g \to q' \qbar' $ splitting, where the gluon
belongs to two dipoles $ag$ and $gb$.
We assume the splitting to have been generated with a normal
$P_{g \to q \qbar}$ splitting function, so there is no rejection
factor to apply.
Using the opening angle $\theta_{q'\qbar'}$ between the new quark and
anti-quark, determine
$\eta^L_{q'\bar q'} = -\eta^R_{q'\bar q'} = |\ln \tan
\theta_{q'\qbar'}/2|$, i.e.\ Eq.~(\ref{eq:etaLRqqbar}).
The $g \to q'\qbar'$ splitting then acts as follows on the segments:
\begin{equation}
  \label{eq:new-transition-qgg}
  [\ldots, C_A, \infty]_{ag} + [-\infty, C_A, \ldots]_{gb}
  \to
  [\ldots, C_A \blue, \eta^L_{q'\bar q'}, C_F,  \infty]_{aq'}
  + [-\infty, C_F, \eta^R_{q'\bar q'}, \black C_A, \ldots]_{\qbar' b}\,.
\end{equation}
In a situation where the insertion of $\eta^L_{q'\bar q'}$ would lead
to a disordered sequence of transition points, remove the last two
segments from the $aq'$ dipole, as in
Eq.~(\ref{eq:new-transition-qgg-gone-bad-soln}), and analogously
remove the first two segments from the $\qbar' b$ dipole if
$\eta^R_{q'\bar q'}$ is disordered with respect to the other
transition points.

\subsection{Use in other showers, e.g.\ Pythia~8}
\label{sec:use-in-pythia}

The technique outlined above can be applied to almost any dipole or
antenna shower.
The one adaptation that is required is to identify a suitable
expression for $\eta_\text{approx}$, i.e.\ the analogue of
Eq.~(\ref{eq:eta-approx}), for that shower.
For example in the Pythia~8
shower~\cite{Sjostrand:2004ef,Sjostrand:2014zea}, for a dipole
$ab$ where the emitter has momentum $\ptilde_b$ and the spectator
$\ptilde_a$, the expression for $\eta_\text{approx}$ is
\begin{equation}
  \label{eq:eta-approx-py8}
  \eta_\text{approx}^{(ab)} = \left\{
    \begin{array}{lll}
      \eta_\text{approx}^b &=
          \ln 2\ptilde_b.Q - \frac12\ln (Q^2 p_{\perp\text{evol}}^2)
          +\ln(1-z)\,,
          \qquad & \eta_\text{approx}^b \ge -\eta_\text{approx}^a\,,
      \\[5pt]
      \eta_\text{approx}^a &=
          -\ln \frac{2\ptilde_a.Q}{2\ptilde_a.\ptilde_b} 
          - \frac12\ln\frac{p_{\perp\text{evol}^2}}{Q^2}
          +\ln(1-z)\,,
          \qquad & \eta_\text{approx}^b < -\eta_\text{approx}^a\,,
    \end{array}
  \right.
\end{equation}
where $p_{\perp\text{evol}}$ is the shower evolution variable (a
transverse momentum), the event momentum is $Q$, and $1-z$ is the
fraction of $\ptilde_b$'s energy carried away by the gluon in the
dipole centre-of-mass frame.
One can verify that when an emission is soft and collinear to one of
$\ptilde_b$ or $\ptilde_a$, this gives the correct signed rapidity
with respect to the closer of $b$ or $a$ in the event centre-of-mass
frame.
\logbook{}{There is an $\order{1}$ region of the $\ln1-z$ variable where the
  emission is soft (and potentially collinear) where there is a finite
  density of emission and the formula is incorrect, specifically for
  $\ln 1-z \simeq \frac12\ln \frac{p_{\perp\text{evol}}^2}{Q^2}$.
  However this only causes an issue if there is an existing transition
  point in the vicinity of the corresponding $\eta_\text{approx}$.
  The requirement of a transition point there and of the candidate gluon
  emission being there means that the effect of this mismatch should be
  \NNDL.}
%

\section{A solution with nested ordered double-soft (NODS) corrections}
\label{sec:ME-solution}

The approach of section \ref{sec:transition-points} had two elements:
one was the determination of an effective colour factor for each new
emission; the other was to establish whether a new emission should be
attributed to a $C_F$ segment from the point of view of the
colour-factor identification for subsequent emissions. 

In this section we consider an approach that retains the second of
these elements, but replaces the binary colour-factor choice ($C_F$
v.\ $C_A/2$) with a local matrix-element correction that reproduces the
full-$\nc$ radiation pattern for configurations involving a pair of
energy-ordered soft gluons that are close in rapidity in the Lund
diagram, but with all other emissions well separated in rapidity from
them and from each other.
The procedure will actually give the correct full-$\nc$ matrix element
even when there are multiple such pairs around, as long as each is
well separated in rapidity from all others (in the same sense as in
our discussion at the beginning of
section~\ref{ref:angular-ordering+lund}).

As in the previous section, we will start by explaining our general
approach in the case of a simple subset of event structures in
section~\ref{sec:NODS-primary-only}.
Next, in section~\ref{sec:me-general-events}, we will consider issues
that arise for more general event structures.
We will then give our complete algorithm in section~\ref{sec:me-algorithm}.

\subsection{Angular-ordered primary-only events}
\label{sec:NODS-primary-only}

To understand the NODS procedure, consider a situation with a single
$\qbar q$ pair and gluons $g_1 \ldots g_n$, each of which is primary in
the Lund-diagram sense and well separated in rapidity from all the
other gluons,
with an ordering in Lund-diagram primary rapidity of
$\eta_1 \ll \eta_2 \ll \ldots \ll \eta_n$.
This configuration will dominantly be associated with a leading-colour
dipole structure $\qbar 1$, $12$, $23$, $\ldots$, $nq$ (up to
corrections suppressed by powers of $e^{-|\eta_i - \eta_j|}$),
\logbook{4caa373d5742c}{Rok mentioned he confirmed our expectation
  that it's actually powers of $e^{-2|\eta_i - \eta_j|}$}%
which we can represent as
\begin{equation}
  \label{eq:primary-dipole-sequence}
  \scalebox{0.9}{\includegraphics{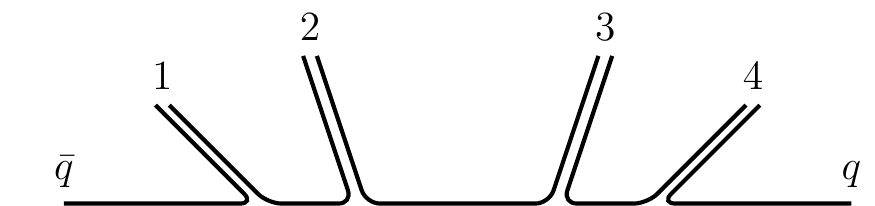}}
\end{equation}
using $n=4$ for concreteness.
The corresponding leading-colour squared matrix element for emission
of a soft gluon with momentum $k$ is\footnote{Strictly, this is the
  ratio of the squared matrix element for production of $\qbar q$ plus
  $n+1$ gluons, to the squared matrix element for $\qbar q$ plus $n$
  gluons.}
\begin{equation}
  \label{eq:LC-ME}
  |M_\text{LC}|^2 = 
   \frac{C_A}{2}\left[ (\qbar 1) + (12) + (23) + \ldots + (nq) \right]
  \,,
\end{equation}
where we have introduced the shorthand
\begin{equation}
  \label{eq:dipole-shorthand}
  (ab) \equiv 16\pi \as \frac{2p_a.p_b}{(2 p_a.k)(2k.p_b)} \propto
  \frac{1-\cos\theta_{ab}}{(1-\cos\theta_{ak})(1-\cos\theta_{kb})}\,,
\end{equation}
and assigned a $C_A/2$ colour factor to each dipole.
In the specific limit that we are considering, i.e.\ primary emissions
that are all well separated in rapidity, the full-colour matrix
element reduces to
\begin{equation}
  \label{eq:full-colour-ME}
  |M|^2 = |M_\text{LC}|^2  + \left(C_F - \frac{C_A}{2}\right) (\qbar q)\,.
\end{equation}
One simple approach to reproducing Eq.~(\ref{eq:full-colour-ME}) would
be to introduce an acceptance probability for each emission of
\begin{equation}
  \label{eq:p-accept-full}
  p^\text{accept} = \frac{|M|^2}{|M_\text{LC}|^2}\,.
\end{equation}
We will introduce a general shorthand for such expressions
\begin{equation}
  \label{eq:paccept-shorthand}
  p^\text{accept}(\qbar, 1, 2, \ldots, n, q) \equiv
  1 + \left(\frac{2C_F - C_A}{C_A}\right)
  \frac{(\qbar q)}{(\qbar 1) + (12) + \ldots + (nq)}\,.
\end{equation}
One approach along these lines was proposed in Ref.~\cite{Giele:2011cb}.
A downside of any approach that uses the full set of momenta in the
acceptance is that for an $N$-particle event it would involve
evaluating $\order{N}$ dot products for each emission, leading to a
contribution to the total showering time that scales as $N^2$.
This is not necessarily prohibitive; for example the implementation of
the Pythia~8 showering algorithm scales as $N^2$, and showers with
global recoil such as PanGlobal also scale as $N^2$ in their current
implementation.
However it turns out to be possible to formulate an expression for
$p_\text{accept}$ that maintains the same accuracy but can be
evaluated in $\order{1}$ time. 

To understand how this can be done, we consider a dipole in the middle
of the chain, say $i,i+1$, where both $i$ and $i+1$ are gluons.
Specifically for this dipole we are free to use
\begin{subequations}
  \label{eq:both-ME-ij}
  \begin{align}
    \label{eq:LC-ME-ij}
    |M_{\text{LC},i,i+1}|^2 &= \frac{C_A}{2}\left[ (i-1,i) + (i,i+1) + (i+1,i+2)\right]\,,
                           \\
    \label{eq:ME-ij}
    |M_{i,i+1}|^2 &= |M_{\text{LC},i,i+1}|^2 + \left(C_F -
                    \frac{C_A}{2}\right) (i-1,i+2)\,,
    \\
    p^\text{accept}_{i,i+1} &= \frac{|M_{i,i+1}|^2}{|M_{\text{LC},i,i+1}|^2}
                              \equiv p^\text{accept}(i-1,i,i+1,i+2)\,,
  \end{align}
\end{subequations}
rather than the full $p^\text{accept}(\qbar, 1, 2, \ldots, n, q)$.
In $|M_{\text{LC}}|^2$ we are justified in dropping all dipoles
$\qbar 1$, \ldots, $(i-2,i-1)$, and $(i+2,i+3)$, \ldots, $nq$,
because of the requirement that all emissions are well separated in
rapidity.
To see this, imagine that $i$ is at negative rapidity
($\theta_{\qbar i} \ll 1$), $i+1$ is at positive rapidity
($\theta_{i+1,q} \ll 1$).
Consider an emitted gluon $k$ such that $\theta_{ik} \simeq
\theta_{\qbar i}$.
Then, because of our requirement that all existing emissions should be
well separated in rapidity, we have
$\theta_{i-2,i-1} \ll \theta_{i-2,k} \simeq \theta_{i-1,k}$.
Examining Eq.~(\ref{eq:dipole-shorthand}) one can then see that the
$(i-2,i-1)$ term is negligible compared to the terms included in
$|M_{\text{LC},i,i+1}|^2$, which is simply a consequence of the fact
that a small-angle dipole does not substantially emit at large angles.
Similarly for the $(i+2,i+3)$ term, as well as all the other terms
that we have neglected in Eq.~(\ref{eq:LC-ME-ij}).
Now we turn to Eq.~(\ref{eq:ME-ij}): here we have replaced a $\qbar q$
with $i-1,i+2$, which is justified since
$\theta_{i-1,k}- \theta_{\qbar,k} \ll \theta_{\qbar k}$ and
$\theta_{k,i+2} - \theta_{k,q} \ll \theta_{kq}$ for any momentum $k$
that is likely to be radiated by the $(i,i+1)$ dipole.

Note that with the truncation adopted in Eq.~(\ref{eq:LC-ME-ij}), it
would not be sensible to use the $(\qbar q)$ emission factor in the
$(C_F - C_A/2)$ term of Eq.~(\ref{eq:ME-ij}), because such a term
would have a negative divergence for $k$ exactly collinear to $\qbar$
or $q$ that would not be compensated for by any of the terms in the
$|M_{\text{LC},i,i+1}|^2$ contribution.
In contrast, one can show that the acceptance probability as written
Eq.~(\ref{eq:both-ME-ij}) is bounded to be in the range
\begin{equation}
  \label{eq:paccept-ij-bounds}
  1 - 3\left(\frac{C_A - 2C_F}{C_A}\right)
  = 1 - \frac{3}{\nc^2} \le p^\text{accept}_{i,i+1} \le 1\,,
\end{equation}
i.e.\ for the physical value of $\nc$ it is always positive definite
and so straightforward to use in event generation.\footnote{Without
  loss of generality, for massless particles, one can always rotate
  and boost an event such 
  that $i-1$ and $i+2$ are along the negative and positive $z$ axes
  and $k$ is along the $-x$ direction.
  The configuration that gives the minimum is then one where $i$ and
  $i+1$ have their momenta along the $(\sin\theta, 0, \pm\cos\theta)$
  directions with $\tan \frac{\theta}{2}=\frac{1}{2}$.
}

We have given the justification for writing Eq.~(\ref{eq:both-ME-ij})
in a specific frame, one in which it is manifest that one can replace
$\qbar$ with $i-1$ and $q$ with $i+2$.
However the underlying expressions are Lorentz invariant, and the
validity of those replacements is ultimately ensured by the condition
that all existing emissions are on the primary Lund plane and
separated by large differences in rapidity.

We also need to consider dipoles at the end of the chain, for example
the $\qbar1$ dipole in Eq.~(\ref{eq:primary-dipole-sequence}).
In such a case, we simply write
\begin{equation}
  \label{eq:both-ME-qbarq}
  p^\text{accept}_{\qbar,1} = p^\text{accept}(\qbar,1,2)\,,
\end{equation}
and analogously at the other extremity.

In what follows, when we write 
\begin{equation}
  \label{eq:both-ME-auxiliary}
  p^\text{accept}_{i,i+1} = p^\text{accept}(i-1;i,i+1;i+2)\,,
\end{equation}
it is useful to introduce the following terminology: $i-1$ is the
\emph{auxiliary} momentum at the colour-anti-triplet ($i$) end of the
dipole, and $i+2$ is the auxiliary at the colour-triplet ($i+1$) end of
the dipole.
We have emphasised that $i-1$ and $i+2$ are auxiliaries by separating
their indices from the others with semi-colons.
One can then represent the event of Eq.~(\ref{eq:primary-dipole-sequence}) as
\begin{equation}
  \label{eq:primary-dipole-auxiliaries}
  \scalebox{0.7}{\includegraphics[valign=c]{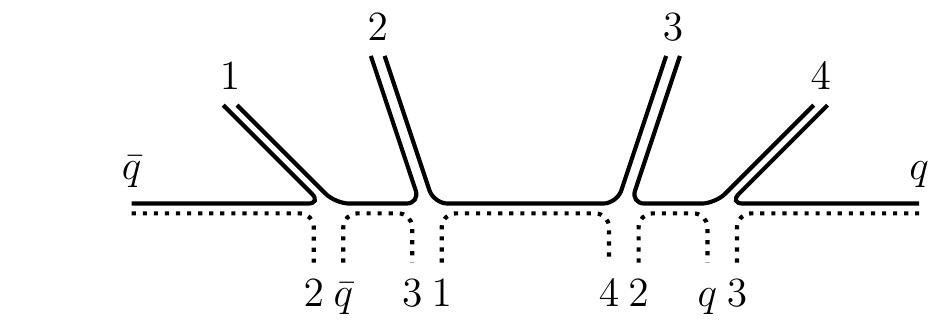}}
\end{equation}
where the dotted line beneath each solid black dipole represents the
$-1/(2\nc)$ dipole that we use in the matrix-element correction.
It shows, for example, that the $23$ dipole has auxiliaries $1$ and
$4$, leading to an acceptance factor $p^\text{accept}(1;2,3;4)$.

\subsection{Considerations for general events}
\label{sec:me-general-events}

Most events do not consist just of angular-ordered primary emissions
discussed so far: primary gluons can themselves emit secondary gluons,
and any gluon may split to $q\bar q$.

The first case that we consider is emission from a region that would
be associated with a $C_F$ colour factor in the segment approach,
e.g.\ the emission of $5$ from the $23$ dipole,
\begin{equation}
  \label{eq:primary-dipole-CFemsn}
  \scalebox{0.7}{\includegraphics[valign=c]{diagrams/primary-dipoles-auxiliaries.pdf}}\quad
  \longrightarrow
  \scalebox{0.7}{\includegraphics[valign=c]{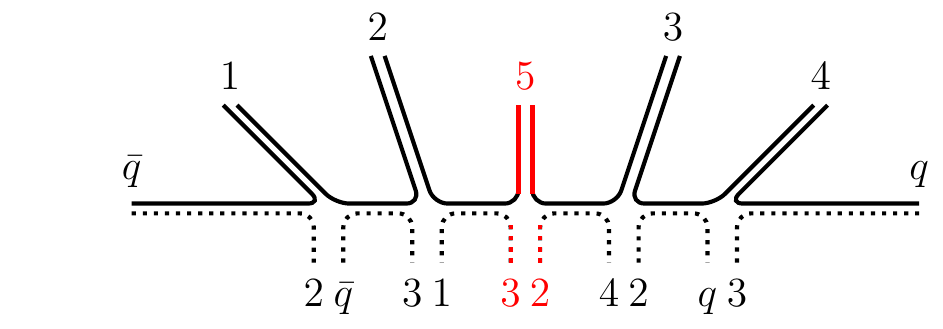}}
\end{equation}
When $5$ is well-separated in rapidity from both $2$ and $3$, the
acceptance factor for emission of $5$ will reduce to $2C_F/C_A$.
The new $25$ dipole retains the left-hand auxiliary ($1$) of the
parent dipole and acquires a new right-hand auxiliary ($3$),
corresponding to the triplet end of the parent dipole.
Similarly, the new $53$ dipole acquires a new left-hand auxiliary
($2$), corresponding to the anti-triplet end of the parent dipole, and
retains the parent's right-hand auxiliary ($4$).

Next, we consider a collinear $g \to gg$ splitting.
Imagine that we have the configuration in
Eq.~(\ref{eq:primary-dipole-auxiliaries}) and that dipole $23$
branches,  with its $2$ end splitting to gluons $5$ and $6$
in a collinear configuration
$\theta_{56} \ll \theta_{15} \simeq \theta_{16}$, illustrated as
follows
\begin{equation}
  \label{eq:primary-dipole-splitting-to-gg}
  \scalebox{0.7}{\includegraphics[valign=c]{diagrams/primary-dipoles-auxiliaries.pdf}}\quad
  \longrightarrow
  \scalebox{0.7}{\includegraphics[valign=c]{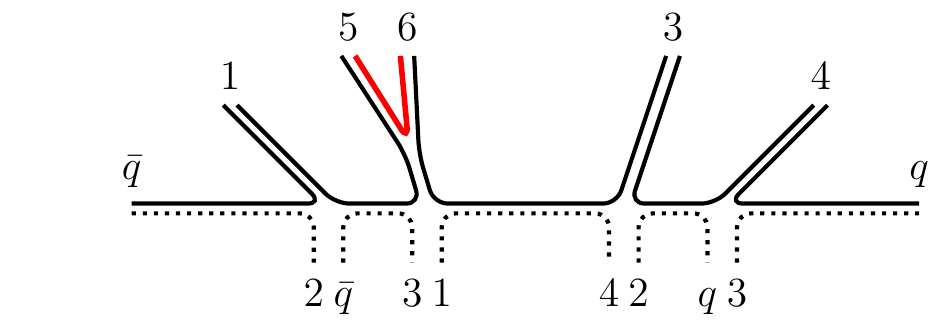}}
\end{equation}
where the new $56$ dipole is highlighted in red.
The $15$ dipole, which is the successor of the $12$ dipole retains the
$12$ dipole's auxiliaries.
Similarly the $63$ dipole retains the $23$ dipole's auxiliaries.
The new $56$ dipole, since it is produced far in Lund-plane rapidity
from any region with a $2C_F/C_A$ correction,
does not need any
auxiliaries, in the same way that it has no transitions points (and so
no $C_F$ segments) in the approach of
section~\ref{sec:transition-points}.

For configurations that are not strongly ordered, e.g.\ emission of
$5$ from the $23$ dipole at an angle that is commensurate with
$\theta_{21}$, the acceptance factor is unambiguous (and correct) in
our approach.
However there is an ambiguity in the assignment of auxiliaries to the
child dipoles, which translates to an ambiguity in the acceptance for
a subsequent third emission at similar angles, a configuration where
we do not aim for full-$\nc$ accuracy.
We resolve that ambiguity as follows: as well as retaining information
on auxiliaries, we also retain information on segments and their
transitions, as in the approach of
section~\ref{sec:transition-points}.
If an emission occurs in a $C_F$ segment then we assign new
auxiliaries as in Eq.~(\ref{eq:primary-dipole-CFemsn}).
Instead, if it occurs in a $C_A/2$ segment, we do not introduce any
new auxiliaries, as in Eq.~(\ref{eq:primary-dipole-splitting-to-gg}).

Finally, we consider a $g \to q' \qbar'$ splitting (for which we
always assign a normal $P_{g\to q\qbar}$ splitting function and use a colour
acceptance factor of $1$).
Consider $2 \to q'\qbar'$,
\begin{equation}
  \label{eq:primary-dipole-splitting-to-qqbar}
  \scalebox{0.7}{\includegraphics[valign=c]{diagrams/primary-dipoles-auxiliaries.pdf}}\quad
  \longrightarrow
  \scalebox{0.7}{\includegraphics[valign=c]{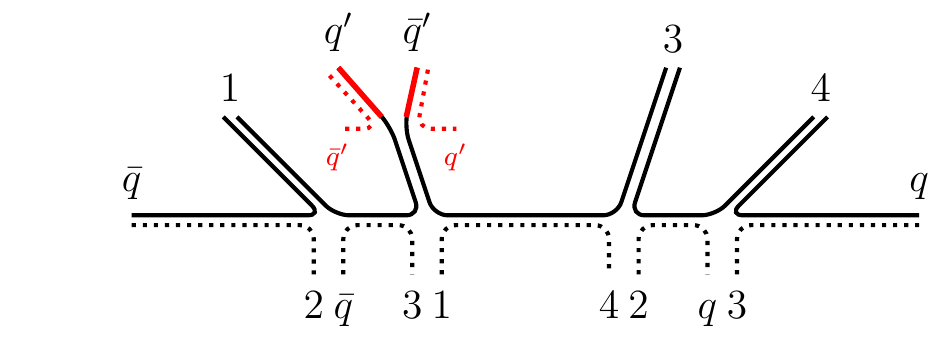}}
\end{equation}
where we have highlighted the new $q' \qbar'$ pair in red.
To justify the choice for the new auxiliary variables, we consider the
radiation of a soft gluon after the $2\to q'\bar{q}'$ splitting.
In the segmented approach of section \ref{sec:transition-points} we
would have said that that the $\qbar' 3$ dipole now has two $C_F$ segments.
In our NODS approach, we introduce an acceptance factor for
each $C_F$-like segment in the dipole and determine the overall
acceptance for the dipole as the product of the individual acceptance
factors. E.g.\ for the $\qbar'3$ dipole we write
\begin{equation}
  \label{eq:two-acceptance}
  p^\text{accept}_{\qbar',3} = p^\text{accept}(\qbar',3;q')\, p^\text{accept}(1;\qbar',3;4)\,,
\end{equation}
where the first $C_F$-like segment starting from the $\qbar'$ is
associated with a single auxiliary ($q'$), at its right-hand
end,\footnote{The correctness of the $p^\text{accept}(\qbar',3;q')$
  factor can be seen using arguments similar to those of
  section~\ref{sec:NODS-primary-only}.
  In the event centre-of-mass frame, angular ordering implies
  $\theta_{q' \qbar'} \ll1$, i.e.\ we only need to consider collinear
  $g \to q' \qbar'$ splitting.
  We imagine boosting the event to a frame in which
  $\theta_{q' \qbar'}$ is of order 1.
  In the angular-ordered limit, all other particles will then be
  constricted along a single collinear direction, which is well
  separated from the $q'$ and $\qbar'$ directions.
  Next, we temporarily imagine that a $C_A/2$ dipole stretches
  between the $\qbar$ and $q$, so as to avoid complications with their
  actual (quark-like) colour structure.
  Then, the full acceptance factor correction associated with the
  $q' \qbar'$ $C_F$ colour structure would be
  $p^\text{accept}(\qbar', 3, 4, q, \qbar, 1, q')$.
  For emissions along the $\qbar'$ end of the $\qbar'3$ dipole, all terms $(34)$,
  $(4q)$, \ldots, $(\qbar 1)$, are irrelevant.
  Furthermore, one can replace $(1q')$ with $(3q')$, since $1$ and $3$
  are collinear and the $(1q')$ term is relevant only when the
  emission is emission is far in angle from $1$.
  Thus the acceptance factor can be reduced to
  $p^\text{accept}(\qbar',3;q')$. 
}
while the second $C_F$ segment retains the auxiliaries of the parent
$23$ dipole.
For strongly angular-ordered configurations, i.e.\
$\theta_{q'\qbar'} \ll \theta_{q'1}$, only one of the two
$p^\text{accept}(\ldots)$ factors in Eq.~(\ref{eq:two-acceptance})
will ever differ substantially from $1$.
For configurations where strong angular ordering does not hold, for
which we do not aim to achieve full-$\nc$ accuracy, there will be
phase-space regions for a subsequent emission $k$ where both factors
will be below $1$.

\subsection{Full algorithm}
\label{sec:me-algorithm}

We are now ready to specify our full matrix-element-based NODS
colour-handling algorithm.
It retains the core framework of the segmented approach, specified in
section~\ref{sec:transition-algorithm}, with the following
augmentations:
\begin{enumerate}
\item Where the extremity of a segment has finite rapidity, it is
  associated with an auxiliary, which we denote $\bar a$ towards the
  anti-triplet end of the segment and $a$ towards the triplet end.
\item Emission acceptance:
  \begin{enumerate}
  \item The acceptance for gluon emission from a dipole $ij$ is the
    product of individual acceptance factors for each of the dipole's
    $C_F$ segments.
    The individual acceptance factor for a given $C_F$ segment is
    \begin{equation}
      \label{eq:individual-acceptance}
      p^\text{accept}_{\text{segment}} = p^\text{accept}([\bar a]; i, j; [a]) \,,
    \end{equation}
    where $p^\text{accept}([\bar a]; i, j; [a])$ is evaluated using
    Eq.~(\ref{eq:paccept-shorthand}).
    The auxiliary momenta are in square brackets to indicate that if the
    segment does not have the corresponding auxiliary, because its
    extremity stretches to infinite rapidity, $p^\text{accept}$ is to be
    evaluated without that auxiliary.
    For example, in the case with no auxiliaries at either end, the
    acceptance reduces to $p^\text{accept}(i, j) \equiv 2C_F/C_A$.
  \item For a $g \to q'\bar q'$ splitting, the colour acceptance
    factor is set to 1.
  \end{enumerate}
\item Auxiliaries update:
  \begin{enumerate}
  \item If a gluon emission $k$ from an $ij$ dipole occurs in a $C_F$ segment
    with auxiliaries $\bar a$ and $a$, then the corresponding $C_F$
    segment in the child $ik$ dipole has auxiliaries $\bar a$ and $j$,
    while that in the child $kj$ dipole has auxiliaries $i$ and
    $a$. (Where an auxiliary is absent in the parent dipole because the
    segment stretches to infinite rapidity, it remains absent in the
    child dipole).
  \item If a gluon emission from an $ij$ dipole occurs in a $C_A$
    segment, the child dipoles' $C_F$ segments retain the auxiliaries of the
    parent dipole.
  \item For $g \to q'\bar q'$ splitting, the new $C_F$ segment in the
    anti-triplet-$q'$ dipole acquires the $\qbar'$ as its anti-triplet
    (left) end auxiliary, while the new segment in the $\qbar'$-triplet
    dipole acquires $q'$ as its triplet (right-hand) end auxiliary.
  \item In those special cases of the algorithm of
    section~\ref{sec:transition-algorithm} where two segments are
    removed from the extremity of a sequence, the corresponding
    auxiliaries are also to be removed.
  \end{enumerate}
\end{enumerate}
When storing the auxiliary for a segment, there is some freedom: for
example one can choose the auxiliary's momentum at the time that it is
associated with the segment, or its (possibly different) momentum at a
later stage in the event when one is evaluating
Eq.~(\ref{eq:individual-acceptance}).
In practice we make a third choice: when gluon emission causes a
left-hand auxiliary to be first added to a segment-sequence of an $ij$
dipole, we store the difference in direction between the auxiliary and
the anti-triplet end of the dipole,
$\delta_{i\bar a} = d_{\bar a} - d_i$, where $d_{i}$ is the 3-vector
direction of $i$.
For a right-hand auxiliary point, we store the difference in direction
between the auxiliary and the triplet end of the dipole
$\delta_{ja} = d_{a} - d_j$.
When we later come to need an auxiliary momentum to evaluate
Eq.~(\ref{eq:individual-acceptance}) in an $\ell m$ dipole that
descends from the original $ij$, we reconstruct auxiliary directions
using differences with respect to the $\ell$ and $m$ directions at
that stage of the event,
\begin{equation}
  \label{eq:individual-acceptance-diffs}
  p^\text{accept}_{\text{segment}} =
  p^\text{accept}(\delta_{i\bar a} + d_\ell; \ell, m; \delta_{ja} + d_m) \,.
\end{equation}
For a $g\to q'\qbar'$ splitting, in the $C_F$ segment on the
anti-triplet-$q'$ dipole we store its left-hand auxiliary ($\qbar'$)
direction difference relative to the $q'$ (i.e.\ triplet) direction,
$\delta_{q'\qbar'}$ so that when that segment is eventually evaluated
in the context of emission from a descendent $\ell m$ dipole we use
\begin{equation}
  \label{eq:individual-acceptance-diffs-qqbar}
  p^\text{accept}_{\text{segment}} =
  p^\text{accept}(\delta_{q'\qbar'} + d_m; \ell, m; [\ldots]) \,,
\end{equation}
and analogously for the $\qbar'$-triplet dipole.
We are not aware of a reason for preferring one of these approaches
over any other, however that based on direction differences has the
advantage of being easiest to use for certain of our tests in
section~\ref{sec:numerical-tests} and so it is our main choice.

\section{Colour-factor from emitter (CFFE) algorithms}
\label{sec:cffe-algs}

When examining the numerical behaviour of our algorithms in the next
sections, it will be useful to have an illustration of the behaviour
of existing algorithms for assigning colour, in order to gauge the
practical impact of our proposals.

In standard dipole showers (e.g.\ the Pythia~8
shower~\cite{Sjostrand:2004ef} or the Dire~v1
shower~\cite{Hoche:2015sya}), each dipole is split into two elements,
according to which end of the dipole is deemed the ``emitter''.
Roughly speaking, each element accounts for the rapidity phase-space
that extends from zero rapidity in the dipole centre-of-mass frame to the
extremity of rapidity phase-space at the emitter end.
For gluon emission, each element acquires the full-colour splitting
function associated with the flavour of the emitter, i.e.\ yielding
$P_{gq}(z)=2C_F/z$ ($P_{gg}(z)=C_A/z$) for emission of a soft gluon
with momentum fraction $z \ll 1$ if the emitter is a quark (gluon).
Accordingly, we refer to this approach as the ``colour-factor from
emitter'' (CFFE) approach.
Ref.~\cite{Dasgupta:2018nvj} showed that the CFFE approach yields
wrong \DLNLC terms starting from second order, $\as^2 L^4$, for some
observables, e.g.\ the thrust.

One can also examine the CFFE approach in PanScales-like showers,
where the phase-space is again divided between the two elements of the
dipole, at an angle that is roughly equidistant between the emitter
and spectator ends of the dipole, in the \emph{event} centre-of-mass
frame.\footnote{For soft and collinear emissions, this is similar to
  the colour assignment discussed in Ref.~\cite{Forshaw:2020wrq}.}
Using the approach of Ref.~\cite{Dasgupta:2018nvj}, one can show that
the method still yields incorrect \DLNLC terms starting from second
order, $\as^2 L^4$, for some observables.

One could fix the second-order issue by dividing the dipole in a way
that mimics the transition points that we have discussed in
section~\ref{sec:transition-points}.
However, at third order, it becomes impossible to obtain the correct
answer for general double-logarithmic observables with a single
transition point for the dipole.
One can see this by examining Fig.~\ref{fig:lund-diagrams-simple},
where the (blue)
$g_2g_1$ dipole requires two transition points, one from $C_A/2$ to
$C_F$ and another back to $C_A/2$.

\section{Matrix element tests} \label{sec:ME-tests}

In order to demonstrate how the methods presented above effectively incorporate
subleading-colour effects, we examine parton shower results at fixed order for
$2 \to 3+g$ and $2 \to 4+g$ final states ($e^+ e^- \to X + g$), where
$X \in \lbrace \bar{q} g_1 q, \, \bar{q} q' \bar{q}' q,\, \bar{q} g_1
g_2 q \rbrace$ and $g$ is an additional soft gluon that is radiated
from system $X$.
We compare the
parton-shower~(PS) result $d\sigma_{\mathrm{PS}} / d\eta d\psi$
($\sigma_\text{PS}$ for brevity) to the exact ratio of full-colour squared matrix
elements $d\sigma_{\mathrm{FC}} / d\eta d\psi = d\Phi_g |M_{X+g}|^2 / |M_{X}|^2$.
Here, $\eta$ and $\psi$ are the rapidity and azimuthal variables
as obtained from a Lund declustering procedure (see below).
We do this for each of the segment, NODS and CFFE parton-shower colour
schemes.

In sections \ref{sec:ME-qgq} and \ref{sec:ME-qqqqg}, we show differential results for soft-collinear
configurations, where all partons in $X$ are strongly ordered in energy and
rapidity, and the additional soft gluon $g$ can be at angles commensurate with
any other particle (while still strongly ordered in energy compared to them).
To test \NDLFC accuracy, in section~\ref{sec:me-integral} we compare the
integrated soft-gluon emission rate $I_{\text{PS}}$, determined
numerically from $d\sigma_{\mathrm{PS}} / d\eta d\psi$, to the
expected value $I_{\text{FC}}$, see Eq.~(\ref{eq:ang-ordered}). We do
this for three different configurations in which the gluon $g_1$ is
either soft and collinear, hard and collinear, or soft and large-angle
with respect to the initial $q\bar q$ dipole.

Note that the approach that we develop below for matrix-element tests
could also be adapted to provide interesting measurements within
Lund-diagram type jet substructure analyses.

\subsection{Differential matrix element: $2 \to 3 + g$ configuration}
\label{sec:ME-qgq}

All differential plots shown below are produced with the
PanGlobal shower, with 
$\beta=0$ (see~\cite{Dasgupta:2020fwr} and
Eq.~(\ref{eq:evolution-variable})).
To set up the event ($e^+ e^- \to \bar{q} g_1 q$, see
Figs.~\ref{fig:lund-diagram-g1} and \ref{fig:subleading-config}), we let the parton shower 
split the initial $\bar{q} q$ dipole, which emits a gluon $g_1$ with
predefined kinematics, corresponding in our soft-collinear case to 
\begin{equation}
z_{g_1} = 10^{-8},\quad \eta_{g_1} = 5,\quad \psi_{g_1} = \pi\,,
\label{eq:me-setup-g1}
\end{equation}
in the event frame.
The emission of the additional gluon, $g$, is then also performed by the parton shower:
we fix the value of the evolution variable $\ln v$ ($\equiv \ln k_t$
for the PanGlobal $\beta=0$ shower),
while sampling over 
the two remaining shower degrees of freedom $\bar \eta$ and $\phi$.
We cluster the event with the Cambridge/Aachen
algorithm~\cite{Dokshitzer:1997in,Wobisch:1998wt}, and work backwards
through the clustering history to determine the effective Lund
diagram~\cite{Dreyer:2018nbf} (we call this Lund declustering).
We then log the event weight in a histogram in variables
$\left( \eta, \psi\right)$ defined with respect to the emission's
parent Lund leaf.
We fill separate histograms for emission on the primary and
secondary Lund leaves.
Starting from the same initial $\bar{q} g_1 q$ configuration, we also
sample the exact full-colour analytic tree-level matrix element, see
Eq.~(\ref{eq:full-colour-ME}), neglecting any recoil from the last
emission (since the impact of the emitted soft gluon $g$ is
negligible).
Finally, we take the ratio of the parton shower and exact full-colour
histograms.

\begin{figure} \centering
  \includegraphics[width=.6\textwidth,page=6]{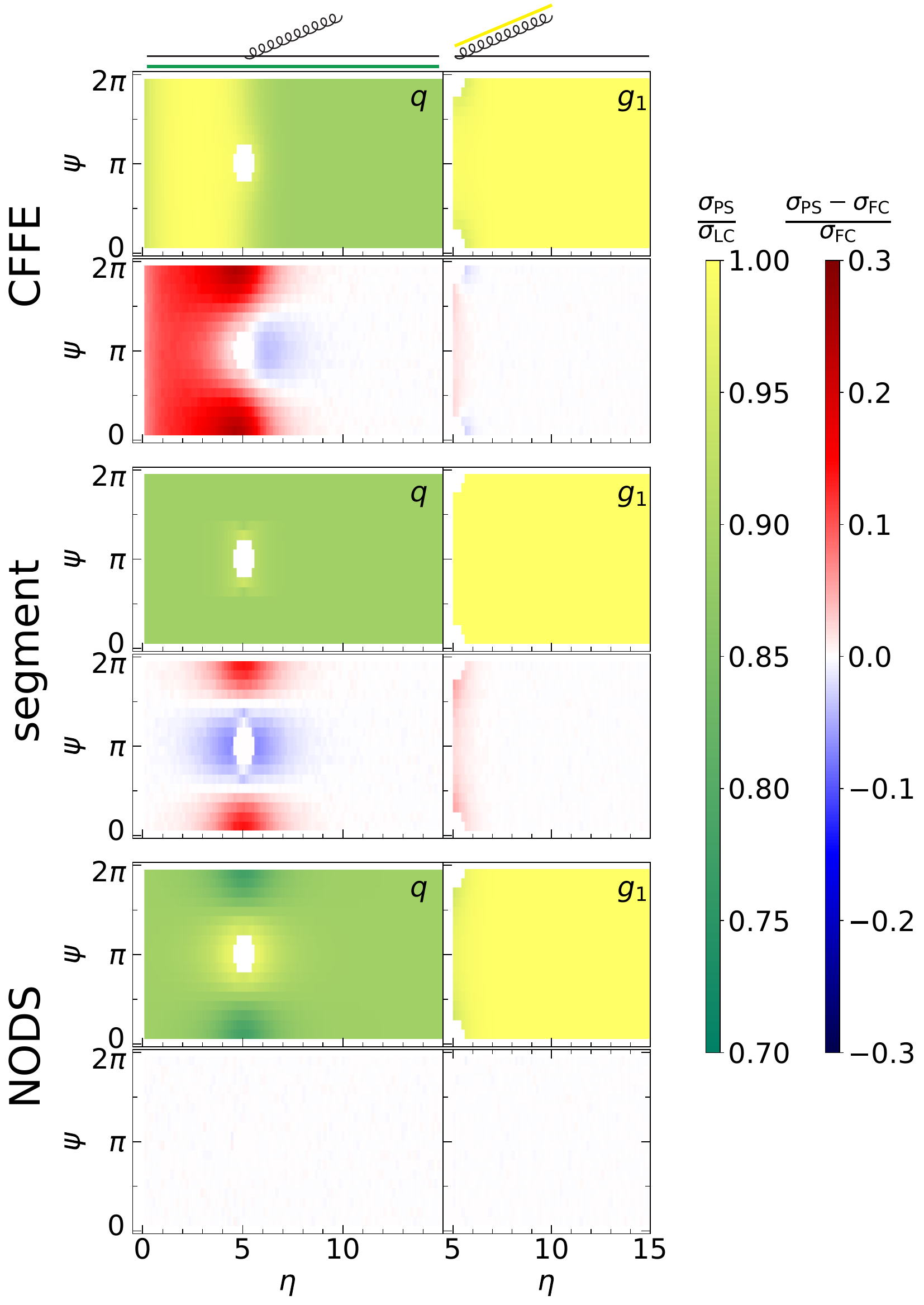}
  \caption{%
    Tree-level fixed-order expansion of the parton shower density for
    emission of a soft gluon from a $\bar{q} g_1 q$ system (as
    represented in Fig.~\ref{fig:lund-diagram-g1}).  %
    The results are shown for the CFFE (top), segment (middle) and
    NODS (bottom) colour schemes, as implemented for the PanGlobal
    shower with $\beta = 0$.
    The first row in each pair of plots depicts the ratio
    $\sigma_\text{PS} / \sigma _\text{LC}$ (green--yellow colour bar).
    The second row
    shows the relative deviation $(\sigma_{\rm PS} - \sigma_{\rm FC}) / \sigma_{\rm
      FC}$ between the parton shower and the analytic matrix element (blue--red colour
    bar).
    The left-hand panels corresponds to the
    primary-emission phase-space region, while the right-hand panels
    corresponds to emission
    from $g_1$, as emphasised in the diagrams at the top.
  }
\label{fig:me-results-qgq}
\end{figure}

The results are collected in Fig.~\ref{fig:me-results-qgq} for the
CFFE (section~\ref{sec:cffe-algs}), segment
(section~\ref{sec:transition-points}) and NODS
(section~\ref{sec:ME-solution}) colour schemes.
The labels on each panel, $q$ and $g_1$, refer to the parton with
which the gluon $g$ is clustered by the C/A algorithm (i.e.\ the
emitter according to the Lund declustering sequence).
At the top of each column, a diagram indicates the corresponding part
of the event that the Lund rapidity $\eta$ refers to (primary Lund
plane, along the quark direction, or secondary plane along the gluon
direction).
Note that in the following, we do not show differential
results at the $\bar{q}$-end since, in this collinear configuration,
the emission correctly gets an overall factor of $C_F$ for all colour
schemes that we consider (to within corrections suppressed by
powers of $e^{-|\eta - \eta_{g_1}|}$).
To help see the effective colour
factor applied by the parton shower, for each method the upper row  in
Fig.~\ref{fig:me-results-qgq} depicts ratios, $\sigma_\text{PS} / \sigma _\text{LC}$, of the various
parton-shower colour schemes
to the leading-colour result $\sigma_\text{LC}$. For the 
latter, we use a constant colour factor $C_F = C_A / 2 = 3 / 2$ irrespective of 
the position of the emission.
The ratio $\sigma_\text{PS}/\sigma_\text{LC}$ is $1$ (yellow in
Fig.~\ref{fig:me-results-qgq}) when the effective colour factor is
$C_A/2$, while it is $8/9 =2C_F/C_A$ (green) when the effective colour
factor is $C_F$.
In the lower row, we plot
the relative deviation $(\sigma_\text{PS} - \sigma_\text{FC}) /
\sigma_\text{FC}$ between the parton-shower weight and the analytic
full-colour matrix element.
Regions where the shower agrees with the full-colour matrix element
come out in white (to within statistical fluctuations).

The two collinear limits  of interest, $g \parallel q$ and $g \parallel g_1$,
are always mapped to the large-$\eta$ region of the corresponding panels, i.e.\
the $q$-leaf (left column) and $g_1$-leaf (right column). Rapidities cover the
range $\eta_\text{p} < \eta < \infty$, where $\eta_\text{p}$ corresponds to the opening angle between
the parent parton and the next particle in the declustering sequence (with
$\eta_\text{p} = 0$ for radiation on the primary plane).
Holes correspond to the point where the emission moves across leaves,
from a primary to a secondary plane (from the $q$-leaf to the
$g_1$-leaf, in our example). 

Let us examine the three colour methods in turn:
\begin{itemize}
\item
  The CFFE method incorrectly assigns a colour factor $C_A/2$ to
  the region $0 < \eta < \eta_{g_1} = 5$.
  The fact that it extends from zero up to $\eta_{g_1}$ means that the
  wrong colour factor is being used for the emission of the soft gluon
  $g$ in a double-logarithmically enhanced region (for fixed
  kinematics of the soft gluon $g$, the double log arises from the
  integration over the transverse momentum and rapidity of
  $g_1$). This ultimately will lead to incorrect subleading $\nc$
  contributions to \LL and \DL terms.

\item
  Instead, with the segment method, colour factors are clearly separated
  into a $C_F$ region for emissions belonging to the $q$-leaf, and a
  $C_A/2$ region for those belonging to the $g_1$-leaf, as expected from
  Fig.~\ref{fig:lund-diagram-g1}.
  A residual deviation from the exact matrix element is present in a
  region of rapidity localised around $\eta_{g_1}$, and reaches the
  level of $+15\%$ for $\psi = 0$, as shown in the lower row of plots
  (blue-red colour scale).
  If one integrates over azimuthal angle and rapidity, the blue and red
  regions will compensate each other.
  In section~\ref{sec:me-integral} below, we shall verify that that
  compensation is exact for large $\eta_{g_1}$, so that the method
  reproduces the correct total rate of soft emission,
  Eq.~(\ref{eq:ang-ordered}), as needed for \NDLFC accuracy in
  observables such as multiplicities.
  \logbook{5b908a08}{logbook/2020-02-27-full-colour-LL (sec. 3.2)
    explores why part of the $C_A/2$ colour factor ``leaks out" of the
    gluon hole.
    The effect depends on the kinematic map. However, all PanScales
    showers are expected to behave the same way - as described in the
    logbook.}   
  
\item
  The NODS procedure reproduces the full
  squared tree-level matrix element (up to statistical fluctuations associated
  with our Monte Carlo sampling), as it should, since in this kinematic
  region the method is effectively using that full tree-level result to
  correct the leading-colour shower matrix element.
\end{itemize}

\subsection{Differential matrix element: $2 \to 4 + g$ configurations}
\label{sec:ME-qqqqg}

\begin{figure} 
\centering
  \captionsetup[subfigure]{oneside,margin={.2\textwidth,0cm}}
  \begin{subfigure}{0.49\textwidth}
    \includegraphics[height=.55\textheight,page=4]{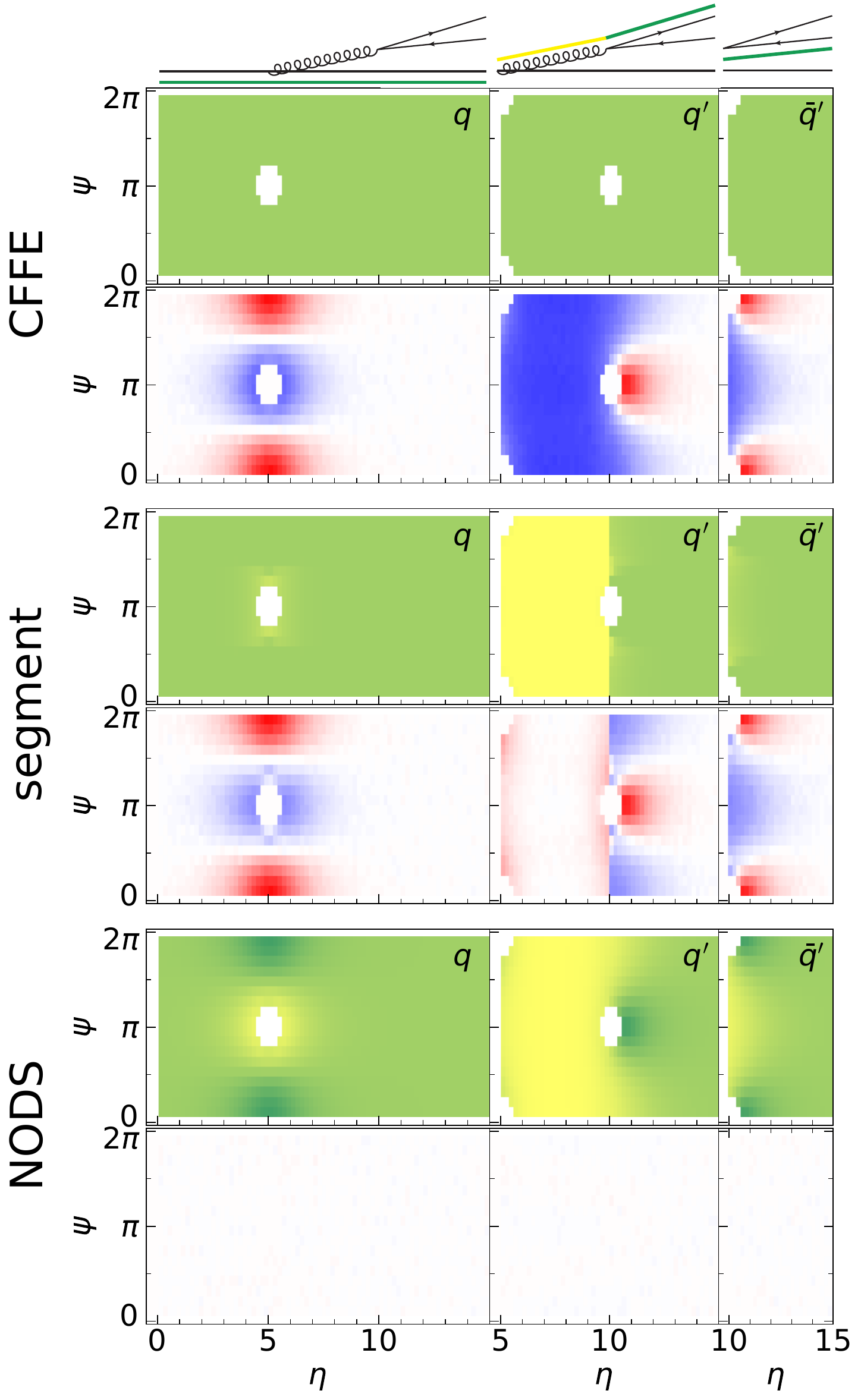}%
    \caption{$\bar{q}q' \bar{q}' q + g$}
    \label{fig:me-results-qqqqg}
  \end{subfigure}%
  \captionsetup[subfigure]{oneside,margin={-.2\textwidth,0cm}}
  \begin{subfigure}{0.49\textwidth}
    \includegraphics[height=.55\textheight,page=5]{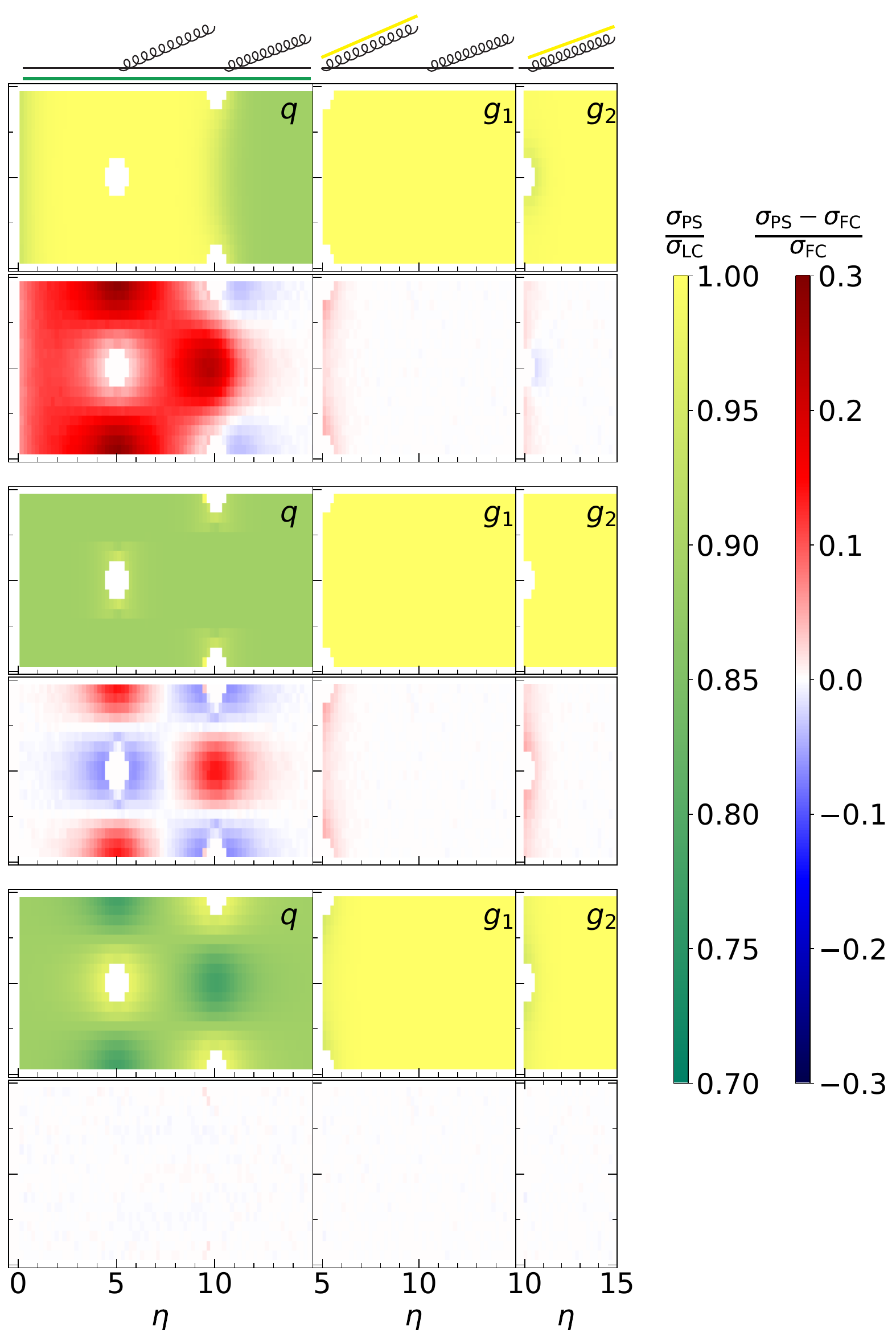}
    \caption{$\bar{q} g_1 g_2 q + g$}
    \label{fig:me-results-qqggg}
  \end{subfigure}
\caption{Colour factor assignment and relative deviation to the squared tree-level
  matrix element, as in Fig.~\ref{fig:me-results-qgq}, for the $2 \to 4+g$
  configurations (a) $\bar{q}q' \bar{q}' q + g$  and (b) $\bar{q} g_1
  g_2 q + g$, corresponding respectively to the Lund diagrams in
  Figs.~\ref{fig:lund-diagram-qqbar} and
  \ref{fig:lund-diagrams-simple} (with $g_2$ there moved to the right of
  $g_1$).
  The results have been obtained with the $\beta=0$ PanGlobal shower
  algorithm.  }
\label{fig:me-results-qqqqg-qqggg}
\end{figure}

Next, we consider tests for  $2 \to 4 + g$ configurations, first for
$e^+ e^- \to \bar{q} q' \bar{q}' q + g$ and then for $e^+ e^- \to \bar{q} g_1
g_2 q + g$. A first gluon $g_1$ is emitted off the $\bar{q}q$ dipole with the
same kinematics as in Eq.~(\ref{eq:me-setup-g1}). The second splitting is
performed with
\begin{subequations}
  \begin{align}
    \bar{q} q' \bar{q}' q\; {\rm configuration}:\qquad
    &z_{\bar{q}'} = 1/4, &\eta_{q'\bar{q}'} = 10,\qquad &\psi = 0, &
                                                                    \hspace{15pt}
    \\
    \bar{q} g_1 g_2 q\; {\rm configuration}:\qquad
    &z_{g_2} = 10^{-16}, &\eta_{g_2} = 10,\qquad &\psi =
    0\,. &\hspace{15pt}
  \end{align}
\end{subequations}
These configurations are such that the second splitting happens at a
much smaller angle than the first gluon emission. For the first
configuration ($g_1\to \bar q' q'$), we choose a $z$ fraction
reflecting the absence of soft enhancements. For the second
configuration (emission of $g_2$ from
the quark, well separated in rapidity from $g_1$) we focus on a
case where $g_2$ is much softer than $g_1$, though the
conclusions are unchanged if we take $g_1$ and $g_2$ to have
commensurate transverse momenta.%
\logbook{0c731a74}{see ../../2020-eeshower/analyses/test-twodcolour-integral/paper-plots/tests-commensurate-kt/kt_g1-equal-kt_g2.pdf}
Results are displayed in Fig.~\ref{fig:me-results-qqqqg-qqggg}. They
have features similar  to those of Fig.~\ref{fig:me-results-qgq}, albeit with a
richer structure.

The $\bar{q} q' \bar{q}' q$ case is
shown in Fig.~\ref{fig:me-results-qqqqg}.
The Lund diagram for the corresponding phase space, and expected
colour assignments, was shown in Fig.~\ref{fig:lund-diagram-qqbar}.
There are several possible
ways for the additional gluon to cluster with the rest of the
event.
It is useful to organise them from a Lund-plane
viewpoint. The three different possibilities correspond to the left,
centre and right panels in each plot of
Fig.~\ref{fig:me-results-qqqqg}. 
First, primary emissions (left-hand panel, labelled $q$) include cases
where $g_2$ clusters with either $\bar{q}$, $q$, or the full
$g_1(\to q'\qbar') q$ system, corresponding respectively to
$\eta<0$ (not shown), $\eta\gtrsim 5$ and $0<\eta\lesssim 5$.
Next, secondary emissions (middle panel, labelled $q'$) include
clusterings with either $q'$ (the harder of $q'$ and $\bar q'$) or
the $g_1\to q'\bar q'$ system.
These correspond to
the regions $\eta\gtrsim 10$ and $5\lesssim \eta\lesssim 10$
respectively.
Finally, tertiary emissions (right-hand panel, labelled $\bar q'$)
correspond to clusterings of the soft gluon $g$ with $\bar q'$.
The hole observed in the primary (secondary) plane corresponds to the
region where emissions are clustered in the secondary (tertiary)
plane.
The correct colour assignment involves a factor $C_F$
everywhere except in the region $5 \lesssim \eta \lesssim 10$
of the $q'$-labelled
plot, where the emission of $g$ is sensitive to the net
colour-octet charge of the whole $g_1 \to q'\qbar'$ system.
The CFFE method tracks neither the intermediate particles nor
segments, and therefore blindly applies a factor $C_F$ across
phase-space.
Our segment and NODS schemes, in contrast, display the
correct behaviour along the intermediate gluon segment.
In the case of
the segment method, azimuthal deviations from the exact matrix element
are observed to be of similar size as shown above in
Fig.~\ref{fig:me-results-qgq}.
Note also the discontinuity in the segment-method $\qbar'$
panel at $\eta=10$.
This is a consequence of our choice to make discontinuous
transitions in the segment approach.
Similar features are present in the other plots, but are less
immediately visible because they coincide with the phase-space boundaries
between primary and secondary Lund planes.

Next, we turn to the $\bar{q} g_1 g_2 q$ configuration
(Fig.\ \ref{fig:me-results-qqggg}), where the gluons $g_1$ and $g_2$ are 
primary emissions, and the holes both appear on the (primary)
$q$-leaf.
The corresponding Lund diagram would be
Fig.~\ref{fig:lund-diagrams-simple} with $g_2$ there moved to the
right of $g_1$.
In this case, the three panels correspond to primary emissions,
emissions from the secondary $g_1$ Lund leaf and emissions from the
secondary $g_2$ leaf.
In the CFFE scheme, the region on the $q$-leaf extending from
$\eta = 0$ to the gluon's position $\eta = \eta_{g_1} = 10$, is
assigned a wrong colour factor of $C_A/2$.
For the segment and ME methods, the same region gets a correct factor of
$C_F$. In all cases a factor of $C_A/2$ is correctly applied to emissions on the $g_1$-
and $g_2$-leaves.

Note that the discrepancies in the CFFE approach are log-enhanced, and
therefore as we discussed in section~\ref{sec:ME-qgq}, can affect
subleading-colour contributions to DL terms.
For the segment method, the discrepancies are localised around the
rapidities of the emission in the parent event and they are designed
to integrate to zero.
We will explicitly test (and confirm) this below in
section~\ref{sec:me-integral}.

The results in this section demonstrate that even though our segment
and NODS corrections only ever consider the
structure of double-emission matrix elements, their iteration
reproduces the full-colour soft matrix elements for higher
multiplicities, in the appropriate angular-ordered limits. 

For a study of the NODS method for parent events that are not
angular-ordered, see Appendix~\ref{sec:ME-tests-non-ang-ordered},
which shows that discrepancies do then arise relative to the correct
full-colour matrix element, as is to be expected.
For some configurations they reach $15\%$. 

\subsection{Integrated results} \label{sec:me-integral}

\begin{figure} \centering
  \includegraphics[width=1.\textwidth,page=1]{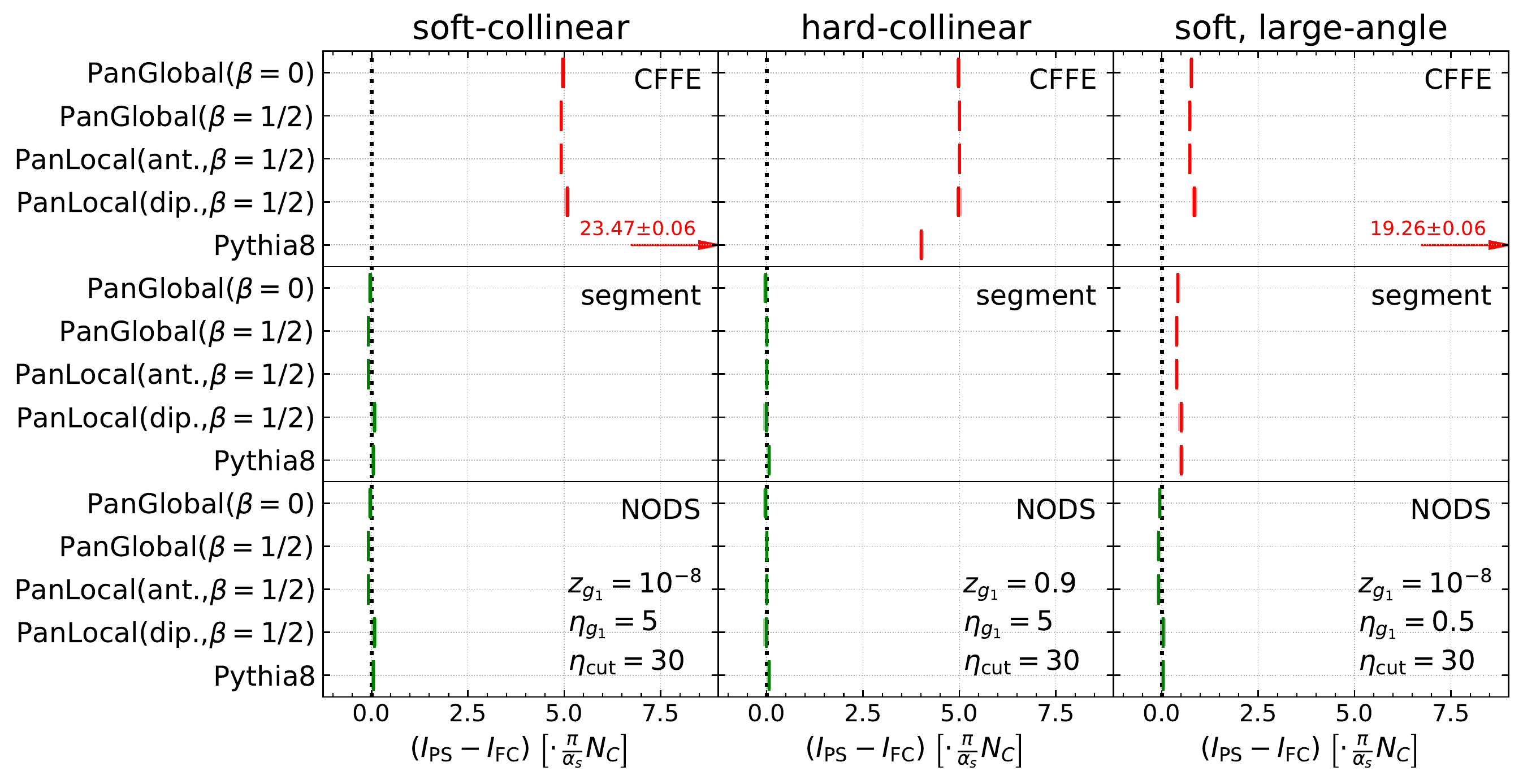}
  \caption{Normalised deviation of the integrated rate of
    parton-shower soft 
    emission (from a $\qbar g_1 q$ system), relative to analytic
    full-colour result, i.e.\ $\frac{\pi}{\alpha_s}\nc
    (I_\text{PS}-I_\text{FC})$, with $I_\text{FC}$ as given in
    Eq.~(\ref{eq:IFC-any-g1}).
    The results are shown for several parton showers and colour
    schemes for emission from a $\qbar g_1 q$ system.
    The kinematics of $g_1$ 
    are indicated in each column, and the integral bounds are set to
    $\eta_{\cut} = 30$.
    The results are colour-coded green and red, respectively,
    according to whether the result is consistent with zero, or not. 
  }
\label{fig:me-results-integral}
\end{figure}

As noted in section~\ref{sec:transition-eta-defs}, for observables
such as the multiplicity, a shower that gives the correct full-colour
rapidity and azimuth integral for the rate of soft emissions is
expected to be correct at \NDLFC accuracy.
Here we evaluate such integrals explicitly.
Keeping in mind Eqs.~(\ref{eq:ang-ordered}) and
(\ref{eq:integral-shower}), we compute the parton shower's integrated
rate of emission, $I_\PS$, for $\bar{q} g_1 q + g$, with the collinear
cut-off set to $\eta_{\rm cut} = 30$
(i.e.\ we integrate over the whole rapidity and azimuth phase-space,
but exclude cones of half-angle corresponding to $\eta_{\rm cut}$
around each parton).
Fig.~\ref{fig:me-results-integral} summarises the integral results for
the CFFE, segment, and NODS colour schemes, for each of three
different kinematic regimes for $g_1$: a soft-collinear regime, a
hard-collinear regime and a soft large-angle regime (cf.\ values of
$z_{g_1}$ and $\eta_{g_1}$ as labelled on the plots).
The known full-colour result is:
\begin{equation}
  \label{eq:IFC-any-g1}
  I_\text{FC} = \frac{2\as\, \Delta\!\ln E}{\pi} \left[ 2 C_F
 \eta_{\cut} + C_A \left(\eta_{\cut} - \eta_{g_1} - \ln
 \left(1+e^{-2\eta_{g_1}}\right)\right) \right],
\end{equation}
which extends Eq.~(\ref{eq:ang-ordered}) beyond the small-angle limit
for $g_1$
(i.e.\ Eq.~(\ref{eq:IFC-any-g1}) is valid for any $\eta_{g_1}$).
The plot shows the difference between the shower result and
Eq.~(\ref{eq:IFC-any-g1}), multiplied by
\begin{equation}
  \label{eq:I-normalisation-factor}
 \mathcal{N} = \frac{\pi}{2\alpha_s} \frac{1}{C_A/2-C_F} = \frac{\pi}{\alpha_s}\nc\,.
\end{equation}
This normalisation is chosen so that
$\mathcal{N}(I_\text{PS}-I_\text{FC})$  is
equivalent to the 
effective net extent in rapidity over which there is a $C_A/2$ versus
$C_F$ discrepancy.

For soft- and hard-collinear configurations (left and middle columns), one
observes a deviation in the integral computed with the CFFE scheme due to the
mis-assignment of a factor $C_A/2$ on the $q$-leaf, see
Fig.~\ref{fig:me-results-qgq}. The difference between the PanScales shower
family and the Pythia~8 parton shower is explained by the different separation of dipole 
elements mentioned in section~\ref{sec:cffe-algs}. In comparison, the segment
and NODS colour schemes reproduce the expression expected from angular ordering
for all the parton showers we considered.

The last configuration that we examine is when $g_1$ is soft and at
large angles (rightmost column of Fig.~\ref{fig:me-results-integral}).
In the case of the CFFE scheme, there remains a disagreement for all
parton showers, although it is again smaller for the PanScales family
than for Pythia~8.

Interestingly, the segment method also shows
a residual deviation from the analytic expression.
This expected deviation stems from our use of the small-angle limit to
determine transition points between segments also in the large-angle
region.
It can be
calculated explicitly for PanScales showers following the same
procedure as in section~\ref{sec:transition-eta-defs} for arbitrary
values of $\theta_{g_1q}$:
\begin{equation}
  \label{eq:segment-artefact}
  I_\text{PS}^{(\text{segment})} - I_\text{FC} =
  \frac{\as}{\pi} \left(C_A - 2C_F\right)
  \left\{
  \begin{array}{lll}
    \eta_{g_1} + \ln(1 + e^{-2\eta_{g_1}} ),
    &&
    |\eta_{g_1}| < \eta_t,
    \\
    \frac32 \ln(1 + e^{-2|\eta_{g_1}\!|}),
    &&
    |\eta_{g_1}| > \eta_t,
  \end{array}
  \right.
\end{equation}
where $\eta_t = \frac12\ln\frac{1+\sqrt{5}}{2} \simeq 0.241$ solves
$\eta_t = \frac12\ln(1 + 
e^{-2\eta_t})$.
The fact that Eq.~(\ref{eq:segment-artefact}) is of order $\as/\nc$
for $\eta_{g_1}$ close to zero, together with its exponential decrease
as $\eta_{g_1}$ increases, ensures that its impact on observables such
as the multiplicity is \NNDLNLC.

As expected, the NODS scheme reproduces the full-colour analytic
integral even when $g_1$ is at large angles.

\section{Numerical tests for observables}
\label{sec:numerical-tests}

In this section, we follow the approach proposed in
Ref.~\cite{Dasgupta:2020fwr} for testing the logarithmic structure of
showers with specific observables in $e^+e^-$ collisions.
Such tests implicitly probe virtual as well as real corrections and so
provide a powerful verification of the correct overall assembly of the
different shower elements.
As the reader will recall from the introduction, the tests can be
separated into two broad classes, according to whether one is probing
a double-logarithmic structure (\DL, \NDL, etc.) or an exponentiated
(\LL, \NLL, etc.) structure.
We start by outlining the structure of the tests, keeping in
mind that we will adjust the details for each specific observable.

For the double-logarithmic case,
Eq.~(\ref{eq:accuracy-non-exponentiating}), when we study some event variable
$V(\as,L)$, we will fix
\begin{equation}
  \label{eq:xi}
  \xi = \as L^2\,,
\end{equation}
and examine the behaviour of $V(\as,L)$ in the limit $\as \to 0$
(with $L$ scaling as $-1/\sqrt{\as}$).
For example to test N$^k$\DL accuracy we will study a quantity such
as\footnote{We could use a variant of~(\ref{eq:NkDL-generic-test})
  where the denominator is taken as
  $V_{\text{N}^k\text{\DL}}-V_{\text{N}^{k-1}\text{\DL}}$, except when
  it vanishes.
  This has the advantage of providing a meaningful relative deviation
  at N$^k$\DL for situations where $\delta V_{\text{N}^k\text{\DL}}$
  does not converge to zero as $\alpha_s\to 0$, and it is the choice
  that we will adopt for some of our multiplicity tests below.}
\begin{equation}
  \label{eq:NkDL-generic-test}
  \delta V_\text{N$^k$\DL} =
  \lim_{\as \to 0} \left(
    \frac{V_\text{PS}(\as, -\sqrt{\xi/\as})  -
      V_\text{N$^k$\DL}(\as, -\sqrt{\xi/\as})}{\as^{k/2}}
  \right), 
\end{equation}
where $V_\text{N$^k$\DL}$ is the known N$^k$\DL prediction from
resummation and
$V_\text{PS}$ is the result from the parton shower.
For a parton shower that is correct to N$^k$\DL accuracy,
$\delta V_\text{N$^k$\DL}$ should be zero.
Values of $\xi$ for different momentum ranges are shown in
table~\ref{tab:xi-lambda-tau}.
In practice we will often use $\xi = \as L^2 = 5$, which is towards
the upper end of the phenomenologically relevant combinations of $\as$
and $L$ accessible at the LHC.
We perform such studies for multiplicities
(section~\ref{sec:multiplicity}) and event shapes
(section~\ref{sec:evshp-DL}).

\begin{table}
  \centering
  \begin{tabular}{cccccc}
    \toprule
    $Q$ [GeV] & $\as(Q)$ & $p_{t,\min}$ [GeV] & $ \xi = \as L^2$  & $\lambda = \as L$ & $\tau$
    \\\midrule    
$ 91.2$ &  $    0.1181$ & $  1.0$   & $       2.4$ & $  -0.53$ & $      0.27$ \\
$ 91.2$ &  $    0.1181$ & $  3.0$   & $       1.4$ & $  -0.40$ & $      0.18$ \\
$ 91.2$ &  $    0.1181$ & $  5.0$   & $       1.0$ & $  -0.34$ & $      0.14$ \\
\midrule
$ 1000$ &  $    0.0886$ & $  1.0$   & $       4.2$ & $  -0.61$ & $      0.36$ \\
$ 1000$ &  $    0.0886$ & $  3.0$   & $       3.0$ & $  -0.51$ & $      0.26$ \\
$ 1000$ &  $    0.0886$ & $  5.0$   & $       2.5$ & $  -0.47$ & $      0.22$ \\
\midrule
$ 4000$ &  $    0.0777$ & $  1.0$   & $       5.3$ & $  -0.64$ & $      0.40$ \\
$ 4000$ &  $    0.0777$ & $  3.0$   & $       4.0$ & $  -0.56$ & $      0.30$ \\
$ 4000$ &  $    0.0777$ & $  5.0$   & $       3.5$ & $  -0.52$ & $      0.26$ \\
\midrule
$20000$ &  $    0.0680$ & $  1.0$   & $       6.7$ & $  -0.67$ & $      0.45$ \\
$20000$ &  $    0.0680$ & $  3.0$   & $       5.3$ & $  -0.60$ & $      0.34$ \\
$20000$ &  $    0.0680$ & $  5.0$   & $       4.7$ & $  -0.56$ & $      0.30$ \\
    \bottomrule
  \end{tabular}
  \caption{Values of $\xi = \as L^2$, $\lambda = \as L$ and $\tau$
    (defined in Eq.~(\ref{eq:HU-tau-def})) for various upper ($Q$) and
    lower ($p_{t,\min}$) momentum scales.
    The coupling itself is in a 5-loop variable flavour number
    scheme~\cite{Chetyrkin:2005ia,Baikov:2016tgj,Luthe:2016ima,Herzog:2017ohr},
    while $\tau$ is evaluated for 1-loop evolution with $n_f = 5$. }
  \label{tab:xi-lambda-tau}
\end{table}

For observables whose logarithmic prediction exponentiates,
Eq.~(\ref{eq:accuracy-exponentiating}), we can
study $\ln V(\as, L)$, taking the limit of $\as \to 0$ with fixed
\begin{equation}
  \label{eq:lambda}
  \lambda = \as L\,.
\end{equation}
To test N$^k$\LL accuracy we can examine
\begin{equation}
  \label{eq:NkLL-generic-test}
  \delta \ln V_\text{N$^k$\LL} =
  \lim_{\as \to 0} \left(
    \frac{\ln V_\text{PS}(\as, \lambda/\as)  -
      \ln V_\text{N$^k$\LL}(\as, \lambda/\as)}{\as^{k-1}}
    \right), 
\end{equation}
where $V_\text{N$^k$\LL}$ is the known N$^k$\LL prediction from resummation.
For a shower that is correct to N$^k$\LL accuracy,
$\delta \ln V_\text{N$^k$\LL}$ should be zero.
For a shower that is incorrect at \LL accuracy, the unnormalised
difference between $\ln V_\text{PS}$ and $\ln V_\text{\LL}$ diverges
for $\as \to 0$, hence Eq.~(\ref{eq:NkLL-generic-test}) multiplies
that difference by $\as$ to give a finite, non-zero
result.\footnote{Alternatively, we may examine the $\as \to 0$ limit
  of
  $[\ln V_\text{PS}(\as, \lambda/\as)]/[\ln V_\text{LL}(\as,
  \lambda/\as)]-1$, which is of interest in some cases with LL
  discrepancies, because it gives a direct measure of the relative discrepancy in
  the logarithm of the Sudakov factor.
} 
In practice we will often use $\lambda = \as L = -0.5$.
This corresponds to a slightly narrower range of logarithm than our
choice for $\xi$, in part to help mitigate some of the technical
difficulties of the $\as \to 0$ limit.
We perform such studies for event shapes (section~\ref{sec:evshp-NLL})
and non-global logarithms (section~\ref{sec:NGL}).

Generation with very small $\as$ and fixed $\xi$ or $\lambda$ is often
difficult.
Many of the techniques that we use were outlined in the
supplemental material to Ref.~\cite{Dasgupta:2020fwr}.
For the work presented here we added three main new advances:
\begin{enumerate}
\item We implemented a weighted generation technique that is
  equivalent to evolving multiple replicas of an event, discarding a
  replica when it emits into a region of phase-space that we wish to
  veto, and then adjusting the number of replicas and their weights so
  as to continue generating with the original effective number of
  replicas (cf.\ section~3 of Ref.~\cite{Lonnblad:2012hz}).
  For the combinations of $\as$, shower and event-shape that were most
  challenging in Ref.~\cite{Dasgupta:2020fwr}, this enabled us to save
  about an order of magnitude in computing time, associated with
  accessing regions with very strong Sudakov suppression.
  It also enabled us to reach small $\as$ values that were simply not
  feasible in Ref.~\cite{Dasgupta:2020fwr}, facilitating the
  extrapolation to $\as = 0$.
  
\item We adjusted the shower implementation so that it can track
  differences in directions between neighbouring particles in the
  dipole chain.
  This works around issues that arise in normal shower implementations
  where it becomes difficult to determine angles between particles
  (and dot products, etc.)\ when those angles go below machine
  precision $\epsilon$.
  This, together with the next point, was especially useful in
  allowing for smaller $\as$ and larger values of the (absolute)
  logarithm in double-logarithmic tests, though it also facilitated
  cutoff dependence tests in the \NLL event-shape studies.
  It has a small $\sim 30\%$ speed penalty, and some implementation
  overhead, but avoids the need for the double-double and quad-double
  types~\cite{hida2000quad} used, with substantially larger speed penalties, in
  Ref.~\cite{Dasgupta:2020fwr}.
  
\item To allow for momenta across such disparate scales that the
  logarithm of the ratio of scales is truly large
  ($|L| \gtrsim \frac12\log \mu \simeq 354 $ where
  $\mu\simeq 1.8\times10^{308}$ is the maximum number that can be
  represented in double precision), we implemented a new floating type
  that supplements a normal double-precision number with a 64-bit
  integer to store the exponent, replacing the usual 11 exponent bits
  in a double precision number.
  This came with a speed penalty relative to double precision numbers
  of about $\times4$, but was substantially faster than solutions we
  investigated based on the MPFR~\cite{fousse2007mpfr} or Boost
  multiprecision libraries.
  It was particularly valuable for double-logarithmic tests and useful
  also for tests of non-global observables.
\end{enumerate}
Throughout we will run with the physical colour factors, $\nc=3$,
$C_F = (\nc^2-1)/(2\nc) = 4/3$, $C_A = \nc$ and $n_f=5$ (except for
leading-$\nc$ comparison results, where we use $C_F = \frac12 C_A = \frac32$).

\subsection{Particle multiplicity}
\label{sec:multiplicity}

One of the most powerful tests of a parton shower is its prediction
for the particle or subjet multiplicity.
To reproduce the NDL multiplicity requires correct modelling of a
variety of aspects of a parton shower, including nested $g \to gg$ and
$g\to q\bar q$ splittings.
Ref.~\cite{Dasgupta:2020fwr} demonstrated agreement between a range of
dipole showers and \NDLLC predictions.
We expect the segment and NODS schemes of
sections~\ref{sec:transition-points} and \ref{sec:ME-solution} to
bring \NDLFC agreement.
For example, in the segment method, the general pattern of $C_F$
dipole segments associated with the Born $\qbar q$ pair should
guarantee \DLFC terms, while the specific choices of transition points
for the segments, together with the $C_F$ segments for $g \to
q'\qbar'$ splittings, should ensure \NDLFC accuracy.

Strictly speaking, the particle multiplicity is infrared and collinear
(IRC) unsafe.
A closely related, but IRC-safe, quantity is the subjet
multiplicity~\cite{Catani:1991pm} in the $k_t$ jet
algorithm~\cite{Catani:1991hj}.
Ref.~\cite{Dasgupta:2020fwr} directly computed that subjet
multiplicity comparing to the results of Ref.~\cite{Catani:1991pm}.
Here we take a slightly different approach that avoids the need for
jet clustering, but can still be directly compared to the results of
Ref.~\cite{Catani:1991pm}.
We run the shower as normal, but set the running coupling to zero
below some threshold transverse momentum $k_{t,\text{cut}}$.
At \NDL accuracy, the multiplicity with such a procedure should be
identical to that with a clustering threshold for the subjets of
$y_\text{cut} = k_{t,\text{cut}}^2/Q^2$.
Only starting from \NNDL do we expect differences to arise between the
multiplicity for a shower definition of a $k_t$ cutoff and the $k_t$
algorithm definition.\footnote{Those differences can for example arise
  because of details of the specific definition of transverse momentum
  in the collinear limit, and because of effects related to the jet
  algorithm's clustering properties.}
We have explicitly verified that this is the case at leading colour.
 
\begin{figure}
  \centering
  \includegraphics[width=0.6\textwidth,page=1]{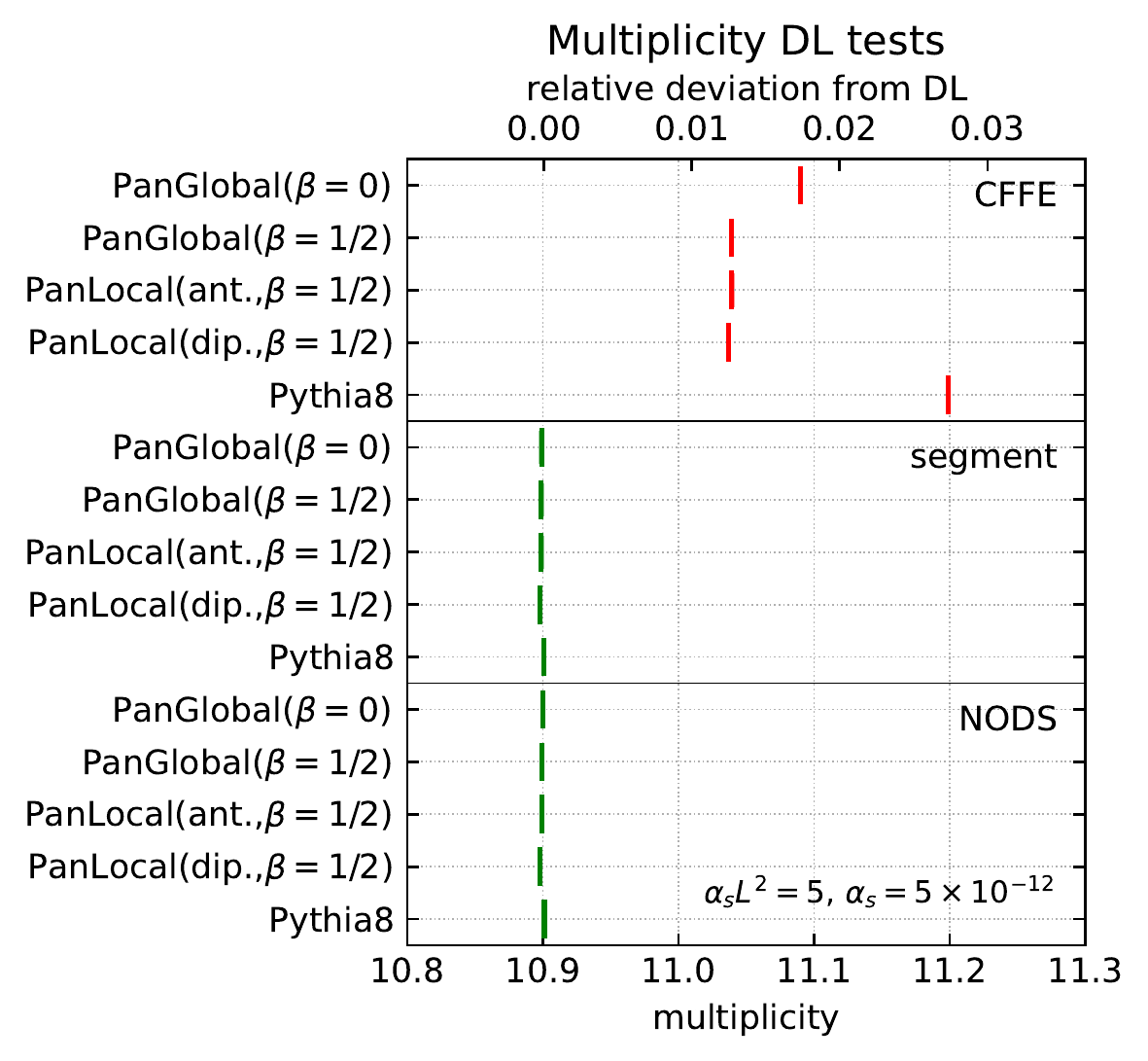}
  \caption{
    DL shower multiplicity tests, showing results for the parton multiplicity
    at $\xi = \as L^2 = 5$ in various parton showers and colour
    schemes, for a sufficiently small value
    of $\as$ that \NDL terms can be neglected.
    The DL expectation from Eq.~(\ref{eq:NDL-multiplicity}) is 
    $N_\text{\DL} \simeq 10.9008$.
  }
  \label{fig:multiplicity}
\end{figure}

The analytic expression for the multiplicity in $e^+e^- \to q\qbar$
events, up to and including \NDL
terms, can be straightforwardly extracted from
Ref.~\cite{Catani:1991pm},
\begin{align}
  N & = 2\frac{C_F}{C_A}
      \left(1+\frac{b(2B-1)}{8\pi}\sqrt{\alpha_s\xi}\right)
      \cosh\left(\sqrt{\frac{2C_A}{\pi}\xi}\right)
      + 2\left(1-\frac{C_F}{C_A}\right)
      \left(1+\frac{n_f C_F}{16\pi C_A}\sqrt{\alpha_s\xi}\right)
      \nonumber\\
    & + \sqrt{\frac{\alpha_s\xi}{32\pi C_A}}\frac{2C_F}{C_A}
      \left[6C_A + \left(\frac{5}{2}-3B-\frac{C_A\xi}{\pi}\right)b\right]
      \sinh\left(\sqrt{\frac{2C_A}{\pi}\xi}\right),      
      \label{eq:NDL-multiplicity}
\end{align}
in the approximation that $|L| \gg 1$, with $b=\frac{11C_A-2n_f}{3}$
and $B=1+\frac{8n_f(C_A-C_F)}{3bC_A}$. 
Fig.~\ref{fig:multiplicity} shows the multiplicity $N$ for
$\xi = 5$ at a value of $\as = 5\times 10^{{-}12}$, which is sufficiently tiny that
one can neglect subleading corrections to within an accuracy of
$\sqrt{\as} \simeq 2.24\times 10^{{-}6}$.
The figure compares several shower algorithms (the main PanScales
showers and our implementation of the Pythia~8 shower), and three
colour schemes, with the analytic \DL result.
We see that the CFFE scheme, the default in many
showers including Pythia~8, differs by up to $\sim 3\%$ from the
analytic result.
A $3\%$ difference is not a huge effect, but it remains a
subleading-$\nc$ \DL difference and gives a measure of the practical
impact of such contributions, i.e.\ $\sim 1/(3 \nc^2)$.
For comparison, the differential matrix element plots,
Figs.~\ref{fig:me-results-qgq} and \ref{fig:me-results-qqqqg-qqggg},
showed $\sim 10\%$ differences for the CFFE approach in extended phase-space regions.
Our new segment and NODS approaches coincide well with the \DL
analytic multiplicity result for all showers, including when applied
to the Pythia~8 shower.

\begin{figure}
  \centering
  \begin{subfigure}{0.49\textwidth}
    \includegraphics[width=\textwidth,page=1]{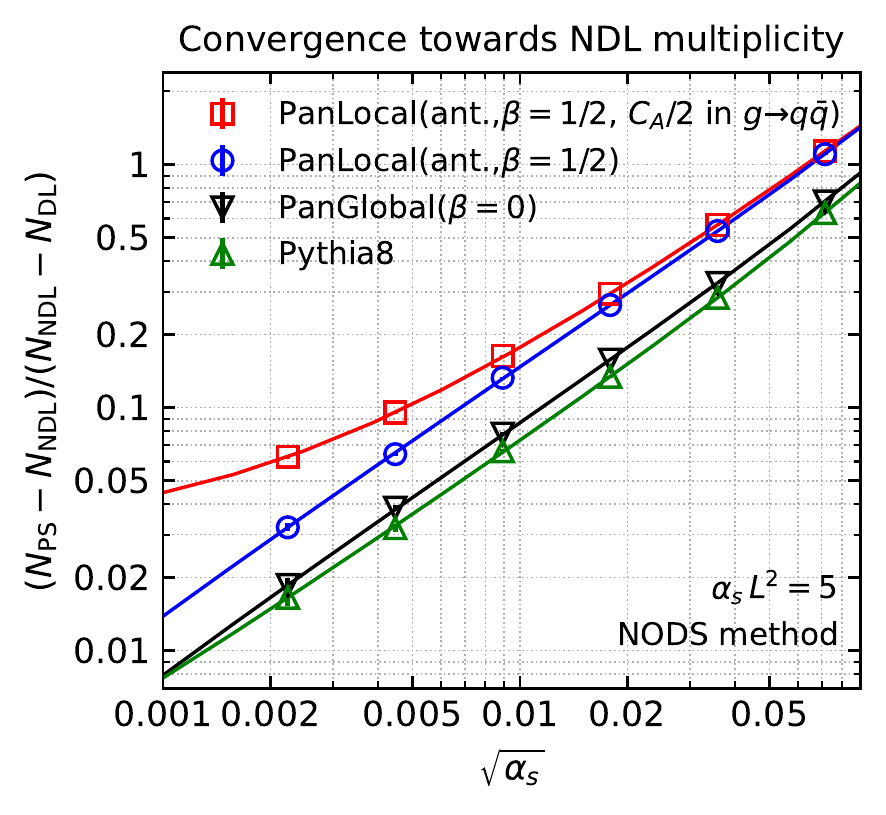}\hfill%
    \caption{}
    \label{fig:NLL-mult-v-alphas}
  \end{subfigure}%
  \begin{subfigure}{0.5\textwidth}
    \includegraphics[width=\textwidth,page=1]{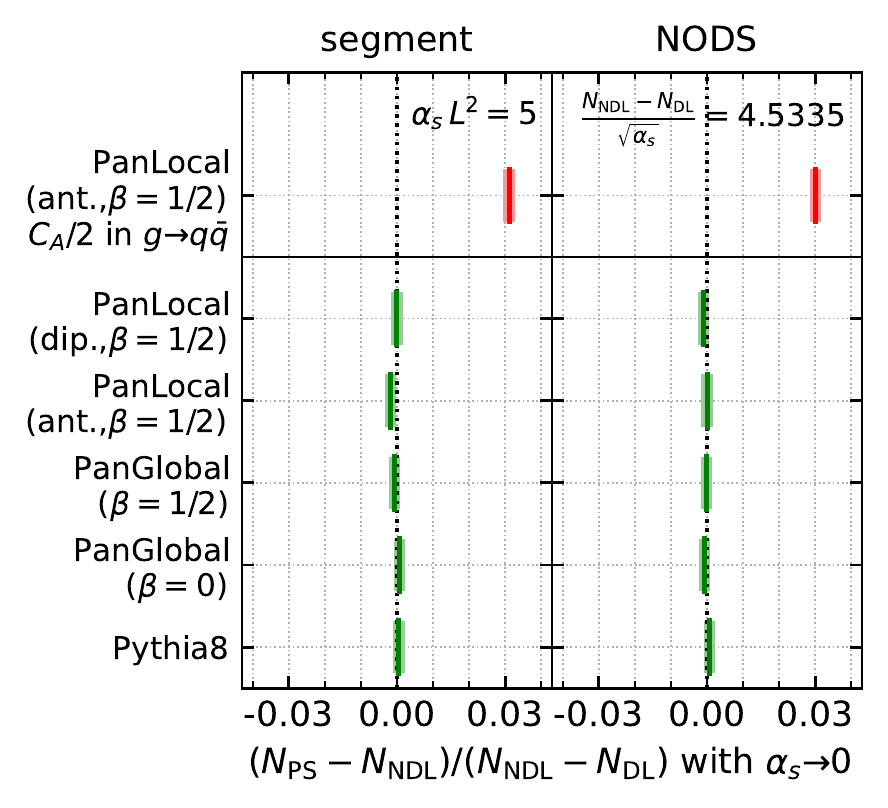}
    \caption{}
    \label{fig:NLL-mult-extrap}
  \end{subfigure}
  \caption{
    NDL shower multiplicity tests.
    (a)
    Results for $\frac{N_\text{PS}-N_\text{\NDL}}{N_\text{\NDL}-N_\text{\DL}}$,
    for a variety of parton showers in the NODS colour scheme, as a
    function of $\sqrt{\as}$, showing that they vanish as $\as \to
    0$, as required to achieve full-colour \NDL accuracy.
    The curves depict a fit  that is a polynomial in powers of $\sqrt{\as}$.
    (b) Explicit extrapolation of
    $\frac{N_\text{PS}-N_\text{\NDL}}{N_\text{\NDL}-N_\text{\DL}}$
     to $\as = 0$, for a range
     of showers, with the segment and NODS colour schemes.
    In both plots we also illustrate the \NDL-level discrepancy that arises in a 
    scheme where
    quarks from $g \to q\bar q$ splittings erroneously continue to emit with a
    $C_A/2$ colour factor.
  }
  \label{fig:NLL-multiplicity}
\end{figure}

To test the \NDL multiplicity terms, we need to examine
$\delta N_\text{\NDL}$, as defined in Eq.~(\ref{eq:NkDL-generic-test}).
For this comparison, we need $\as$ to be large enough for $\sqrt{\as}$
terms in the multiplicity to be visible after summing over a finite
number of Monte Carlo events, while also small enough that we can
safely extrapolate away $\order{\as}$ terms.
To this end, Fig.~\ref{fig:NLL-mult-v-alphas} shows
\begin{equation}
  \label{eq:mult-observable}
  \frac{N_\text{PS}-N_\text{\NDL}}{N_\text{\NDL}-N_\text{\DL}}
   = \frac{N_\text{PS}-N_\text{\DL}}{N_\text{\NDL}-N_\text{\DL}}-1,
\end{equation}
a quantity conceptually similar to $\delta N_\text{\NDL}$ in
Eq.~(\ref{eq:NkDL-generic-test}) (since
$N_\text{\NDL}-N_\text{\DL}\propto\sqrt{\as}$), as a function of
$\sqrt{\as}$.
It shows results for the NODS method for several showers and one sees
that most of the results go to zero as $\as \to 0$, with the
exception of the red curve, to which we return below (the segment
method, not shown, gives almost identical results).
The result of the actual $\as \to 0$ extrapolation (using a cubic
polynomial extrapolation based on
$\alpha_s=\{0.000005, 0.00032, 0.00128, 0.00512\}$) is shown in
Fig.~\ref{fig:NLL-mult-extrap} for each shower and for each of the
segment and NODS methods.
All results are consistent with the (full-colour) \NDL result, to
within the small statistical errors.

Fig.~\ref{fig:NLL-multiplicity} also includes results (red points)
obtained using a deliberately erroneous prescription that omits the
insertion of new $C_F$ segments following $g\to q\bar q$ branchings,
resulting in those quarks emitting with a $C_A/2$ colour
factor. 
While this prescription gives the correct \DLFC result, one should
expect it to fail to reproduce the \NDLFC results, because
$g\to q\qbar$ splittings start to contribute to the multiplicity from
NDL terms onwards, as can be verified by inspecting the $n_f$ terms in
Eq.~(\ref{eq:NDL-multiplicity}).%
\footnote{The prescription bears similarities with that of
  Ref.~\cite{Friberg:1996xc}, which concentrated on corrections of the
  colour factor for primary gluon emissions.}
We see in Fig.~\ref{fig:NLL-mult-v-alphas} that with this incorrect
treatment of $g\to q\bar q$ splittings, the limit $\sqrt{\alpha_s}\to
0$ fails to converge to the \NDL expectation. Instead the extrapolated
\NDL coefficient is $\sim 3\%$ larger than the analytic expectation.

We have also carried out similar tests for the multiplicity in
$H \to gg$ events for all showers shown in
Fig.~\ref{fig:NLL-multiplicity} and found a similar level of agreement
with the full-colour \NDL predictions for both the segment and NODS
colour prescriptions.
\logbook{2ec6e34c}{See ../../2020-eeshower/analyses/multiplicity/test-h2gg.pdf}

\subsection{Event shapes}
\label{sec:evshp}

The main event shapes that we consider here are obtained by
considering all primary Lund declusterings~\cite{Dreyer:2018nbf}, and
for each declustering evaluating,
\begin{equation}
  \label{eq:ui-def}
  u_i^{\beta_\text{obs}}
  \equiv
  \frac{k_{ti} e^{-\beta_\text{obs} |\eta_i|}}{Q},
\end{equation}
where $k_{ti}$ is the transverse momentum of the declustered subjet $i$
with respect to its partner direction, and
$\eta_i = \ln \tan\theta_{i}/2$, with $\theta_i$ the angle of the
declustered subjet with respect to the partner.
The parameter $\beta_\text{obs}$ determines the relative weighting of
different rapidities and we will consider three values,
$\beta_\text{obs} = 0, \frac12, 1$.
We will study two combinations of the $u_i$,
\begin{equation}
  \label{eq:ui-combinations}
  M_{\beta_\text{obs}} \equiv \max_i \left\{u_i^{\beta_\text{obs}}\right\}\,, \qquad
  S_{\beta_\text{obs}} \equiv \sum_i u_i^{\beta_\text{obs}}\,.
\end{equation}
For each event shape, we define $\Sigma(\as,L)$ to be the fraction of
events for which that event shape has a value smaller than a threshold
defined to be $e^{-|L|}$. 
Up to \NLL accuracy, $\Sigma(\as,L)$ for $M_{\beta_\text{obs}=0}$
coincides with that for the Cambridge $\sqrt{y_{23}}$ resolution
scale~\cite{Dokshitzer:1997in}, and $S_{\beta_\text{obs}=1}$ with that
for one minus the thrust (below, in section \ref{sec:evshp-NLL}, we will
consider those explicitly as well).

Ref.~\cite{Dasgupta:2018nvj} showed that the CFFE procedure led to
spurious subleading-$\nc$ terms starting from order $\as^2 L^4$ in
standard dipole showers.
With similar reasoning it is straightforward to show that the issue is
present also for the PanScales showers with the CFFE
approach.
Those issues are caused by mis-attribution of a $C_A/2$ colour factor
to emissions that should be seen as coming from the primary $\qbar q$
pair.
Once that issue is fixed, one should obtain \NLLFC accurate results
for all PanScales showers that already give \NLLLC accuracy.

Given that the CFFE issue arises at order $\as^2 L^4$, one expects to
be able to observe it numerically in both \DL and \LL-style tests.
As we shall see below, numerical \LL and \NLL tests
require us to prune the shower branchings, so as to keep
multiplicities under control in the limit $\as \to 0$ for fixed
$\lambda = \as L$.
That pruning can be delicate in situations with \DL discrepancies.
Accordingly we first carry out \DL tests, for which we can run
complete, unpruned showers.

When it helps to limit notational ambiguity with respect to $\beta_\text{obs}$,
in some cases below we will write $\beta_\text{PS}$ instead of $\beta$
for the parameter that determines the shower ordering variable (cf.\ 
Eq.~(\ref{eq:evolution-variable})).

\subsubsection{Double-logarithmic study}
\label{sec:evshp-DL}

\begin{figure}
  \centering
  \includegraphics[width=0.8\textwidth,page=1]{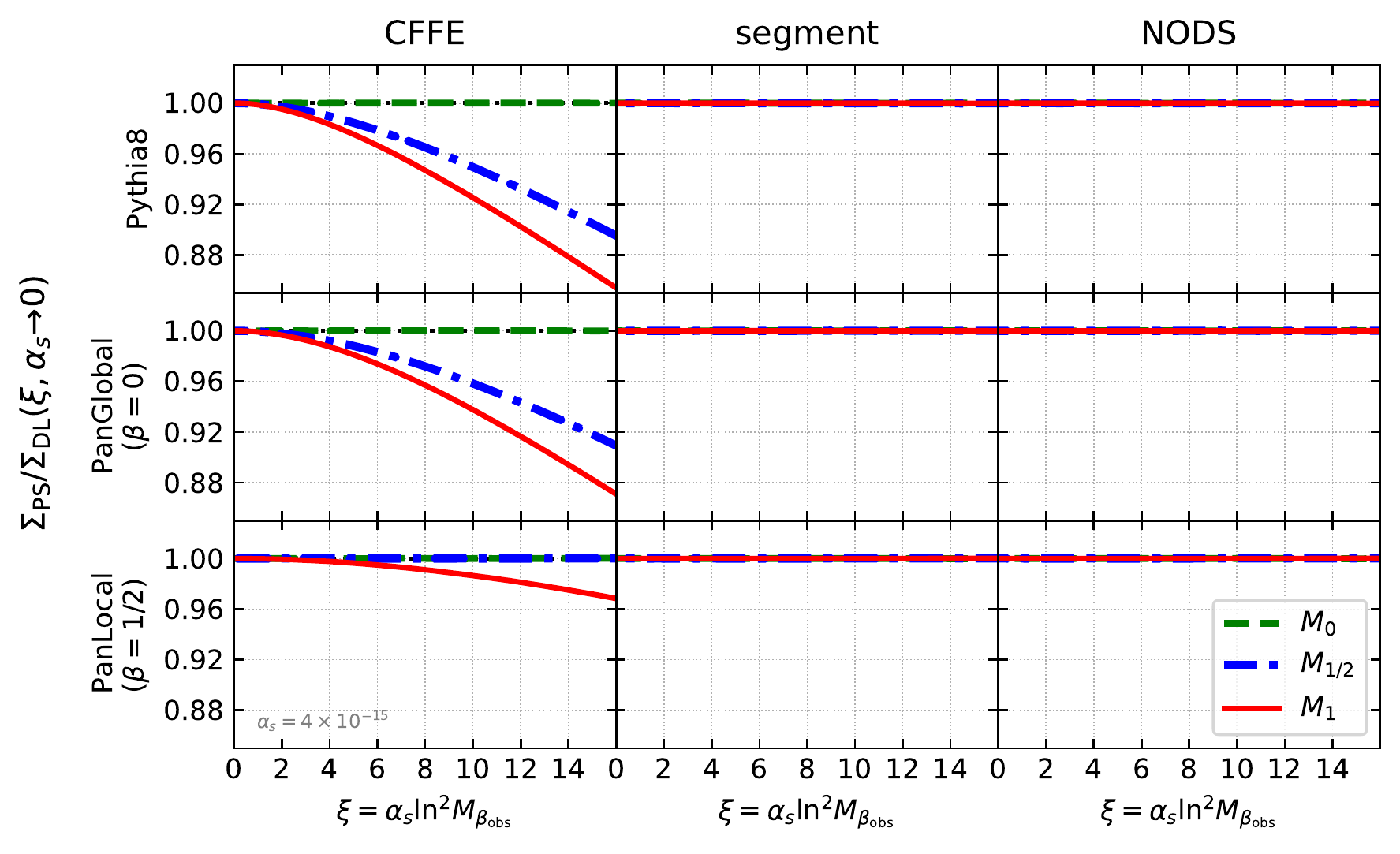}%
  \caption{
    Double-logarithmic tests for event shapes, showing the
    ratio $\Sigma_\text{PS}/\Sigma_\text{DL}$ in the limit of $\as
    \to 0$ for fixed $\xi = \as \ln^2 M_{\beta_\obs}$, as a function
    of $\xi$.
    The $M_{\beta_\obs}$ observables are defined in
    Eq.~(\ref{eq:ui-combinations}), with $M_0$ equivalent to the
    Cambridge $y_3$ clustering scale, and $M_1$ similar to the thrust
    at DL accuracy.
    Each row shows a different parton-shower algorithm, each
    column a different colour scheme, and each subplot shows
    results for three values of $\beta_\obs$.
    Showers that are \DLFC accurate give a ratio of $1$ for all $\xi$.
    The CFFE scheme shows DL discrepancies for at least some of the
    observables for each shower.
    The segment and NODS schemes agree with the \DLFC expectation.
  }
  \label{fig:DL-evshp}
\end{figure}

To enable us to study the subleading colour double-logarithmic
contributions, Fig.~\ref{fig:DL-evshp} shows
$\Sigma_\PS/\Sigma_\text{DL}$ as a function of $\xi = \as L^2$, in the
limit of $\as \to 0$, for $M_{\beta_\text{obs}}$ observables with
three different values of $\beta_\text{obs}$.
It includes results for two PanScales showers and for Pythia~8, for
all three colour schemes.
To understand the plots, it is useful to recall that the PanScales
showers are ordered in a variable
$\sim k_t e^{-\beta_\text{PS} |\eta|}$ (we show PanGlobal with
$\beta_\text{PS}=0$ and PanLocal with $\beta_\text{PS}=1/2$).
We expect (and see) that double logarithmic discrepancies are present
in the CFFE scheme for any
$\beta_\text{obs} > \beta_\text{PS}$.%
\footnote{For $\beta_\text{obs} \le \beta_\text{PS}$, a given contour of
  fixed $M_{\beta_\text{obs}}$ is traversed by the shower in order of
  increasing absolute rapidity, which means that relevant emissions
  occur either on the primary $\qbar q$ dipole, or on a $\qbar g$ or
  $g q$ dipole.
  Taking the example of a gluon emitted at some rapidity $y_g>0$ and
  forming a $gq$ dipole, emissions that are closer to the quark than
  to the gluon, i.e.\ along the quark with rapidity $y>y_g$, are
  emitted with the correct $C_F$ colour factor.
  Since the only emissions that can increase the event shape value are
  those that occur at rapidities larger than previous gluon emissions,
  one obtains the correct full-colour double logarithmic result.
  \label{fn:CFFE-PanScales-expectations}
}
For Pythia~8, which is $k_t$ ordered, we expect \DL discrepancies to
be present in the CFFE scheme for any $\beta_\text{obs} \neq 0$, though
our tests only consider $\beta_\text{obs} \ge 0$.
Discrepancies at \DL accuracy are clearly visible as deviations of the
curves from $1$.
In the physically accessible region, $\xi \lesssim 7$, those
discrepancies are fairly modest, no more than $2{-}3$ percent.
The results for the segment and NODS colour scheme are all consistent
with $1$.
While shown only for a subset of showers and for the
$M_{\beta_\text{obs}}$ observables, the results for other PanScales
showers are the same for any given $\beta_\text{PS}$ and for the
$S_{\beta_\text{obs}}$ observables.

In the case of the PanScales showers, one can also calculate the
analytic form of the CFFE discrepancies, cf.\
Appendix~\ref{sec:DL-evshp-analytics}.
The analytic results agree well with the observed shower
discrepancies.
It is interesting to consider the large-$\xi$ limit of the
discrepancy, which will be relevant also in understanding the results
of the next subsection.
For example, for the thrust determined with any of the PanScales
showers with $\beta_\text{PS}=0$, we find that
\begin{equation}
  \label{eq:asymptotic-DL-CFFE-evshp}
  \lim_{\xi \to \infty} \lim_{\as \to 0}
  \frac{\ln \Sigma_{\beta_\text{PS}=0}(\as,-\sqrt{\xi/\as})}%
  {\ln \Sigma_{\DL}(\as,-\sqrt{\xi/\as})} 
  = 1 + \frac{1}{2\nc^2 -1}\,.
\end{equation}
However, the approach to this limit is very slow: for $\xi = 5$ the
ratio is about $1.009$, to be compared to an asymptotic value of
$1 + 1/17 \simeq 1.0588$.

\subsubsection{\LL and \NLL studies}
\label{sec:evshp-NLL}

Our next set of tests is to carry out \LL and \NLL studies for the
same set of event shapes discussed above, supplemented with the total
and wide-jet broadenings ($B_T, B_W$)~\cite{Catani:1992jc}, the Cambridge
$y_{23}$ jet resolution parameter~\cite{Dokshitzer:1997in}, 
thrust~\cite{Farhi:1977sg} and fractional energy-energy correlation
moments ($\text{FC}_x$) \cite{Banfi:2004yd}.
We consider the limit $\as \to 0$ for fixed $\lambda = \as L$.

A fundamental difficulty with these \LL and \NLL studies is that the
logarithm of the multiplicity, $\ln N$, scales as $\sqrt{\as L^2}$,
cf.\ Eq.~(\ref{eq:NDL-multiplicity}), and when we fix
$\lambda = \as L$ and take the limit $\as \to 0$, the logarithm of the
multiplicity blows up as $\sqrt{1/\as}$.
Since generation time and memory consumption scale at least in
proportion to $N$, event generation becomes prohibitively expensive
with too small a value of $\as$.

To address this problem we adopt a strategy where, at each stage of
the shower, branchings that are guaranteed to be irrelevant to the
observable are vetoed.
Specifically, for a given $\beta_{\obs}$, we track the maximum value
of
$O_\text{approx} = k_{t,\text{approx}} e^{-\beta_\obs
  |\eta_\text{approx}|}$ that has occurred so far in the showering of
the event (using Eq.~(\ref{eq:eta-approx}) for $\eta_\text{approx}$
and $k_{t,\text{approx}}$ equivalent to the $\kT$ in
Eq.~(\ref{eq:evolution-variable})).
We accept a given new branching only if it has
$O_\text{approx} > e^{-\Delta} O_\text{approx,max}$, with $\Delta$ a
parameter to be chosen appropriately.
The logarithm of the multiplicity is then expected to scale roughly as
$\sqrt{\as \Delta^2} + \ln |\lambda \Delta|$. 
The value of $\Delta$ should be small enough that the multiplicity
remains under control, and large enough that recursively IRC safe
observables~\cite{Banfi:2004yd} are not affected by it, which is
ensured by the requirement $e^{-\Delta} \ll 1$.

For shower--observable combinations where there is a
double-logarithmic subleading-$\nc$ effect, we also need to be aware
that the mechanism generating that effect may only be fully
operational if $\as \Delta^2 \gg 1$, which is in tension with the
constraint on the multiplicity.
In practice it is difficult to ensure that the full DL discrepancy is
captured in LL tests, especially when considering the interplay
between the $\as \to 0$ and $\Delta \to \infty$ limits.
Our approach will instead be to ensure that the combinations of $\as$
and $\Delta$ are such that the presence of any DL issues is correctly
diagnosed in our LL tests, even if we do not reproduce the exact value
of any discrepancy.

\begin{figure}
  \centering
  \includegraphics[width=0.9\textwidth,page=5]{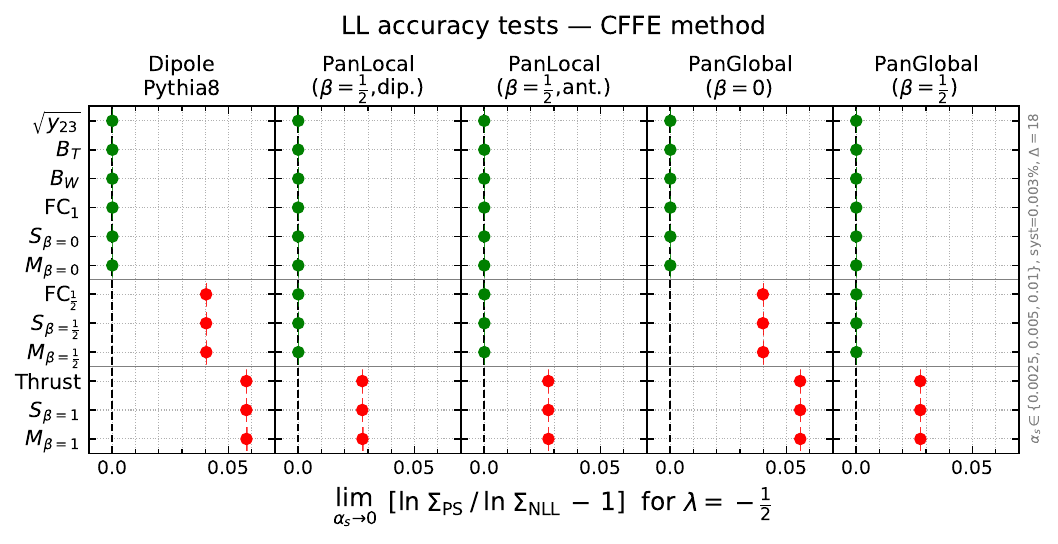}
  \caption{
    \LL test of the CFFE scheme for event shapes, showing the expected
    pattern of discrepancies for the PanScales and Pythia~8 showers.
    For further details, see text.
  }
  \label{fig:global-LL}
\end{figure}

This is illustrated in Fig.~\ref{fig:global-LL}, which shows LL tests
for the CFFE colour approach, using $\Delta = 18$ and a quadratic
polynomial extrapolation to $\as=0$ based on runs at
$\as = \{0.0025,0.005,0.01\}$.
There, one clearly sees that the set of shower--observable
combinations that fails the LL test is consistent with the
expectations from the DL tests in Fig.~\ref{fig:DL-evshp} and from
Appendix~\ref{sec:DL-evshp-analytics}.
However, taking the example of the $\beta_\text{obs}=1$ observables,
for the PanGlobal shower with $\beta_\text{PS} = 0$, one sees a
discrepancy of $0.056$, to be compared to the expectation of $0.068$
in table~\ref{tab:LL-cffe-deviations} of
Appendix~\ref{sec:DL-evshp-analytics}, as obtained when one first takes the limit
$\Delta \to \infty$, and then $\as \to 0$ (this $0.068$ is the running
coupling analogue of Eq.~(\ref{eq:asymptotic-DL-CFFE-evshp})).
\logbook{14310b42}{Studies relating to the poor
  quality of the extrapolation function for LL tests are now logged in
  logbook/2020-02-27-full-colour-LL, in section 13.2 and fig. 31.
  Similarly, variation in (N)LL tests under changing the $\as$ extrapolation
  points are documented in section 13.3 there, figs 32-33 (LL) and
  figs 34-35 (NLL).}

\begin{figure}
  \centering
  \includegraphics[width=0.9\textwidth,page=15]{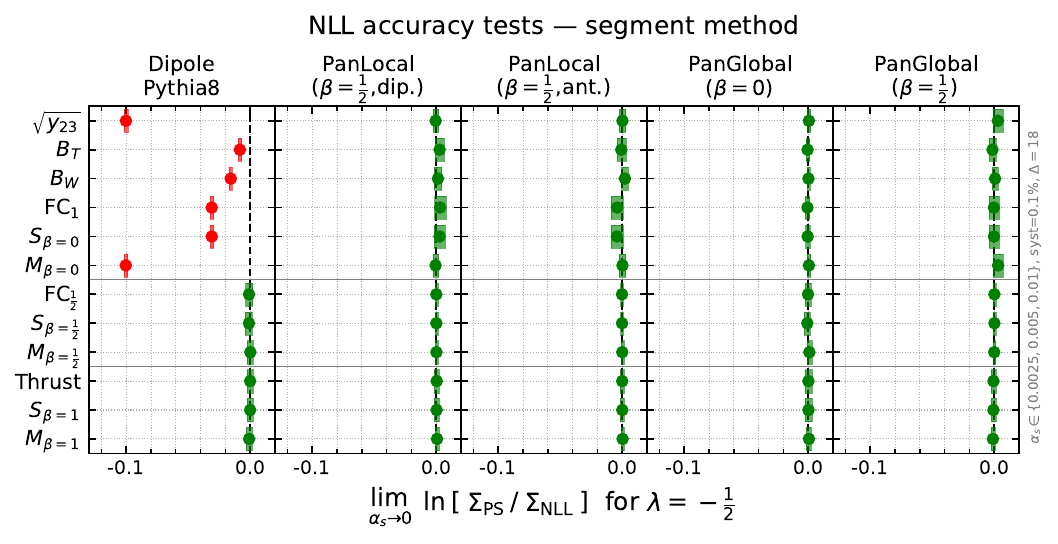}\\
  \includegraphics[width=0.9\textwidth,page=16]{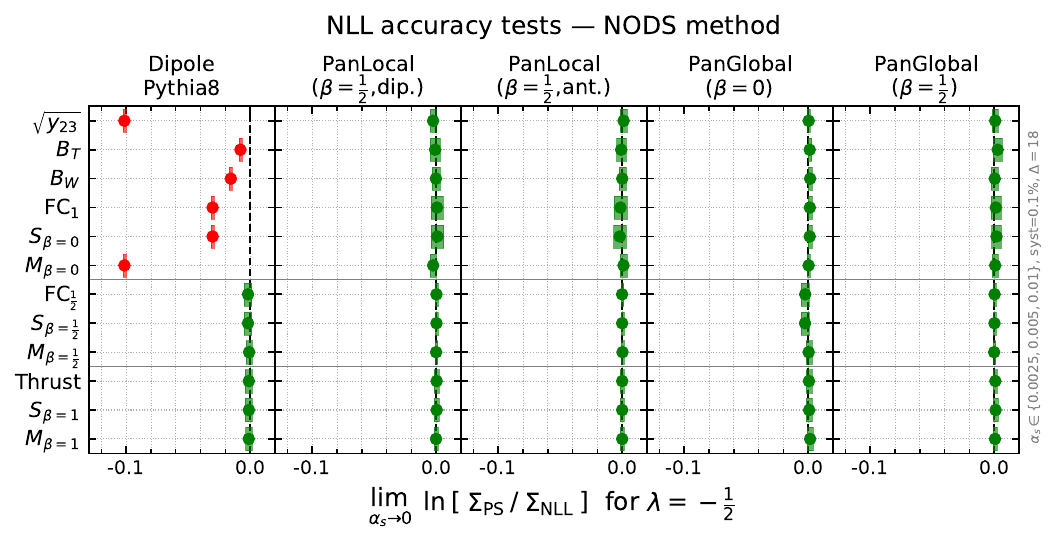}
  \caption{
    \NLL global event-shape tests of the segment and NODS colour
    schemes, showing \NLL agreement for $\beta = 1/2$ PanScales showers
    and for the $\beta = 0$ PanGlobal shower.
    In contrast to the \NLLLC tests of Ref.~\cite{Dasgupta:2020fwr},
    the Pythia~8 $\beta_\obs > 0$ results here are coloured green
    rather than amber, because our colour code does not incorporate
    the information about failure of exponentiation in fixed-order
    tests, tests that we have not explicitly repeated for this
    paper. }
  \label{fig:global-NLL}
\end{figure}

The \LL tests for the segment and NODS colour schemes (not shown in
Fig.~\ref{fig:global-LL}) are consistent with the analytic \LL results
for all observables and showers, including Pythia~8.
Accordingly in Fig.~\ref{fig:global-NLL}, where we show the
full-colour \NLL tests, i.e.\ examining
$\delta \ln \Sigma_\text{\NLL}$, Eq.~(\ref{eq:NkLL-generic-test}), we
include results just for those two schemes.%
\footnote{One can also carry out NLL tests for
  observable--shower combinations that have the correct \LLFC result
  in the CFFE approach.
  For the PanScales showers, where \LLFC results were correct for
  $\beta_\text{obs} \ge \beta_\PS$, one finds that most observables
  are correct at \NLLFC only for $\beta_\text{obs} > \beta_\PS$, with
  the exception of those in the max-type class, which remain correct
  for $\beta_\text{obs} \ge \beta_\PS$.
  \logbook{14310b42}{NLL terms for things that fail at LL: Fig. 30 on
    p.49 of logbook 2020-02-27-full-colour-LL.pdf, or pp.\ 9 and 13 of
    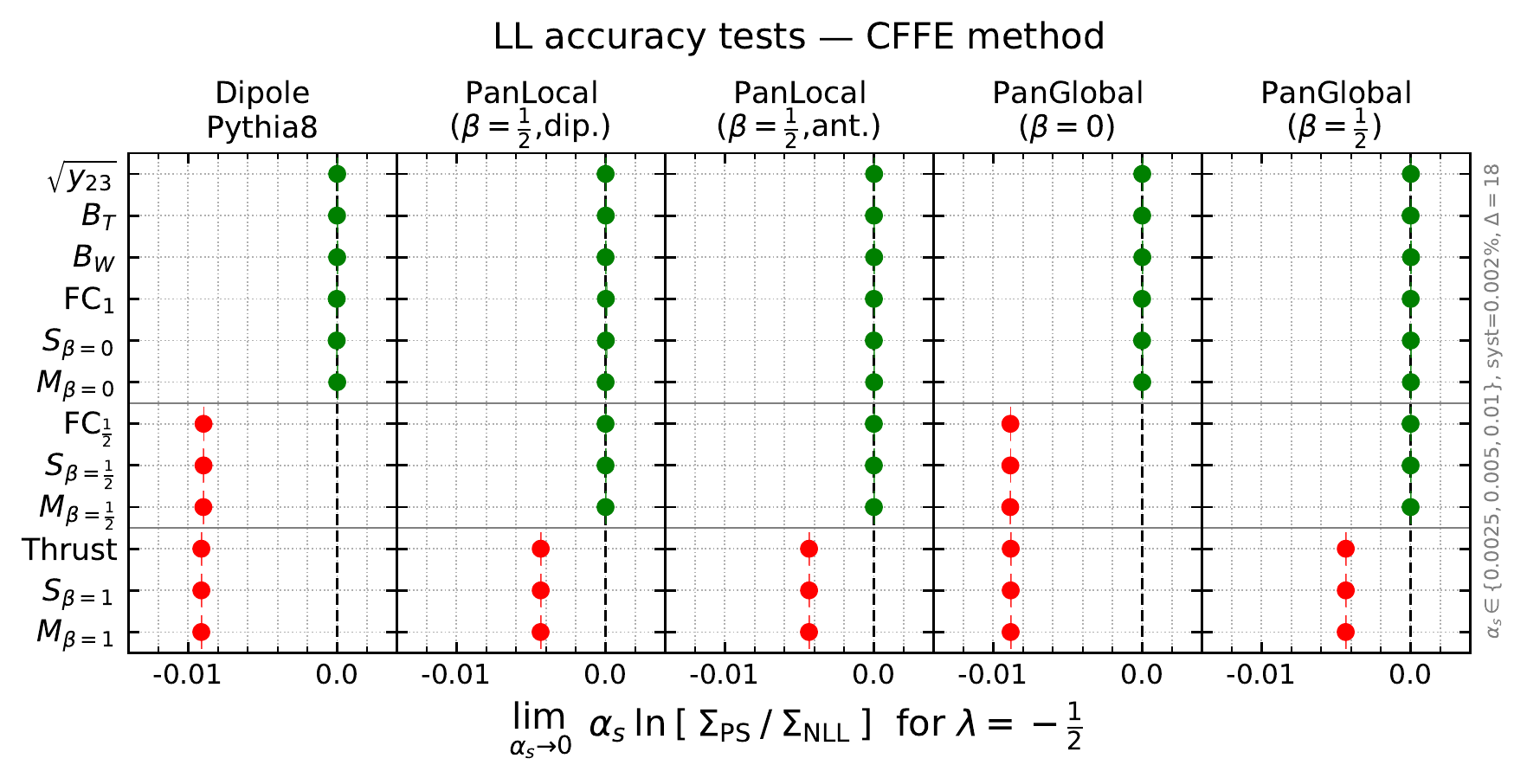}
}
For all showers that were in agreement with \NLL predictions at \NLLLC
in Ref.~\cite{Dasgupta:2020fwr}, we now see that these two new colour
schemes ensure agreement with the \NLLFC predictions.
This is to be expected, because, beyond the recoil issues that were
relevant for \NLLLC accuracy, \NLLFC requires the correct treatment of
the colour factor only for emissions that are widely separated in
rapidity, which the segment and NODS schemes both accomplish
by design.

Note that we have only tested event-shape observables that vanish in
the $2$-jet limit.
One could also envisage testing event shapes such as the thrust minor
and $D$-parameter, which vanish in the limit of 3 narrow jets, and
whose \NLLFC resummations have long been
known~\cite{Banfi:2000si,Banfi:2001pb} for planar events.
Our expectation is that the NODS scheme (but not the segment scheme)
will yield the correct \NLLFC results also for these observables, as
long as the appropriate matching is included for the 3-jet matrix
element, and the segment variables and auxiliary momenta are properly
initialised.
%

\subsection{Energy flow in a rapidity slice}
\label{sec:NGL}

The final observable that we consider is the probability,
$\Sigma(\as,L)$, for the energy in a given central slice of rapidity
to be less than $e^{-|L|} Q$.
Such an observable is of interest because its resummation involves
non-global logarithms, single-logarithmic terms $\as^n L^n$ that involve configurations
with an arbitrary number of (soft, large-angle) gluons in the
neighbourhood of the slice~\cite{Dasgupta:2001sh,Dasgupta:2002bw} (see
also Ref.~\cite{Banfi:2002hw}).
The full-colour resummation for such observables is sensitive to
arbitrarily complex colour correlators, both in the real emissions and
the virtual corrections, which need to be evaluated at amplitude
level.
The resulting subleading-colour single-logarithmic corrections go far beyond the
scope of the colour schemes that we introduced in
sections~\ref{sec:transition-points} and \ref{sec:ME-solution}.
In particular, we expect the segment scheme to be correct at full
colour only up to order $\as L$, and the NODS
scheme to be correct at full colour up to order $\as^2 L^2$.
Recall that leading-colour all-order single-logarithmic accuracy for
PanScales showers was demonstrated in Ref.~\cite{Dasgupta:2020fwr}.

The question we ask in this section is to what extent our schemes
differ from a full-colour computation.
Such a calculation was performed for the energy flow in a slice for
the $Z \to \qbar q$ process, by Hatta and Ueda~\cite{Hatta:2013iba}.
Their approach was based on a refinement of a proposal by
Weigert~\cite{Weigert:2003mm}, whereby the problem was reduced to the
simulation of associated Langevin dynamics in the space of Wilson
lines, and solved on a two-dimensional angular grid.
It was also applied,
with Hagiwara, to the calculation of a hemisphere
observable~\cite{Hagiwara:2015bia} and, very recently, to $H\to gg$
decays and a number of $2\to2$ scattering
processes~\cite{Hatta:2020wre}. 
The approach was formulated in such a way that it included only the
single logarithms, $\as^n L^n$.
Accordingly, in using it as a reference to which to compare our
$\as \to 0$ limit, we can be sure that any difference is exclusively
associated with subleading colour effects.

Two other calculations are based on parton showers at finite $\as$ and
a truncation of the $1/\nc^2$ series: Nagy and Soper examined rapidity
gaps between dijets at hadron
colliders~\cite{Nagy:2019pjp,Nagy:2019rwb}, while De Angelis, Forshaw
and Pl\"atzer examined the energy in a slice for the $Z \to \qbar q$
and $H \to gg$ processes~\cite{DeAngelis:2020rvq}.
Only the processes examined in that latter paper are within the scope
of our work here, but their simulation at finite $\as$ precludes
a meaningful direct comparison, because it would be impossible to know
whether any differences between their results and ours are associated
with subleading-colour effects or instead subleading logarithmic
($\as^n L^{n-1}$) effects.

The specific definition of the energy in a rapidity slice is as
follows: we cluster each event with the $e^+e^-$ Cambridge
algorithm~\cite{Dokshitzer:1997in} with the $y_\text{cut}$ parameter
set to $1$ and then undo one step of the resulting clustering sequence
to obtain two back-to-back jets.
Defining rapidity with respect to those two jet axes, we examine the
total energy contained in the rapidity region $|y| < -\ln \tan
\theta_\text{cut}/2$, with $\theta_\text{cut}$ setting the boundary of
the slice.
We will make the choice $\theta_\text{cut} = \pi/3$, which coincides
with that of Hatta and Ueda~\cite{Hatta:2013iba}.
As in section \ref{sec:evshp-NLL}, if we carry out a full run of the
shower in the $\as \to 0$ limit for fixed $\as L$, parton
multiplicities will grow too large to handle.
Here, we solve this problem by vetoing emissions whose rapidity with
respect to the closer end of the emitting dipole is larger than
$\eta_{\max}$, which by default we take to be $10$ (we will verify
that reducing it to $8$ does not modify the results).

It is commonplace in studies of non-global logarithms to show the
results as a function of
\begin{equation}
  \label{eq:HU-tau-def}
  \tau(\as,L)
  = \int_{Qe^{-|L|}}^Q \frac{dk_t}{k_t} \frac{\as(k_t)}{\pi}
  = \int_L^0 \frac{d\ell}\pi \frac{\as(Q)}{1 + 2b_0 \as(Q) \ell}
  = -\frac{1}{2\pi b_0} \ln \left(1 + 2b_0 \as(Q) L\right)\,,
\end{equation}
with $b_0 = \frac{11C_A - 2n_f}{12\pi}$ and where the result on the
right-hand side is given for a one-loop running, which is the only
contribution that survives in the limit $\as \to 0$ at fixed
$\lambda = \as L$.
Values of $\tau$ for various momentum ranges are shown in
table~\ref{tab:xi-lambda-tau} and, at the LHC, the largest accessible
value is $\tau \simeq 0.4$.

\begin{figure}
  \centering
  \begin{subfigure}{0.48\textwidth}
  \includegraphics[width=\textwidth,page=1]{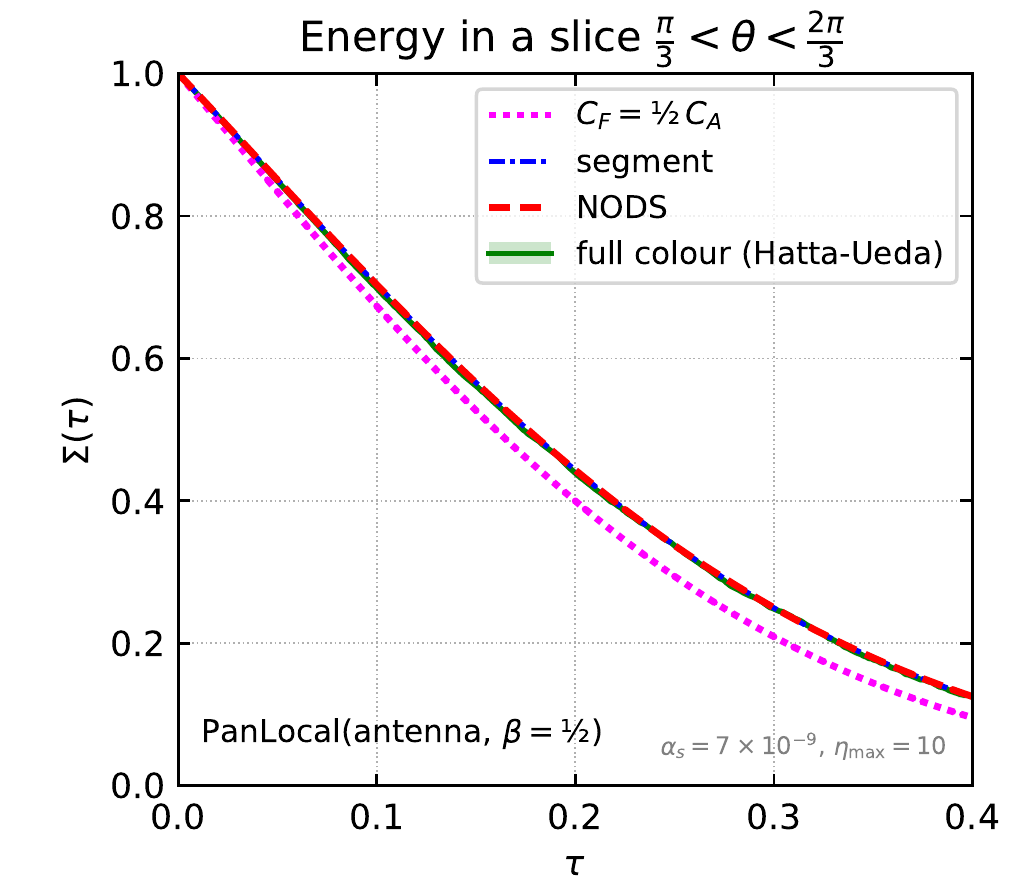}
    \caption{}
    \label{fig:ngl-qqbar-raw}
  \end{subfigure}%
  \begin{subfigure}{0.48\textwidth}
  \includegraphics[width=\textwidth,page=2]{plots/rapidity-slice.pdf}
    \caption{}
    \label{fig:ngl-qqbar-ratio}
  \end{subfigure}
  \caption{
    NLL (single-logarithmic) tests for a non-global observable.
    (a) Fraction of events whose energy flow in a central slice of
    rapidity is less than $e^{-|L|} Q$, shown in the limit $\as \to 0$
    for fixed $\as L$, as a function of $\tau(\as,
    L)$, defined in Eq.~(\ref{eq:HU-tau-def}).
    Our results are shown for the PanScales antenna shower with
    $\beta_\PS=1/2$, with three different 
    colour schemes: leading-$\nc$ (with $C_F=C_A/2=3/2$), segment and
    NODS.
    They are compared to the full-colour Hatta-Ueda
    (``finite-$\nc$ (exact)'')
    result~\cite{Hatta:2013iba}.
    (b) Ratio of the same set of results to the NODS result,
    illustrating apparent consistency of the segment and NODS
    schemes with the Hatta-Ueda result, to within its statistical
    uncertainty.
    The agreement is potentially surprising given that our schemes do not
    achieve \NLLFC ($\as^n L^n$) accuracy for non-global observables.
    The thin band for our results represents  the statistical
    uncertainty added in quadrature to estimates of systematics
    obtained using the difference between our default runs
    ($\eta_{\max}=10$ and $\as = 0.7\times10^{-8}$) and runs with
    $\eta_{\max}=8$ and $\as = 1.4\times10^{-8}$. 
    Our results for other showers with the same colour schemes are
    very similar, as is to be expected. 
    \logbook{e30cad4a}{for those other showers, see plots/zqq-slice-comparison.pdf} }
  \label{fig:ngl-qqbar}
\end{figure}

Fig.~\ref{fig:ngl-qqbar} (left) shows the fraction of events,
$\Sigma(\tau(\as,L))$, whose energy flow in the slice is less than
$e^{-|L|}Q$.
It includes several results: a leading-$\nc$ result with
$C_F = \frac12C_A = 3/2$, the Hatta-Ueda full-$\nc$
prediction,\footnote{We are very grateful to Hatta and Ueda for
  rerunning their code with higher statistics than in their original
  paper and providing us with the corresponding numerical results.}
and our predictions with the segment and NODS methods.
Recall that those methods are not expected to work beyond order
$\lambda$ and $\lambda^2$ respectively.
However in Fig.~\ref{fig:ngl-qqbar} (left) they are indistinguishable
from the full-$\nc$ Hatta-Ueda result.
To further probe this observation, the right hand plot shows ratios to
a reference, which we take to be the PanLocal-antenna $\beta=1/2$ \meas
(the specific choice is largely immaterial, since our aim is to
compare different predictions on this ratio plot).
One sees that the difference between the full-$\nc$ Hatta-Ueda result
and our leading-$\nc$ result is about $23\%$ at $\tau = 0.4$.
Remarkably, both our segment and \meas methods seem to be in good
agreement with the Hatta-Ueda result across the full range of $\tau$:
the whole range is within two standard deviations of the Hatta-Ueda result, and
in much of the range the agreement is within one standard deviation.
Some caution is needed in interpreting these results: firstly, they
correspond to one specific choice of slice size.
Secondly, when using a finite-resolution angular grid (as in the
Hatta-Ueda approach), there are inevitably some residual systematic
effects associated with that finite resolution,
and in this case we cannot exclude the possibility that they are
comparable to the statistical error.\footnote{We base this statement
  on a run of our PanScales showers, assigning each particle to a bin
  on an $80\times60$ grid in $\cos\theta$ and $\psi$ (for extremal
  $\cos\theta$ bins, we choose to map all $\psi$ values to a single
  bin).
  We then accept emissions only if they are in a distinct bin of the
  grid from both the emitter and spectator of the parent dipole.
  Comparing this to our standard continuum-limit runs, we see effects
  of the same order of magnitude as the Hatta-Ueda statistical
  uncertainty.}
A further observation is that the segment and \meas results come out
largely identical, even though the former (latter) is FC-accurate only to $\as
L$ ($\as^2 L^2$).
We attribute this to a partial cancellation that arises after
azimuthal integrations.
In particular, in Appendix~\ref{sec:segment-versus-nods-patches} we
demonstrate that if one breaks the azimuthal symmetry by considering a
patch in azimuth and rapidity instead of a slice (which covers all
azimuths), a difference between the segment and NODS scheme does
appear at the statistically significant level of a couple of percent.

A final comment concerns the same slice observable for $H\to gg$ events.
For a given value of $\tau$, both our segment and \meas methods are identical to the
leading-$\nc$ ($2C_F=C_A=3$) result in the $\as \to 0$, fixed
$\tau$ limit, because $g \to q\bar q$ branchings only affect NNLL
terms, $\as^n L^{n-1}$.
Subleading-$\nc$ effects start for the slice only from order $\as^3 L^3$
onwards, because soft-gluon emission from a $ggg$ system (i.e.\ $\as^2
L^2$) is given
exactly by the sum of emission from three $C_A/2$ dipoles.
Therefore we have the same formal subleading-$\nc$ accuracy as the
\meas method for $Z \to q\bar q$: exact up to $\as^2 L^2$, and only
leading-$\nc$ for terms $\as^3 L^3$ onwards.

Very recently, Hatta and Ueda have shown that an FC calculation for
the slice observable in $H\to gg$ is in good agreement with the
large-$\nc$ result, to within statistical errors~\cite{Hatta:2020wre}.
The findings of Ref.~\cite{DeAngelis:2020rvq} appear to be
consistent with this conclusion.\footnote{Specifically, in its
  Fig.~7a, the subleading colour curve (``all, $d=2$'') is in good
  agreement with the leading-colour curve
  (``$\text{LC}_\text{V+R}$''), after rescaling the latter by $8/9$ to
  account for $\nc^2-1$ rather than $\nc^2$ in the Born $H\to gg$
  colour sum, notably in the region
  $\rho \equiv e^{-|L|} \gtrsim 0.1$, where statistical fluctuations
  seem to be under control at the few percent level.}
Accordingly, the NODS and segment methods agree with the FC results
for $H\to gg$, to within the precision of the predictions for the
latter.

\subsection{Timing assessment}

\begin{figure}
  \centering
  \begin{subfigure}{0.49\textwidth}
    \includegraphics[width=\textwidth,page=2]{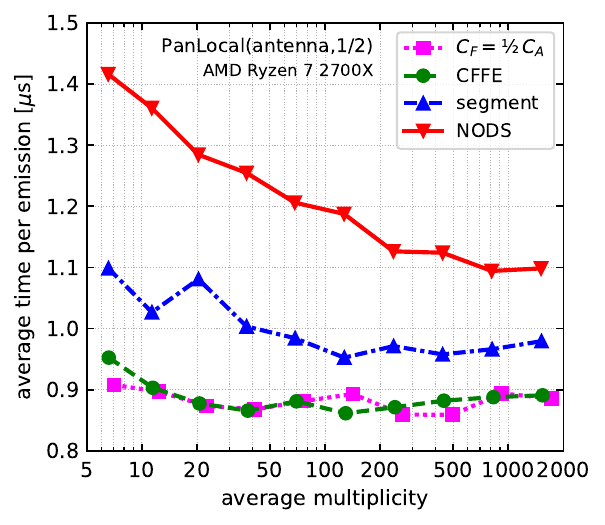}
    \caption{}
    \label{fig:timings-main}
  \end{subfigure}%
  \begin{subfigure}{0.49\textwidth}
    \includegraphics[width=\textwidth,page=5]{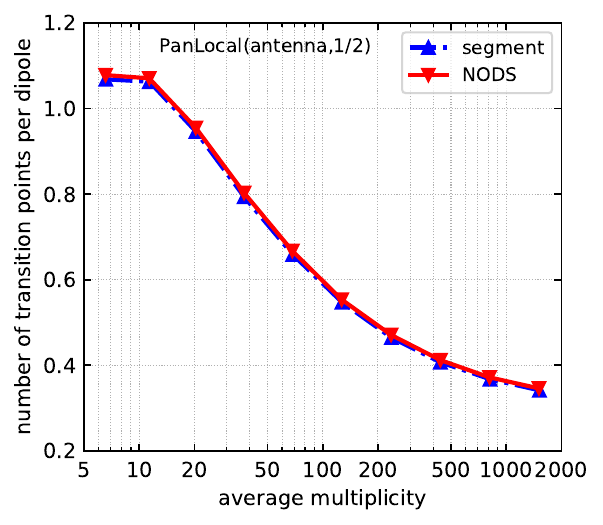}
    \caption{}
    \label{fig:timings-n-segment}
  \end{subfigure}
  \caption{(a) Time per emission plotted as a function
    of the average multiplicity, $N$, in the event sample, for different
    colour schemes.
    To maximise the relative impact of the colour scheme timing
    penalty, results are shown for one of the most optimised of our shower
    setups, the PanLocal antenna shower with $\beta=\frac{1}{2}$, run with
    fixed coupling.
    Additional details are given in the text.
    (b) The average number of transition points per dipole in the
    segment and NODS methods, as a function of the average
    multiplicity.  }\label{fig:timings}
\end{figure}

One of the motivations behind this work is to provide a method to
include logarithmically relevant subleading colour effects in parton
showers that is simple and computationally efficient.
In this section we show that the two methods highlighted in
sections~\ref{sec:transition-points} and \ref{sec:ME-solution} bring
only a modest time penalty.

To check this, we have run the PanLocal shower, in its antenna
variant, setting $\beta=\frac12$, using a fixed-coupling prescription,
$\alpha_s=0.1$, and varying the ordering variable ($\ln v$) cut-off scale
between ${-}24$ and ${-}10$.
At leading-$\nc$, the PanLocal antenna shower is currently one of the
fastest of our showers, and since the time penalty of the colour
schemes is largely independent of the shower, the choice of PanLocal-antenna provides a worst-case
estimate of the relative impact of the colour schemes.
Following a similar logic, we use a fixed coupling for these tests, because our running-coupling
shower implementation is not yet fully optimised, and so would produce
a misleadingly high timing baseline to which to compare the timing
penalties of the new colour schemes.

Fig.~\ref{fig:timings-main} shows the average event showering time
divided by the average number of emissions, for each value of the
$\ln v$ cutoff, plotted as a function of the average multiplicity $N$
for that $\ln v$.
It includes results for four colour schemes.
One sees that the timing per emission for the $C_F = C_A/2 = 3/2$ and
CFFE schemes is about $1\mus$, almost independently of $N$.
The penalty for the segment and \meas schemes is at most $0.2\mus$ and
$0.5\mus$ respectively.\footnote{The ratio on NODS versus shower
  timing is unexpectedly sensitive to the choice of CPU and system,
  and in some cases we have found that the NODS penalty could reach $100\%$ of the CFFE
  showering time, though it was still of the order of a $\mus$ per
  emission.}
It decreases with increasing $N$, which is probably a consequence of
the fact that the number of segments per dipole decreases with
increasing $N$, cf.\ Fig.~\ref{fig:timings-n-segment}.

\section{Conclusions}
\label{sec:concl}

In this work on subleading-colour effects in final-state dipole and
antenna showers, we have paid particular attention to the interplay
between colour and logarithmic accuracy.
Insofar as the accuracy of current parton showers is at best \NLL, we
argued that an essential requirement is subleading-colour accuracy for
\LL terms, with potentially missing subleading-$\nc$ \NLL contributions being on
the same footing as leading-colour \NNLL terms, which are currently
not available in parton showers.

In sections~\ref{sec:transition-points} and \ref{sec:ME-solution},
we outlined two schemes that are
simple and efficient to implement in a range of parton showers (including,
e.g.\ the Pythia~8 shower) and that provide full-colour \LL accuracy (\LLFC).
One of them, the segment method, varies the colour factor along a
dipole, using a Lund-diagram type classification to identify
regions that should have either a $C_F$ or a $C_A/2$ colour factor,
according to whether an angular-ordered picture implies emission from
a quark or a gluon.\footnote{We note that a similar angular-ordered classification
  could be of interest for kinematic maps.}
The other method, dubbed \meas, nests full-colour energy-ordered
double-soft matrix-element corrections for emissions from any dipole
that, according to the segment approach, contains at least one $C_F$
segment.

In practice it was possible to engineer our approaches so as to
provide full colour accuracy beyond the \LL (or \DL) approximation for
a range of observables: all global event-shape variables, as well as
observables that examine particle or jet multiplicities.
For these, both of our schemes achieve \NLLFC or \NDLFC accuracy, as
appropriate for the specific observable (aside from spin correlations,
which do not affect the observables studied here at NLL-FC accuracy,
and whose study we postpone to future work).
This involved care with $g \to q\bar q$ splittings and with the
specific locations of $C_F$ to $C_A/2$ transitions in the segment
method.
Numerically demonstrating the resulting accuracy relied on a range of
techniques for taking the $\as \to 0$ limit of our showers while
maintaining fixed $\as L$ or $\as L^2$, techniques that we have
improved relative to earlier work~\cite{Dasgupta:2020fwr}.

The one context in which we do not achieve \NLLFC accuracy is for
non-global observables (in our definition, \LL for these is zero, and
they start at NLL, i.e.\ $\as^n L^n$).
There we achieve full-colour accuracy only for a finite set of $n$:
for our segment method, $n\le 1$, and for our \meas method, $n \le 2$.
Remarkably, despite this limitation, our schemes still give good
numerical agreement with the full-colour \NLL results for the energy
in a slice, as obtained by Hatta and Ueda~\cite{Hatta:2013iba}, to
within the $1{-}3\%$ statistical error of the latter.
In future work, it would be interesting to gain a better understanding
of why this is the case.

One of the important points related to the computational efficiency of our colour
schemes, is that it makes it straightforward to obtain high
statistical accuracies.
The time penalty for the more sophisticated scheme, NODS,
was below a microsecond per emission.
That has enabled us to draw robust conclusions about non-trivial
sub-leading colour effects, which were typically at the level of a few
percent.
This provides a target for future work on subleading colour schemes,
which should straightforwardly be able to provide the high statistical
precision needed to conclusively determine the size of any subleading
colour effects that exist beyond those accessible in simple schemes
such as those developed here.

\section*{Acknowledgements}

We are very grateful to Yoshitaka Hatta and Takahiro Ueda for
providing their full-colour rapidity slice  predictions with new
higher-precision runs than in Ref.~\cite{Hatta:2013iba} and for
numerous discussions.
We wish to thank Bryan Webber for discussions about the relevance of
Lund diagrams and for sharing unpublished material in
Ref.~\cite{BryanUnpublished}.
GPS is grateful to Thomas Becher for discussions of subleading-colour
effects in non-global logarithms. 
Finally we are grateful to our PanScales collaborators
(Melissa van Beekveld, Mrinal Dasgupta, Frederic Dreyer, Basem
El-Menoufi, Silvia Ferrario Ravasio, Alexander Karlberg, Pier Monni,
Alba Soto Ontoso and Rob Verheyen), for their work on the code, the
underlying philosophy of the approach and comments on this
manuscript.

This work was supported
by a Royal Society Research Professorship
(RP$\backslash$R1$\backslash$180112) (GPS, LS),
by the European Research Council (ERC) under the European Union’s
Horizon 2020 research and innovation programme (grant agreement No.\
788223, PanScales) (KH, RM, GPS, GS) and
by the Science and Technology Facilities Council (STFC) under
grants ST/P000274/1 and ST/T000856/1 (KH) and ST/T000864/1 (GPS).

\appendix

\section{Matrix-element tests for non-angular ordered parent configurations}
\label{sec:ME-tests-non-ang-ordered}

The NODS scheme is not expected to reproduce the full-colour tree-level matrix
element for the emission of a gluon $g$ from configurations with two partons at
commensurate angles. In this appendix we study to what extent the NODS scheme
fails in such configurations, performing comparisons to the tree-level matrix
element for $\bar q g_1 g_2 q + g$.

We start by emitting the first gluon $g_1$ from the $\bar q q$ dipole as in
Eq.~(\ref{eq:me-setup-g1}):
\begin{equation}
z_{g_1} = 10^{-8},\quad \eta_{g_1} = 5,\quad \psi_{g_1} = \pi\,.
\label{eq:me-setup-g1-appendix}
\end{equation}
We emit a second
gluon $g_2$ close to the first gluon in
rapidity, $\eta_{g_2} \approx \eta_{g_1}$. Contrary to the angular-ordered case
presented in section~\ref{sec:ME-qqqqg}, there are now two routes that
contribute to the total rate of emission, i.e.\ gluon $g_2$ can be emitted
from either of the dipoles $(\bar q g_1)$ or $(g_1 q)$.
The two routes have distinct Jacobians relating the $g_2$
shower-generation phase-space to the Lund-diagram phase-space and this
must be accounted for.%
\footnote{In the angular-ordered case $\eta_{g_2} \gg \eta_{g_1}$, the
  relative weight for the emission of $g_2$ from $(\bar q g_1)$ is
  suppressed compared to that from $(g_1 q)$, such that we only needed
  to consider one route when performing the matrix-element tests for
  $2 \to 4+g$ configurations in section~\ref{sec:ME-qqqqg}.}

We solve numerically for the values of the generation variables $(\ln v_2,
\bar \eta_2, \phi_2)_i$, for each dipole $i \in \lbrace (\bar q g_1), (g_1 q)
\rbrace$, such that the gluon $g_2$ is emitted at:
\begin{equation}
z_{g_2} = 10^{-16},\quad \eta_{g_2} = \eta_{g_1} + \Delta \eta_{g_1g_2},\quad \psi_{g_2} = 0\,,
\label{eq:me-setup-g2-appendix}
\end{equation}
in the event centre-of-mass frame.
We evaluate the three-dimensional Jacobian determinant
\begin{equation}
J_i = \left| \frac{\partial (\ln v_2, \bar\eta_2, \phi_2)_i}{\partial
  (\ln E_{g_2}, \eta_{g_2}, \psi_{g_2})} \right|,
\end{equation}
numerically as well.
Starting from the parent configuration $\bar q g_1 g_2 q$,
where $g_2$ has been emitted from dipole $i$, we compute the rate of emission
$\sigma_\text{PS}^{(i)}$ of a further soft gluon $g$, and combine the
contributions from the two routes weighted by their relative Jacobian factors,
\begin{equation}
\sigma_{\text{PS}} = \sum_{i \in \lbrace (\bar q g_1), (g_1 q)
  \rbrace}
\frac{J_i}{J_{(\bar q g_1)} + J_{(g_1 q)}} \sigma_{\text{PS}}^{(i)}\,.
\end{equation}

\begin{figure}
\centering
  \includegraphics[width=.75\textwidth,page=2]{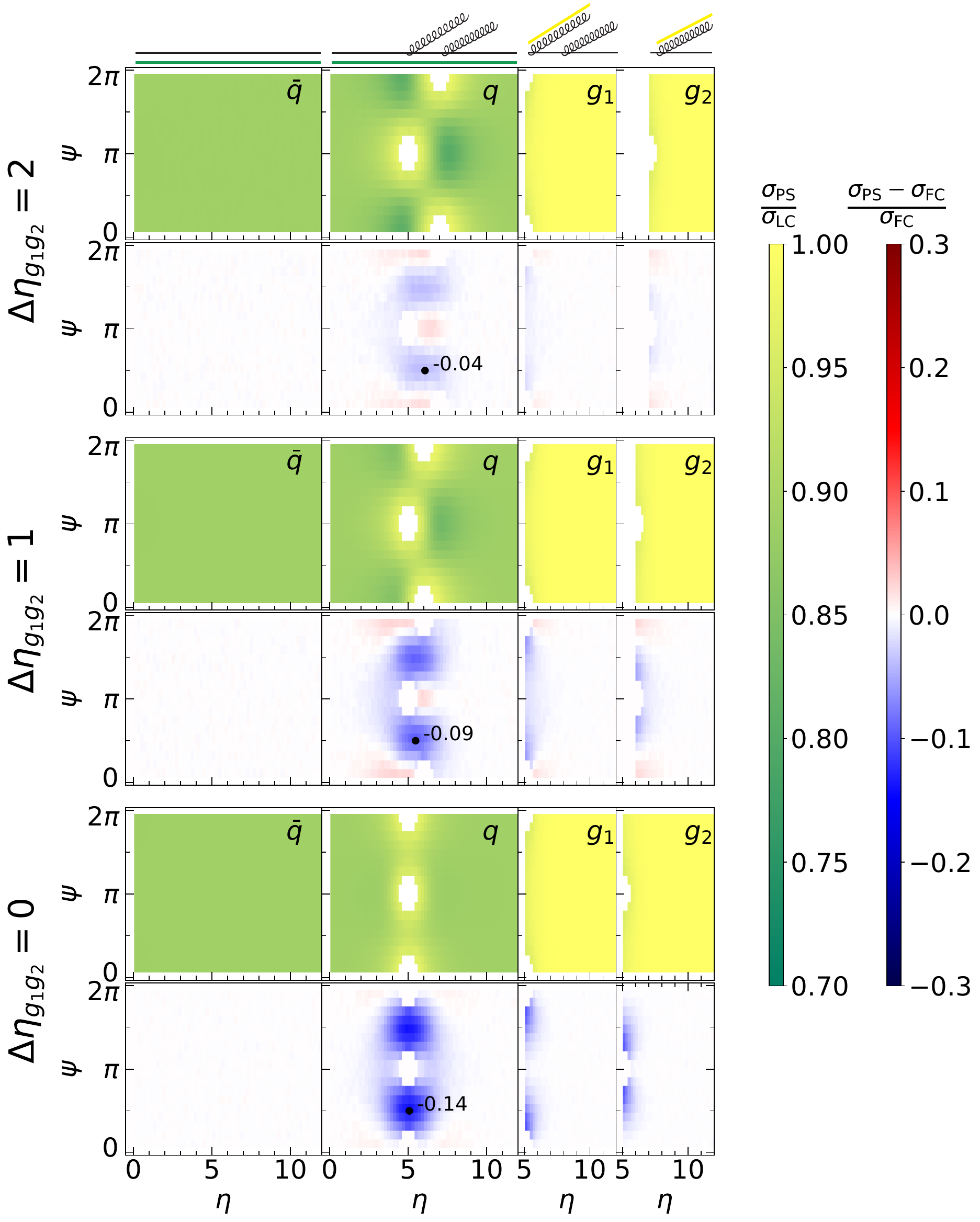}
  \caption{Results from the NODS scheme for the PanGlobal ($\beta = 0$) shower,
in an initial configuration $\bar q q g_1 g_2 + g$ where the two gluons $g_1$
and $g_2$ are not strongly angular-ordered. We show three configurations with
the gluon $g_2$ brought increasingly closer to $g_1$: for a fixed $\eta_{g_1} =
5$, $\psi_{g_1}=\pi$, the second gluon $g_2$ is placed at $\eta_{g_2} = 7$ (top), $\eta_{g_2} = 6$
(middle) and $\eta_{g_2} = 5$ (bottom) at an azimuthal angle $\psi_{g_2} = 0$.
Black dots convey the largest negative deviation between the NODS scheme and the
exact tree-level matrix element.
}
\label{fig:me-results-qqggg-commensurate}
\end{figure}

Results for the NODS scheme are shown in
Fig.~\ref{fig:me-results-qqggg-commensurate} for three values of the
separation of $g_1$ and $g_2$ in rapidity,
$\Delta \eta_{g_1 g_2} = \eta_{g_2} - \eta_{g_1} = \{0,1,2\}$.
The disagreement between the shower rate of emission, $\sigma_\text{PS}$,
and the full-colour tree-level matrix element, $\sigma_\text{FC}$, is observed
to be relatively small for $\Delta\eta_{g_1g_2} = 2$, where the largest negative
deviation is identified by a black dot and reaches $-4\%$ at $\psi_g = \pi/2$ on
the $q$-leaf.
As the second gluon $g_2$ moves to an angle close to that
of the first gluon $g_1$, the discrepancy becomes larger, reaching a value of
$-9\%$ for $\Delta\eta_{g_1g_2}=1$, and $-14\%$ for $\Delta \eta_{g_1g_2} = 0$.
Note that the regions where the matrix element is not reproduced are,
as expected, localised around kinematic configurations for which the
three emissions are at commensurate angles. These configurations are
beyond the scope of the accuracy we are aiming for.

The initial 4-parton $\qbar g_1 g_2 q$ configurations studied here differ from
the hadron-collider $2\to2$ processes such as those studied in
Ref.~\cite{Hatta:2020wre}, in that $g_1$ and $g_2$ are both soft.
Still, our observation that non-trivial subleading-colour effects
decrease as $g_1$ and $g_2$ become more separated (which is to be
expected based on coherence arguments), may be relevant also more
widely.
Specifically, they suggest that if one seeks to maximise
subleading-colour effects in hadron-collider processes, it may be of
interest to consider $2\to2$ configurations where the outgoing jets
are both at the same rapidity.
%

\section{Analytical results for CFFE subleading-$\nc$ effects in
  event shapes}
\label{sec:DL-evshp-analytics}

Here, we develop a (semi-)analytical approach to determining the
expected \DL and \LL deviations in event shapes for the CFFE colour
scheme as applied to the PanScales family of showers.
The intent is to help provide an independent validation and
understanding of some of the numerical results shown in
section~\ref{sec:evshp}.
The analysis is common to all of the PanLocal-dipole, PanLocal-antenna
and PanGlobal showers.
In contrast traditional dipole showers such as Pythia~8 are, we
believe, more complex to analyse and,
as can be seen from Fig.~\ref{fig:DL-evshp}, they clearly yield
different CFFE DL results.

\begin{figure}
  \centering
  \includegraphics[width=0.4\textwidth]{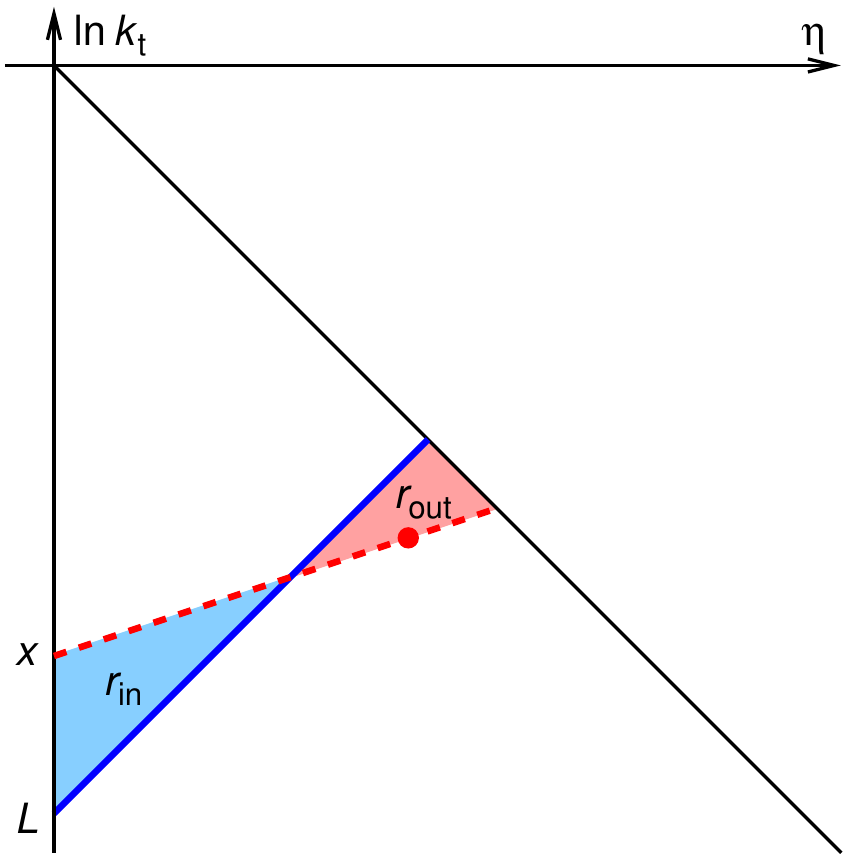}\\
  \caption{Primary Lund plane representation of the configuration
    represented in Eq.~(\ref{eq:cffe-shape-master}), with the solid
    line corresponding to a fixed limit $e^{-|L|}$ for the observable value and the
    dashed-line to a fixed value of the shower ordering variable.
    The emission indicated with a red dot has no direct impact on
    $\Sigma(\as, L)$, but for PanScales
    showers with the CFFE scheme, it changes the colour factor (from 
    $C_F$ to $C_A/2$) for all subsequent primary emissions at smaller
    (positive) rapidities and lower values of the shower ordering
    variable.
    This effectively modifies the Sudakov for $\Sigma(\as, L)$ in the
    region shaded in blue.
  }
  \label{fig:lund-cffe-shape}
\end{figure}

Let us consider an observable with
$\beta_\text{obs}>\beta_\text{PS}$ and compute $\Sigma_\text{PS}(L)$.
Whenever we have an emission at a scale $x=\ln v$ and rapidity $\eta$,
say with $\eta>0$, s.t.\ $x-\beta_\text{obs}\eta<L$, subsequent
emissions (real or virtual) at rapidities between 0 and $\eta$ will
have a factor $C_A/2$ instead of $C_F$.
This gives the following expression
\begin{equation}\label{eq:cffe-shape-master}
  \frac{\Sigma_\text{PS}}{\Sigma_\text{FC}}
  = \left[ e^{-C_Fr_\text{out}(L)} + 
  \int_L^{\frac{1+\beta_\text{PS}}{1+\beta_\text{obs}}L} dx\, C_F\,
  r_\text{out}'(x)
  e^{-C_Fr_\text{out}(x)-\left(\frac{C_A}{2}-C_F\right)r_\text{in}(x)}
  \right]^2,
\end{equation}
where the $r_\text{in}$ and $r_\text{out}$ Sudakov exponents
correspond to the light blue and pink regions in Fig.~\ref{fig:lund-cffe-shape}, respectively.
When writing this expression, the second term describes the case where
the first emission below the observable boundary is at a scale $\ln
v=x$. Compared to the expected result, one gets a veto for emissions
below the boundary with $\ln v>x$ --- the $r_\text{out}$ contribution
--- as well as a change $C_F\to\frac{C_A}{2}$ in the region
corresponding to $r_\text{in}$. (The $r_\text{out}'$ factor is minus
the derivative of $r_\text{out}$ with respect to $x$.)
The first term in the square bracket describes the situation where
there is no emission below the observable boundary that affects the
observable Sudakov.
Finally, the overall square accounts for both hemispheres.

The exponents $r_\text{in}$ and $r_\text{out}$ can be easily computed
including one-loop running coupling effects. Here we just give their
fixed-coupling expressions:
\begin{subequations}
  \begin{align}
  r_\text{out}(x) & \overset{\text{f.c.}}{=} \frac{\alpha_s}{\pi}\frac{(1+\beta_\text{obs})(1+\beta_\text{PS})}{\beta_\text{obs}-\beta_\text{PS}}\left(\frac{x}{1+\beta_\text{PS}}-\frac{L}{1+\beta_\text{obs}}\right)^2,\\
  r_\text{in}(x) & \overset{\text{f.c.}}{=} \frac{\alpha_s}{\pi}\frac{(L-x)^2}{\beta_\text{obs}-\beta_\text{PS}}.
  \end{align}
\end{subequations}
From Eq.~\eqref{eq:cffe-shape-master}, we can consider two
interesting limits. The first one corresponds to the study of \DL in
section~\ref{sec:evshp-DL}, where we fixed $\xi=\alpha_s L^2$ and take
the limit $\alpha_s\to 0$.
In this case, running-coupling effects in $r_\text{in}$ and
$r_\text{out}$ can be neglected as they only bring corrections
proportional to $\alpha_s L=-\sqrt{\alpha_s\xi}\to 0$.
In this case, Eq.~(\ref{eq:cffe-shape-master}) can be computed exactly
and one finds
\begin{equation}\label{eq:dla-cfee-analytic}
 \frac{\Sigma_\text{PS}}{\Sigma_\text{FC}}
  \overset{\text{fixed }\xi}{\underset{\alpha_s\to 0}{=}} \Phi^2,
\end{equation}
with
\begin{align}
 \Phi & =
  e^{-\frac{C_F}{\pi}\frac{\beta_\text{obs}-\beta_\text{PS}}{(1+\beta_\text{obs})(1+\beta_\text{PS})}\xi}
  +
        e^{-\frac{C_F}{\pi}\frac{(c-1)(\beta_\text{obs}-\beta_\text{PS})}{(1+\beta_\text{obs})(\beta_\text{obs}+(c-1)\beta_\text{PS}+c)}\xi}\\
  & \left[
    \Psi\left(-\sqrt{\frac{(1+\beta_\text{PS})(\beta_\text{obs}-\beta_\text{PS})(c-1)^2}{(\beta_\text{obs}+(c-1)\beta_\text{PS}+c)(1+\beta_\text{obs})^2}}\right)
    -\Psi\left(\sqrt{\frac{(\beta_\text{obs}-\beta_\text{PS})}{(1+\beta_\text{PS})(\beta_\text{obs}+(c-1)\beta_\text{PS}+c)}}\right)
  \right],
\nonumber
\end{align}
$c=C_A/(2C_F)$, and
\begin{equation}
\Psi(z) = \frac{1+\beta_\text{obs}}{\beta_\text{obs}+(c-1)\beta_\text{PS}+c}
  e^{-\frac{C_F}{\pi}z^2\xi} - (c-1)\frac{\sqrt{C_F(\beta_\text{obs}-\beta_\text{PS})(1+\beta_\text{PS})\xi}}{(\beta_\text{obs}+(c-1)\beta_\text{PS}+c)^{3/2}}\text{erf}\left(\sqrt{\frac{C_F\xi}{\pi}}z\right).
\end{equation}
\begin{figure}
  \centering
  \includegraphics[width=0.48\textwidth, page=1]{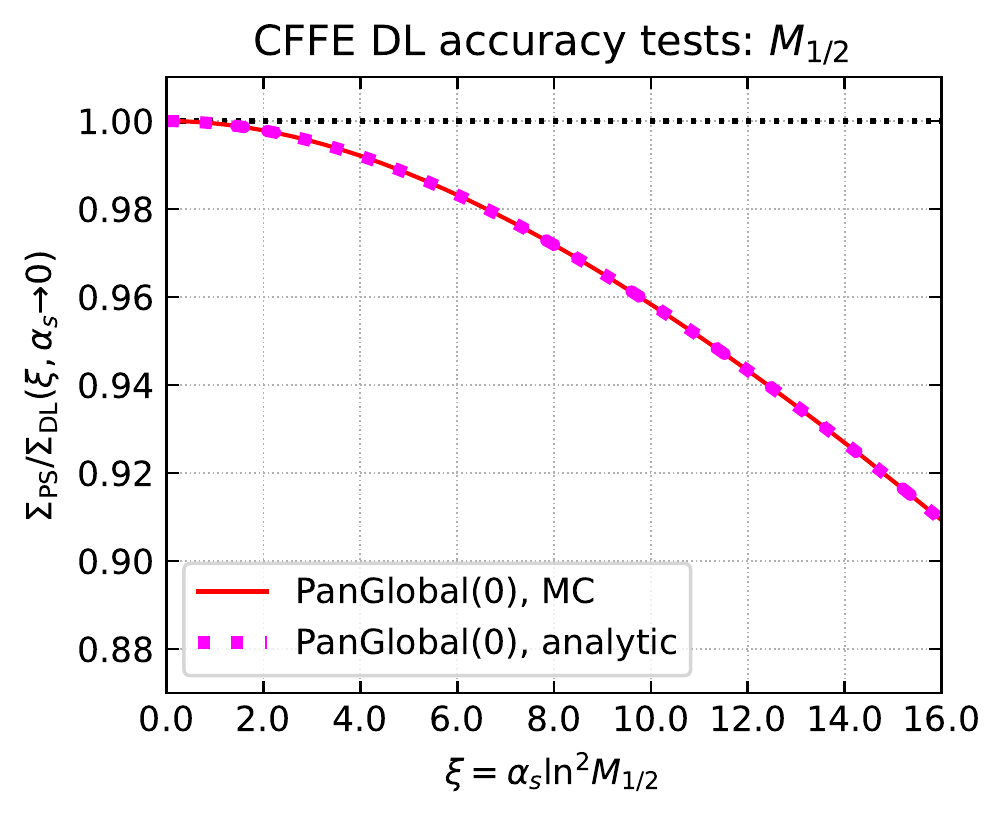}%
  \hfill
  \includegraphics[width=0.48\textwidth, page=2]{plots/dla-log-analytic.pdf}%
  \caption{Demonstration that our analytical calculation of the DL
    discrepancy for the CFFE colour scheme, Eq.~(\ref{eq:dla-cfee-analytic}),
    agrees with the full shower runs in
    the $\as \to 0$ limit, as a function of $\xi = \as L^2$, showing
    the $M_{1/2}$ (left) and $M_1$ (right) observables.}
  \label{fig:dl-cffe-analytic}
\end{figure}%
Eq.~\eqref{eq:dla-cfee-analytic} and the numerical results from
section~\ref{sec:evshp-DL} are compared
in Fig.~\ref{fig:dl-cffe-analytic}, illustrating the perfect
agreement between the two.

\begin{table}
  \centering
  \begin{tabular}{|c|c|c|c|}
    \hline
    &&
    \multicolumn{2}{|c|}{$\ln\Sigma_\text{PS}/\ln\Sigma_\text{FC} - 1$}\\
    $\beta_\text{PS}$ & $\beta_\text{obs}$ & running coupling ($\lambda=-0.5$) & fixed coupling\\
    \hline
     0  & 1/2 & 0.04749 & 0.03846 \\
     0  &  1  & 0.06832 & 0.05882 \\
    1/2 &  1  & 0.03590 & 0.02857 \\
    \hline
  \end{tabular}
  \caption{Value of the LL discrepancy for $\as \to0$ with fixed
    $\lambda = \as L=-0.5$ for various shower--observable combinations
    in the CFFE colour scheme.}\label{tab:LL-cffe-deviations}
\end{table}

The next limit we want to study is the one where we keep
$\lambda=\alpha_sL$ fixed and take $\alpha_s\to 0$, corresponding to
the studies presented in section~\ref{sec:evshp-NLL}.
Here, running-coupling effects can no longer be neglected.
We first change the integration variable in~(\ref{eq:cffe-shape-master})
to $\nu = \alpha_s x$, which should be integrated over a finite range
going between $\lambda$ and
$\frac{1+\beta_\text{PS}}{1+\beta_\text{obs}}\lambda$.
For fixed $\lambda$, the Sudakov exponents are proportional to $1/\alpha_s$ and so
become exponentially suppressed when $\alpha_s$ is taken to 0.
One can thus evaluate~(\ref{eq:cffe-shape-master}) in the saddle-point
approximation which becomes exact in the limit $\alpha_s\to 0$.
If one includes running-coupling effects, the saddle-point equation
can only be solved analytically for $\beta_\text{PS}=0$ and we have
used a numerical evaluation for  $\beta_\text{PS}>0$.
For simplicity, let us quote the analytic result obtained in the
fixed-coupling approximation, where we find
\begin{equation}\label{eq:LL-cffe-deviations}
  \frac{\ln\Sigma_\text{PS}}{\ln\Sigma_\text{FC}} - 1
  \overset{\text{fixed }\lambda}{\underset{\alpha_s\to 0}{=}}
  \frac{(C_A-2C_F)(\beta_\text{obs}-\beta_\text{PS})}{(1+\beta_\text{PS})C_A+2(\beta_\text{obs}-\beta_\text{PS})C_F}\,.
\end{equation}
We note that this result is equivalent to taking the limit
$\xi\to\infty$ in~(\ref{eq:dla-cfee-analytic}). In particular, it
reproduces Eq.~(\ref{eq:asymptotic-DL-CFFE-evshp}).
For completeness, we list in table~\ref{tab:LL-cffe-deviations} the
values of $\frac{\ln\Sigma_\text{PS}}{\ln\Sigma_\text{FC}} - 1$ with
fixed and running coupling.
%

\section{Segment versus NODS results for patches}
\label{sec:segment-versus-nods-patches}

\begin{figure}
  \centering
  \begin{subfigure}{0.49\textwidth}
    \includegraphics[width=\textwidth,page=2]{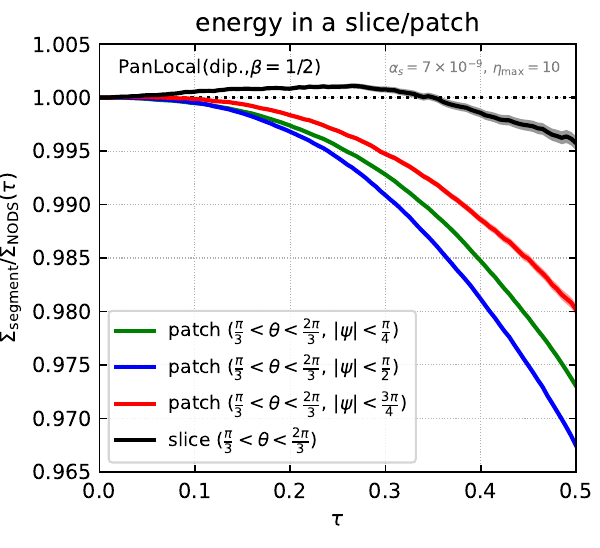}
    \caption{}
    \label{fig:patches-ratio}
  \end{subfigure}%
  \begin{subfigure}{0.49\textwidth}
    \includegraphics[width=\textwidth,page=7]{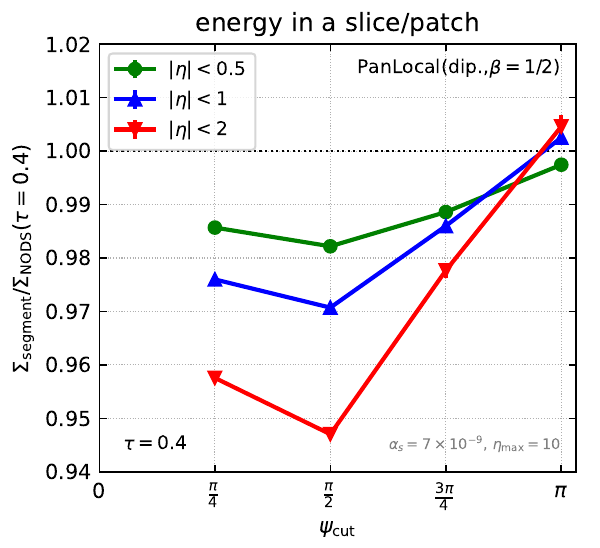}
    \caption{}
    \label{fig:patches-tau04}
  \end{subfigure}
  \caption{(a) Ratio between the segment and NODS methods for the
    energy deposited in a square patch of finite rapidity and
    azimuthal extent, as a function of $\tau$ defined in
    Eq.~(\ref{eq:HU-tau-def}). The bands correspond to the statistical
    uncertainty.
    (b) Same shown as a function of the azimuthal extent of the patch
    for different rapidity widths and for
    $\tau=0.4$.}\label{fig:patches}
\end{figure}

When comparing our subleading-colour results in section~\ref{sec:NGL}
for the energy in a rapidity slice to the full-colour ones obtained by
Hatta and Ueda~\cite{Hatta:2013iba}, the apparent agreement between
the segment and NODS methods might come as a surprise, given that the
NODS method reproduces the full-colour double real energy-ordered soft
matrix element, while the segment method does not.
The difference between the methods is clearly visible in our
matrix-element tests, see e.g.\ Fig.~\ref{fig:me-results-qgq}.
For the energy in a rapidity slice, this means that the segment method
should deviate from the full-$\nc$ result from
${\cal {O}}(\alpha_s^2L^2)$ onwards, one order earlier than the NODS
method for which deviations are expected from
${\cal {O}}(\alpha_s^3L^3)$.

One possible explanation for the similarity between the segment and
NODS methods for the rapidity slice observable is that the
${\cal {O}}(\alpha_s^2L^2)$ deviation in the segment method largely
disappears when integrating over the azimuthal angle of the two
emissions.
To test this hypothesis, we study the energy deposited in a
rectangular patch of both finite rapidity and azimuthal extent,
$|\eta|<\eta_\text{cut}=-\ln\tan\frac{\theta_\text{cut}}{2}$ and $|\psi|<\psi_\text{cut}$, thus breaking
the azimuthal symmetry of the slice.

Fig.~\ref{fig:patches-ratio} shows the ratio between the segment and
NODS methods for a patch of the same rapidity width as used in
section~\ref{sec:NGL} and different
azimuthal extent.
We see that while the difference between the two methods is smaller
than 0.5\% for the rapidity slice ($0.2\%$ at $\tau=0.4$), it can
reach a few percent for a 
rectangular patch with finite $\psi$ extent.
Among the patch extents we studied, $|\psi|<\frac{\pi}{2}$ showed the
largest effect.

Finally, Fig.~\ref{fig:patches-tau04} shows the ratio between the
segment and NODS schemes measured at $\tau=0.4$ for different rapidity
and azimuthal extents.
We notice that the deviation from 1 increases as the rapidity width of
the patch increases, reaching $\sim 5\%$ for a patch with
$|\eta|<2$ and $|\psi|<\frac{\pi}{2}$.
The results here suggest that for future comparisons of non-global
logarithms in methods implementing subleading-$\nc$ effects, it could
be desirable to study not just slices but also patches.

\bibliographystyle{JHEP}
\bibliography{MC}

\end{document}